\documentclass[%
 reprint,
nofootinbib,
 amsmath,amssymb,
 aps,
]{revtex4-2}

\usepackage{graphicx}
\usepackage{dcolumn}
\usepackage{bm}

\usepackage{listings}
\usepackage{tensor}
\usepackage{hyperref}
\usepackage{color,xcolor,colortbl}
\usepackage{soul}
\setstcolor{red}
\usepackage[normalem]{ulem}
\usepackage{orcidlink}
\usepackage[capitalise]{cleveref}

\newcommand{\mightyseven}{M87$^\ast$} 

\crefformat{equation}{Eq.~(#2#1#3)}
\crefformat{table}{Tab.~(#2#1#3)}
\crefformat{figure}{Fig.~(#2#1#3)}
\crefformat{appendix}{App.~(#2#1#3)}
\crefformat{section}{Sec.~(#2#1#3)}
\crefformat{subsection}{Subsec.~(#2#1#3)}

\begin{document}

\preprint{}

\title{A Lensing-Band Approach to Spacetime Constraints}

\author{Alejandro C\'ardenas-Avenda\~no\, \orcidlink{0000-0001-9528-1826}} 
\email{cardenas-avendano@princeton.edu}
\affiliation{
The Princeton Gravity Initiative, Jadwin Hall, Princeton University,
Princeton, New Jersey 08544, U.S.
}
\author{Aaron Held\,\orcidlink{0000-0003-2701-9361}}
\email{aaron.held@phys.ens.fr}
\affiliation{Institut de Physique Théorique Philippe Meyer, Laboratoire de Physique de l’\'Ecole normale sup\'erieure (ENS), Universit\'e PSL, CNRS, Sorbonne Universit\'e, Universit\'e Paris Cité, F-75005 Paris, France}

\begin{abstract}
General relativity’s prediction that all black holes are described by the Kerr metric, irrespective of their size, can now be empirically tested using 
electromagnetic observations of supermassive black holes and gravitational waves from mergers of stellar-mass black holes. 
In this work, we focus on the electromagnetic side of this test and quantify the constraining power of very-long-baseline-interferometry (VLBI) observations of emission generated by hot gas surrounding supermassive black holes.
We demonstrate how to use lensing bands---annular regions on the observer’s screen surrounding the critical curve---to constrain the underlying spacetime geometry. Contingent upon a detection of a lensed VLBI feature, the resulting lensing-band framework allows us to exclude spacetimes for which said feature cannot arise from geodesics that traversed the equatorial plane more than once.
Focusing on the first indirect image and tests of black-hole uniqueness, we employ a parametrized spacetime as a case study.
We find that resolving geometric information that goes beyond the apparent size of the critical curve has the potential to lift degeneracies between different spacetime parameters. 
Our work thereby quantifies a conservative estimate of the constraining power of VLBI measurements and contributes to a larger effort to simultaneously constrain geometry and astrophysics.
\end{abstract}

\maketitle
\interfootnotelinepenalty=10000

\section{Introduction}
\label{sec:intro}

Do all astrophysical black holes source (up to rescaling) the same exterior gravitational field?
In general relativity (GR), the answer is `yes': uniqueness theorems imply that (i) the vacuum exterior of any spherically symmetric source is given by the Schwarzschild solution~\cite{2005GReGr..37.2253J, 1923rmp..book.....B}; (ii) the vacuum exterior of a stationary axisymmetric (as well as uncharged and non-degenerate) black-hole horizon is given by the subextremal Kerr solution~\cite{Robinson:1975bv}; and (iii) the leading-order asymptotics of any stationary and axisymmetric source are unique and agree with the asymptotics of the Kerr solution~\cite{1995CQGra..12..149K}. Taken together, these three theorems guarantee that the vacuum exterior of stationary solutions in GR reproduces the Newtonian limit, allow us to uniquely identify the (asymptotic quantities) mass~$M$ and angular momentum~$J$, and fix the first parametrized post-Newtonian (PPN) parameters to~$\beta_\text{GR}=1$ and~$\gamma_\text{GR}=1$~\cite{Nordtvedt:1968qs,Will:1971zzb,1993tegp.book.....W}.
Physically, the latter two PPN parameters, respectively, quantify the spatial curvature which is ``generated'' per unit rest mass and the degree of ``nonlinearity'' arising in the superposition law for gravity~\cite{Will:2014kxa}.

\begin{figure}
    \centering
    \includegraphics[width=\linewidth]{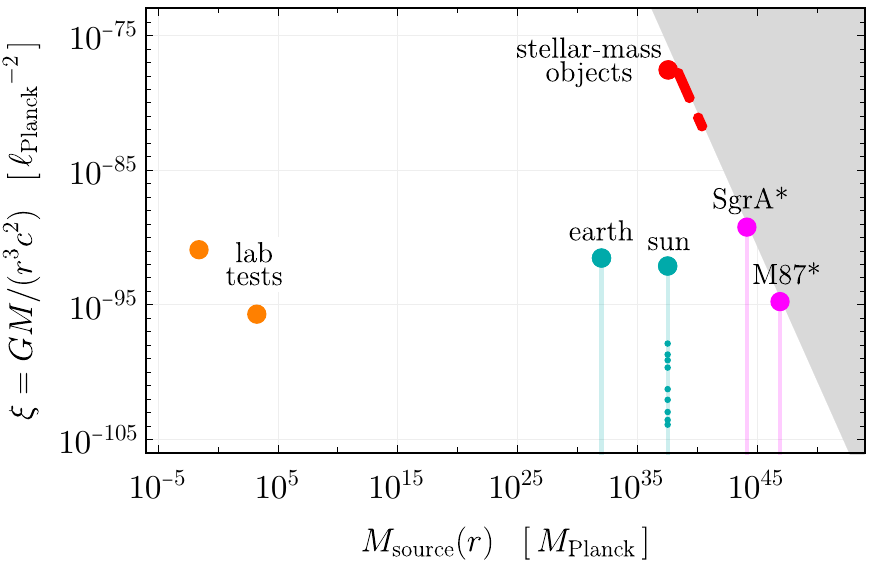}
    \caption{Different types of observations classified by their source mass~$M_\text{source}$ (in units of Planck mass~$M_\text{Planck}$) and their curvature scale~$\xi$ (in units of inverse Planck area~${\ell_\text{Planck}}^{-2}$). The boundary of the gray region indicates the formation of a horizon (according to GR). The vertical lines indicate increasing radial distance to the source, along which light rays and test masses (small dots denote the planets of the Solar System and Pluto) probe the background geometry.
    }
    \label{fig:scales}
\end{figure}

Any observation of a negligibly small test mass in a stationary background spacetime (sourced by a significantly larger mass) puts the assumptions of these uniqueness theorems to the test. But in which regimes, see~Fig.~\ref{fig:scales} as well as Ref.~\cite{Baker:2014zba}, might we find deviations?
On the one hand, we can probe uniqueness at large curvature scales~$\xi\equiv GM/\left(r^3 c^2\right)$: in this case, stellar-mass black holes or neutron stars are the prime astrophysical targets.
On the other hand, we can probe black-hole uniqueness at varying source mass~$M_\text{source}$, whereby we refer to an appropriate notion of mass in a stationary axisymmetric spacetime~\cite{arnowitt1962dynamics,Brown:1992br}. Laboratory experiments currently push the low-$M_\text{source}$ frontier at sub-milligram mass scales~\cite{Geraci:2008hb,Westphal:2020okx} (shown on the far left of Fig.~\ref{fig:scales}). At large $M_\text{source}$, electromagnetic observations of active galactic nuclei (AGN) through, for example, radio very-long baseline interferometry (VLBI)~\cite{EventHorizonTelescope:2019dse,EventHorizonTelescope:2022xnr}, X-rays (see, e.g., Refs.~\cite{Bambi:2016sac,Bambi:2022dtw}), or the near-infrared (see, e.g., Refs.~\cite{Gravity1,GRAVITY:2023cjt}), are currently the only way to explore the galactic regime at horizon scales. In the future, LISA will also be able to explore this regime in the gravitational wave sector~\cite{LISA:2022kgy}.

Extracting information about the underlying geometry from VLBI data is an active area of research and progress is happening fast, see, for example Ref.~\cite{Perlick:2021aok} for a recent review of analytical studies.
The image of a black hole is formed by incident null geodesics on an observer's screen. The characteristic image features are tied to the presence of a horizon \emph{and} the presence of an unstable photon sphere or, more generally, an extended photon shell, see Refs.~\cite{2003GReGr..35.1909T,Perlick2004,Johnson:2019ljv} for its description in the Kerr spacetime. The photon shell can be defined as a spatial region in which closed photon orbits are possible. 
Incident null geodesics on the observer's screen can then be divided into two classes: an exterior image region in which incident geodesics originate from radial asymptotic infinity and an interior image region in which geodesics originate from the horizon. The division is demarcated by the apparent shadow boundary~\cite{Bardeen:1973tla} or, henceforth, critical curve~\cite{Gralla:2019xty}. On either side, the critical curve is surrounded by an infinite series of exponentially stacked annular lensed images of the entire spacetime region interior/exterior to the photon shell~\cite{Darwin1959,Luminet:1979nyg,ohanian1987black,Johnson:2019ljv}.

Since, by definition, no light can escape the horizon, one may intuitively expect a central brightness depression, commonly referred to as ``the shadow.'' The critical curve also contains the direct image of the event horizon itself --- the ``inner shadow''~\cite{Dokuchaev:2018kzk,Dokuchaev:2020wqk,Chael:2021rjo}. In realistic observations, both, ``the shadow'' and even ``the inner shadow'' are not expected to be completely dark due to astrophysical foreground emission. Hence, the distinction between the presence and absence of the horizon is nontrivial~\cite{Vincent:2020dij,EventHorizonTelescope:2022xqj,Eichhorn:2022fcl}.

Going beyond the central brightness depression, the apparent size of the critical curve, and the surrounding exponentially stacked lensed images, have been used to infer a quantitative imprint of the geometry~\cite{EventHorizonTelescope:2019dse,EventHorizonTelescope:2022wkp}. The quantitative precision of this inference is subject to astrophysical uncertainty and has been scrutinized in various studies~\cite{Gralla:2019xty,Gralla2021, Volkel:2020xlc,Bauer:2021atk,Broderick_2022,Lockhart:2022rui}.

The above issue of disentangling geometry and astrophysics (at least in part) arises from the following dichotomy. On the one hand, the critical curve itself depends only on the geometry~\cite{Bardeen:1973tla}. On the other hand, the full observable image will clearly depend on the astrophysical source. Subsequent lensed images approximate the critical curve exponentially well but carry some residual astrophysical dependence~\cite{Gralla:2020srx}. Recent progress towards quantifying this dependence is based on a precise theoretical characterization of gravitational lensing in the Kerr spacetime~\cite{Luminet:1979nyg,ohanian1987black,Gralla:2019ceu,Johnson:2019ljv}. We will refer to the resulting definition of an ($n$-fold) lensed image as the $n$-th photon ring, where the indexing $n+1$ is counting the maximum number of times an emitted photon that arrives to the observer plane has crossed the equatorial plane. Thus, for example, $n=0$ corresponds to the direct (lensed) image, i.e., photons that arrive to the observer's screen and have crossed the equatorial plane at most once.

In particular, this has led to a proposed null test of GR, capable to achieve a sub-percent level of accuracy~\cite{Gralla:2020srx}. The proposed test consists of measuring the shape of the second photon ring ($n=2$), which is (exponentially) close to critical curve, on multiple baseline angles and check whether or not it follows the GR prediction for the Kerr metric~\cite{Gralla:2020yvo}. The level of accuracy can be achieved due to the small width-to-diameter ratio of the ($n=2$) photon ring. This null test works remarkably well even for moderate inclinations ($i\lesssim45^\circ$)~\cite{Paugnat:2022qzy}.

On the other hand, the first ($n=1$) photon ring has a larger width-to-diameter ratio than subsequent rings, and consequently the definition of its diameter in the image plane is ambiguous. Nevertheless, the first photon ring \emph{does} admit an angle-dependent diameter in \emph{visibility space}~\cite{Cardenas-Avendano:2023dzo}. In other words, GR predicts a particular functional form for the angle-dependent diameter for the critical curve that, surprisingly, provides an interferometric ring diameter of the first photon ring close, to a few  percent, to the critical curve. Therefore, detections of orbiting ($n\geq1$) photons provide a window into strong gravity and enable higher-precision probes of the Kerr geometry at large-$M_\text{source}$.

Previous works beyond GR have mostly relied on the assumption of black-hole uniqueness and have then explored deviations which are, nonetheless, unconstrained by other observations (see, for example, Refs.~\cite{EventHorizonTelescope:2020qrl,Younsi:2021dxe, Ayzenberg:2022twz, Staelens:2023jgr}). This assumption effectively collapses the entire parameter space in Fig.~\ref{fig:scales} and assumes that gravity acts universally.
There are several reasons to, nevertheless, remain agnostic. First, GR itself produces event horizons which can be reached by either going to sufficiently large curvature (stellar-mass black holes) or to sufficiently large source mass (AGNs). Similarly, deviations from GR could be tied to nonlinearities~\cite{Babichev:2013usa} or even to nonlocal quantities. Second, theories beyond GR can have multiple (stable) branches of black-hole solutions, see e.g., Refs.~\cite{Doneva:2017bvd,Held:2022abx}.
Different mass scales may be prone to end up in one or another branch. Indeed, the formation history of AGNs is not well understood~\cite{Fabian:2012xr,Volonteri:2021sfo}. Third, and maybe most importantly, whenever we are granted new observational opportunities, it is good practice to best not be guided by any theoretical prejudice.

Given the above status of the field, our work is motivated by three objectives.
Our first objective is to provide a generic framework to constrain spacetime parameters, including the ones related to black-hole uniqueness, with VLBI observations.
Such uniqueness tests for AGN have been recently been considered in spherical symmetry in Ref.~\cite{EventHorizonTelescope:2022urf}. Here, we provide the means to extend them beyond spherical symmetry. Our second objective is to quantify the constraining power of the first indirect image beyond null tests~\cite{Cardenas-Avendano:2023dzo}, and extend this work to deviations from the Kerr paradigm. Our third objective is to disentangle the above constraints on the underlying spacetime geometry from uncertainty about the astrophysical accretion process. Previous studies have attempted to tackle this issue by large computation-intensive joint inference~\cite{Lara:2021zth,Nampalliwar:2021oqr,Kocherlakota:2022jnz,Ayzenberg:2022twz,Nampalliwar:2022smp}, by calibration factors~\cite{EventHorizonTelescope:2020qrl,EventHorizonTelescope:2021dqv,Younsi:2021dxe, EventHorizonTelescope:2022xqj,Vagnozzi:2022moj,Salehi:2023eqy},
or by means of the second lensing band~\cite{Staelens:2023jgr}. Here, we extend the latter approach and propose a generic scheme which circumvents modelling or calibrating the impact of astrophysical uncertainties.

Focusing on geometric lensing bands~\cite{Paugnat:2022qzy,Cardenas-Avendano:2022csp}, annular regions on the observer’s screen surrounding the critical curve, allows us to progress on all three of the above objectives simultaneously. 
By extending previous works within GR~\cite{Gralla:2020srx,Cardenas-Avendano:2023dzo} and beyond GR~\cite{Wielgus:2021peu,Staelens:2023jgr}, we will show how lensing bands deform when parametrized deviations from the Kerr spacetime grow sufficiently large. Parameter regions that result in lensing bands which cannot contain an observed lensed image are thereby excluded. Upon confident detection of an associated VLBI feature, this translates to independent constraints of the respective deformation parameters, e.g., the leading PPN parameters $\beta_\text{M87}$ and $\gamma_\text{M87}$, for supermassive black holes. A comparison of, e.g., $(\beta_\text{M87},\gamma_\text{M87})$ and $(\beta_\text{GR},\gamma_\text{GR})$ then provides a test of the uniqueness of black holes at up-to-now unexplored source-mass scales. 

The rest of this paper is organized as follows. In Sec.~\ref{sec:framework}, we present the framework by reviewing the geometric concept of lensing bands (Subsec.~\ref{sec:lensing-bands}) and discussing some observational avenues to extract
a lensed emission region from VLBI measurements (Subsec.~\ref{sec:observing-lensed-emission-region}).
With these two inputs, we present a generic scheme to translate a confident observation of lensed emission in the image plane into constraints on the underlying geometry.
Since in the present work we are not using real data for the VLBI feature, we will refer to all quantitative statements as ``\emph{projected constraints}.'' Our results aim to quantify the capability of VLBI observations to impose constraints. As a proof of concept, we then focus on VLBI observations of \mightyseven~\cite{EventHorizonTelescope:2019dse,Broderick_2022}.
In Sec.~\ref{sec:Kerr}, we benchmark the framework by recovering the Kerr parameters (mass and spin) and the inclination to the source.
In Sec.~\ref{sec:parametric} we apply the lensing-band framework to a parametric deviation of the Kerr metric, whose parameters are associated with black-hole non-uniqueness. \emph{Projected constraints} on specific spacetimes can be obtained by expressing said spacetime within the parametrization and comparing the respective coefficients.
We summarize our results and discuss the prospects of future work in Sec.~\ref{sec:discussion}. 
Throughout the paper, we use geometric units in which $G_{\rm N}=c=1$, and the $(-,+,+,+)$ metric signature.

\section{The lensing-band framework}
\label{sec:framework}

The main motivation behind our work is to constrain spacetime geometries based on an inconsistency with an observed ``($n$-fold) lensed'' image of an astrophysical emission source. In this section, we will provide the detailed definitions and ingredients of the method.

The proposed lensing band framework requires: (i) a precise (geometric) definition of what is meant by the ``($n$-fold) lensed'' image; (ii) 
a confident detection of a persistent VLBI ``feature'';
and (iii) the assumption that this persistent VLBI feature can be explained as the defined ($n$-fold) lensed image.

Given the definition of a ``lensed image'' that we choose to work with (see below), the precise exclusion statement, which we explore quantitatively in the subsequent sections, can be expressed as follows:
\begin{quote}\textit{
    Upon a confident detection of a persistent VLBI feature, we exclude spacetimes for which this feature cannot arise from geodesics that traversed the equatorial plane more than once. 
}\end{quote}
We expect it to be straightforward to apply our framework to any other definition of a ``lensed image'' and to any other assumed/observed VLBI feature.

If one defines the ``photon ring'' as the lensed image associated with the above lensing-band definition, and assumes that a VLBI measurement has detected said photon ring, then the above statement can be rephrased as follows:
\begin{quote}\textit{
    We identify spacetimes which are excluded by a confident detection of the photon ring.
}\end{quote}
The resulting constraints are then rigorous and depend only on the level of confidence in the above input definitions and assumptions.

In Subsec.~\ref{sec:lensing-bands}, we review several available definitions of the term ``lensed'' image or ``($n$-fold) lensed'' image. We compare these definitions and, in particular, review the henceforth-used notion of equatorial lensing bands. In Subsec.\ref{sec:observing-lensed-emission-region}, we briefly review some observational avenues toward confidently detecting the associated VLBI feature. Relying only on these two inputs, in Subsec.~\ref{sec:scheme}, we present an algorithm for excluding parameter regions in families of spacetimes such as the Kerr spacetime.

\subsection{On the definition of a lensed image}
\label{sec:lensing-bands}

Lensing bands demarcate the region in an observers' image plane in which the ($n$-fold) lensed image of the emission can be incident. Clearly, the very definition of any such lensing bands rests on the underlying definition of what one precisely refers to as ``($n$-fold) lensed image.''

In general one may relate 
lensing to the deflection angle between the ingoing and the outgoing portion of the ray (see, for example, Ref.~\cite{Iyer:2009wa} for how to obtain the deflection angle in the Kerr spacetime). 

When there is a sufficiently compact central object, such as a black hole, geodesics can orbit the central object more than once---or infinitely many times in the presence of a photon shell. Therefore, it is desirable to have a general definition quantifying this $n$-fold orbital motion.
In addition, it is desirable to exploit the symmetries associated to stationarity and axisymmetry. Together with the assumption of reflection symmetry, this singles out the equatorial plane as a plane with respect to which orbital motion may be defined precisely.

Related concepts were discussed in Refs.~\cite{Gralla:2019xty,Johnson:2019ljv}, and a precise definition with respect to crossings of the equatorial plane was made rigorous in Refs.~\cite{Gralla:2019ceu,Gralla:2019drh}.

\subsubsection{Black hole lensing bands}
\label{sec:lensing-bands_def}
In any stationary and axisymmetric spacetime, the equatorial plane is singled out since it is perpendicular to the axis of symmetry. It is thus useful to tie a notion of a lensed image to the equatorial plane\footnote{Other definitions of what one may refer to as a geometric lensing band are possible. For instance, one may also count crossings (or turning points) with respect to other planes such as the one orthogonal to the observer's line of sight, see, for example the discussion around Fig.~1 in Ref.~\cite{Eichhorn:2022bbn}.}. The $n^\text{th}$-order lensing band may, therefore, be defined as the region in the image plane corresponding to geodesics that have crossed the equatorial plane up to $n+1$ times~\cite{Paugnat:2022qzy,Cardenas-Avendano:2022csp}. In Boyer--Lindquist coordinates $(t,r,\theta,\phi)$, this definition simply corresponds to counting how many times the condition $\theta=\pi/2$ (or equivalently $\cos\theta=0$) is met.
In the context of comparing to alternative definitions below, we will refer to the above as the ``crossing definition'' of lensing bands. Throughout the paper, whenever we use the term lensing bands without further specification, we refer to the ``crossing definition.''

In particular, the zeroth-order lensing band then corresponds to all light rays that reach the observer's screen with a cut-out portion corresponding to the apparent location of the event horizon in the observer's screen. Therefore, the zeroth-order lensing band covers the entire image plane apart from the direct image of the event horizon, commonly referred to as the inner shadow~\cite{Dokuchaev:2018kzk,Dokuchaev:2020wqk,Chael:2021rjo}, and corresponds to the direct, and therefore weakly lensed, image of the accretion structure. In contrast, already the first-order lensing band covers only a finite region within the image plane. Its inner (outer) boundary corresponds to rays originating from the location of the event horizon (from equatorial radial infinity) and arriving at the screen after crossing the equatorial plane once more. Each successive higher-order lensing band is contained within the former and, for $n\rightarrow\infty$, the $n^\text{th}$-order lensing band exponentially converges to the critical curve
~\cite{Bardeen:1973tla}.

\subsubsection{Alternative definitions}
\label{sec:alternative_definitions}
%
\begin{figure}
    \centering
    \includegraphics[width=\linewidth]{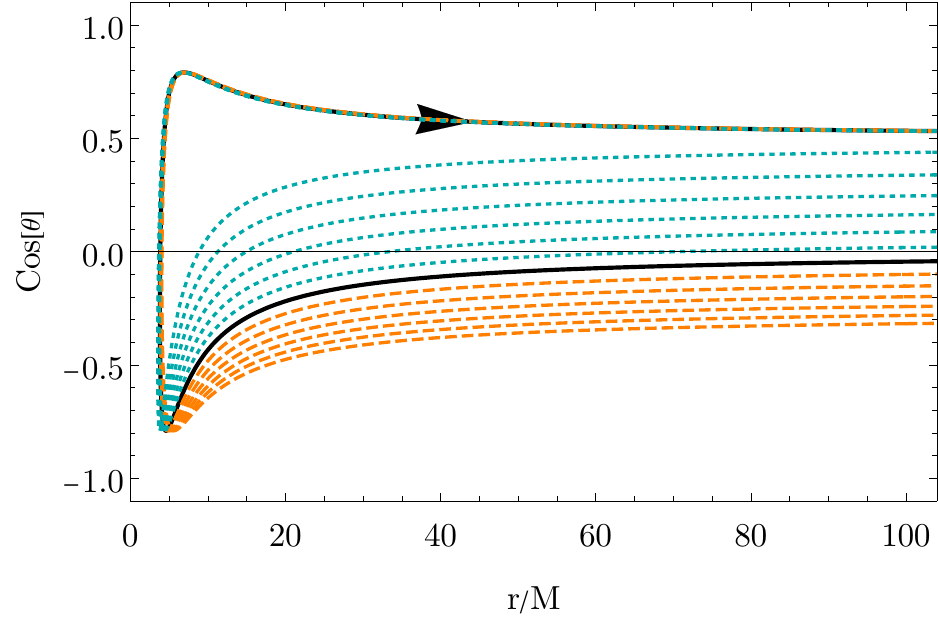}
    \caption{
    To exemplify the inequivalence of different geometric definitions of lensing bands, we show the Boyer--Lindquist coordinates $(r,\theta)$ of example trajectories in the Schwarzschild spacetime, viewed at inclination $i=17^{\circ}$ and for screen coordinates $(\alpha,\beta)=(x,x)$ in the vicinity of $x\approx 4\,M$. 
    All the shown trajectories have two turning points (i.e., $m=1$) and, therefore, are part of the first lensing band according to the turning-point definition. 
    In contrast, the trajectories that cross the equatorial plane only once (i.e., $n=0$), denoted with dashed orange lines, are not part of the first lensing band according to the equatorial-plane-crossing definition. Conversely, the trajectories that cross the equatorial plane twice (i.e., $n=1$), plotted with dotted cyan lines, are part of the first lensing band according to the equatorial-plane-crossing definition.
    The critical trajectory, for which $\cos(\theta)\rightarrow 0$ for $r\rightarrow\infty$, is marked as the black (continuous) curve, on which the arrow points towards the screen.
    }
    \label{fig:crossing-vs-turning}
\end{figure}
Several other (sometimes closely related) geometric definitions of lensing bands have been discussed in the literature, see, e.g., Refs.~\cite[Sec.~4.3]{Perlick2004}, \cite[Fig.~1]{Eichhorn:2022bbn}, or~\cite[App.~A]{Staelens:2023jgr}. In particular, this includes a definition via turning points~\cite{Johnson:2019ljv} in the polar coordinate $\theta$: one may alternatively define the $m^\text{th}$-order lensing band as the image region in which incident geodesics exhibit $m+1$ turning points in $\theta$, i.e., for which the condition $d\cos(\theta)/d\lambda=0$ (with $\lambda$ denoting the affine parameter along the geodesic)
is met $m+1$ times along the geodesic. We will refer to this as the ``turning-point definition'' of the lensing bands.

The ``crossing definition'' and the ``turning-point definition'' are related but are not equivalent. To be specific, it was shown that all geodesics in Kerr spacetime must cross the equatorial plane between two turning points~\cite{Gralla:2019drh}. In stationary and axisymmetric spacetimes beyond vacuum GR, the above relation can be broken, see for example, Fig.~6 of Ref.~\cite{Eichhorn:2021iwq}. That said, even in the Kerr spacetime, geodesics which start (or end) at radial asymptotic infinity may or may not cross the equatorial plane while coming in (or escaping from) radial asymptotic infinity, as shown in Fig.~\ref{fig:crossing-vs-turning}. Hence, while a lensing band obtained via the ``turning-point definition'' will always encompass the respective lensing band obtained with the ``crossing definition'', the inverse statement is not generally true, and the two lensing band boundaries do not coincide.

Consequently, for an optically thin and geometrically thick accretion disk, emission from the far half of the disk transits the equatorial plane once more than emission from the near half of the disk (where near and far are defined with respect to the observer's position). The tails of the associated emission, i.e., emission which originates from $r\approx r_{0}$ or $r\approx \infty$ but on the near side of the equatorial plane, can thus ``leak" from the $n^\text{th}$ into the $(n-1)^\text{th}$ equatorial lensing band.

Two comments are in order. First, this effect is increasingly relevant for non-equatorial emission from either close to the horizon or close to asymptotic infinity. Second, even for rather exotic geometrically thick disks with an emission peak close to $r\approx r_{0}$ or $r\approx \infty$, the ($n$-fold) lensed emission (according to the ``turning point definition'') must always lie within the $(n-1)^\text{th}$-order lensing-band (according to the ``crossing definition'').

Nevertheless, the above discussion highlights that the very definition of what one refers to as lensed emission may be related to the question of an astrophysics-independent confident detection of the associated VLBI feature. 
In practice, different geometric notions of a lensing band discussed may serve as an estimate for a systematic error in the identification of a VLBI feature with the lensing band. As the purpose of the present work is merely to introduce and demonstrate the lensing-band framework, we defer such quantitative estimates of a potential systematic error to future work.

\subsubsection{Numerical approximation via bisection}
\label{sec:lensing-bands_bisection}

Due to their simple geometric definition, the effort to obtain lensing bands is comparable to the effort to obtain the critical curve. For any algebraically special spacetime in which geodesic motion is separable, we expect that the lensing bands can thus be expressed by closed-form one-dimensional integrals, see Ref.~\cite[App.~A]{Paugnat:2022qzy} for the Kerr case. In the non-separable case, each lensing-band boundary can be obtained by numerical bisection. In complete analogy to numerically obtaining the critical curve, this amounts to bisecting while numerically solving the geodesic equation, i.e., a coupled system of ODEs.

For this purpose, we have developed a \texttt{Mathematica} package, \texttt{LBeyondGR}\footnote{The code is publicly available at \href{https://github.com/aaron-hd/LBeyondGR}{https://github.com/aaron-hd/LBeyondGR}.}, to iteratively move the bisection forward along the lensing-band boundary. Thereby, we can reliably obtain even non-star-convex lensing-band boundaries~\cite{Cardenas-Avendano:2022csp}. A description of the algorithm is given in App.~\ref{app:bisection}. A brief summary of the resulting scheme to constrain the underlying geometry is presented in Sec.~\ref{sec:scheme}.

\subsection{On the observation of a lensed emission region}
\label{sec:observing-lensed-emission-region}

In the following Subsecs.~\ref{sec:hybrid-imaging} to~\ref{sec:polarization}, we briefly review three current avenues to observationally extract a lensed emission region from VLBI data. In Subsec.~\ref{sec:concrete-VLBI-input}, we then model a specific lensed emission region with which we obtain the \emph{projected constraints} presented in the remainder of this paper.

\subsubsection{A hybrid imaging algorithm}
\label{sec:hybrid-imaging}

One way to isolate evidence for a lensed image from foreground emission is hybrid imaging. Hybrid imaging combines rasterized image reconstruction with the specific modeling of \emph{expected} features. In Ref.~\cite{Broderick_2022}, for the data obtained in the 2017 EHT observation run of M87$^\ast$~\cite{EventHorizonTelescope:2019dse}, hybrid imaging algorithms~\cite{EventHorizonTelescope:2020eky,Broderick:2020wda} indicated a Bayesian preference for a model fit upon inclusion of an expected thin-ring feature over a model fit without said thin-ring feature. The authors of Ref.~\cite{Broderick_2022} interpreted these results as first evidence of (with the definitions of the present work) $n=1$ lensed emission. We refer the reader to Refs.~\cite{Lockhart:2022rui,Tiede:2022grp} for other analyses where no evidence for the presence of lensed emission was found.

\subsubsection{An interferometric diameter inferred from the visibility amplitude}
\label{sec:interferometric-diameter}

An image consisting of nested rings, as expected from GR, produces a cascade of damped oscillations on progressively longer baselines~\cite{Johnson:2019ljv}. Such decomposition is displayed, for example, in Fig.~1 of Ref.~\cite{Cardenas-Avendano:2023dzo} for $0\leq n\leq 2$. Consequently, the visibility amplitude of such an image will display a ringing pattern whose period of oscillation is related to a characteristic length/diameter in the image domain, modulated by an envelope whose specific falloff is dictated by the ring thickness~\cite{Johnson:2019ljv}. Due to these properties in the visibility domain, a space-based VLBI observation can provide evidence for a lensed emission region within the $\text{LB}^{n\geq1}_\text{VLBI}$, if a ringing pattern in the visibility amplitude is detected. By sufficiently long baselines, we mean that, since the $n=0$ image is also ringlike, the baseline should be long enough to provide evidence of a change in the slope of the envelope of the interferometric signature~\cite{Johnson:2019ljv}. The location, in baseline length, where the direct image ($n=0$) stops dominating the visibility domain depends on the width of the $n=0$ image itself, which is highly dependent on the accretion structure surrounding the black hole~\cite{Gralla:2020srx,Cardenas-Avendano:2023dzo}.

\subsubsection{A phase shift on polarimetric images}
\label{sec:polarization}

Another approach to increase confidence in the detection of a lensed emission region, is given by polarimetric images. Assuming that the astrophysical emission source, and, in particular, its polarization structure, shares the rotational symmetry of the spacetime, one expects that the polarization flips between successive lensing bands~\cite{Himwich:2020msm}. In particular, there is a complex conjugation of the rotationally symmetric Fourier mode, known as $\beta_2$, between the $n = 0$ and $n = 1$ images. Thus, measuring a shift in polarimetric phases on long baselines is another estimator~\cite{Palumbo:2023auc}.

Such a rotationally-symmetric polarization structure is expected to arise from the accretion flow of magnetically arrested disks which are currently preferred by observations of \mightyseven~\cite{EventHorizonTelescope:2023fox}. At moderate (near-face-on) inclination, the above polarimetric signature has been confirmed in simulated interferometric observations~\cite{Palumbo:2023auc}. In particular, the authors construct a gain-robust interferometric quantity and demonstrate that future observations of \mightyseven~have the potential to detect this polarization signature.

\subsubsection{Concrete implementation of an input VLBI feature}
\label{sec:concrete-VLBI-input}

Irrespective of which of the above avenues the reader deems most promising, we demonstrate the potential of such an observation to rule out parameter ranges in an underlying family of spacetime geometries such as the Kerr spacetime. 
As a concrete proof of concept, we assume a lensed emission region $\text{LB}^{n\geq1}_\text{VLBI}$ which matches the (upper prior limit of the) narrow-ring feature reported in Ref.~\cite{Broderick_2022}, i.e., a geometrically circular ring of radius $\theta_\text{ring}=21.7 \pm 0.1\,\mu\text{as}$ and fractional width $\psi=0.05$. We neglect the error in the radius $\theta_\text{ring}$, since it is a subleading contribution. A graphic representation of the resulting lensed emission region $\text{LB}^{n\geq1}_\text{VLBI}$ is given in Fig.~\ref{fig:scheme-example}. 

We emphasize once more that the posteriors in Ref.~\cite{Broderick_2022} are insufficient to constrain the width or shape of the ring-like VLBI feature. In the following, we implicitly assume that joint data analysis of future observations, see, for instance Refs.~\cite{Blackburn:2019bly,Johnson:2023ynn}, can confidently associate such a thin-ring VLBI feature to lensed emission. 

Provided an external measurement of the distance $D$ to the source (in this case to \mightyseven, for example), we can translate between the apparent on-sky angle $\alpha$ (in units of radians or equivalently $\mu$as) to geometric units $r_0=2\,MG_N/c^2$ in the underlying spacetime (with $M$ the asymptotic mass, $G_N$ the Newton constant, and $c$ the speed of light) by
\begin{align}
    \tan(\alpha) \equiv \frac{MG_N}{D\,c^2}\;.
\end{align}
In all statistical inferences, we assume $D=16.8\pm0.8\,\text{Mpc}$ as inferred in Ref.~\cite[App.~I]{EventHorizonTelescope:2019ggy}, cf. also Refs.~\cite{2009ApJ...694..556B, 2010A&A...524A..71B, 2018ApJ...856..126C} for the original measurements. For our exploratory analysis, we neglect the error on $D$ as subleading, recognizing that any error in $D$ fully degenerates with the inferred mass $M$.

\subsection{Resulting scheme and numerical implementation}
\label{sec:scheme}
%
\begin{figure}
    \centering
    \includegraphics[height=0.48\linewidth]{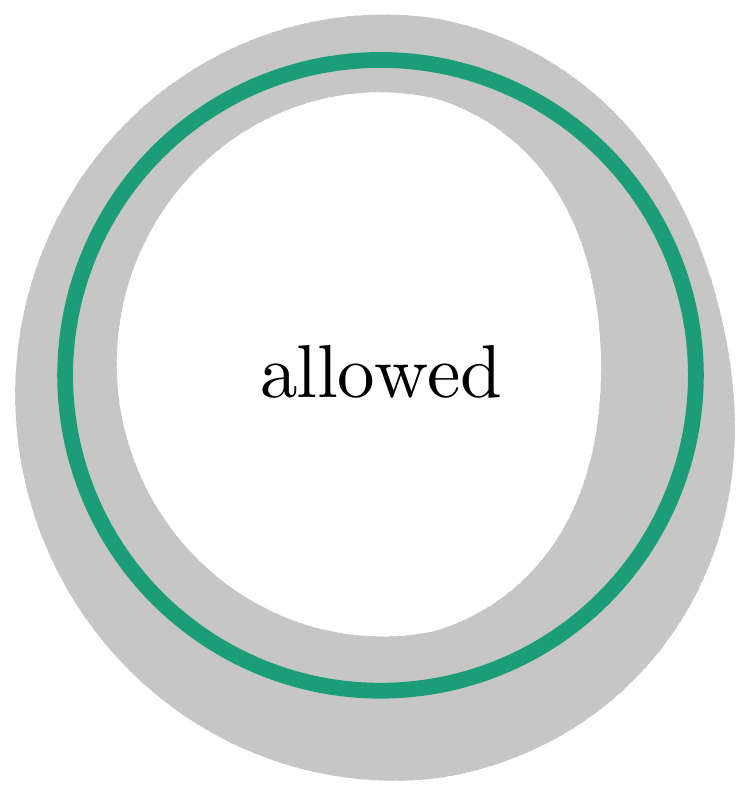}
    \hfill\vrule\hfill
    \includegraphics[height=0.48\linewidth]{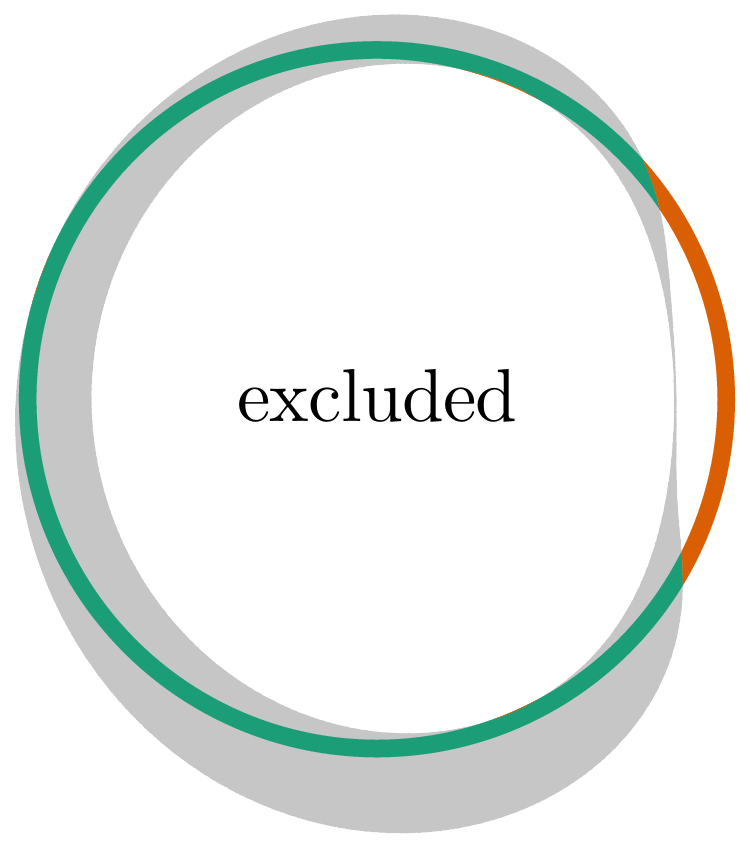}
    \caption{Two schematic examples of $n=1$ lensing bands $\text{LB}^{n=1}(\mathfrak{g})$ (light-shaded regions), for which the respective spacetime $\mathfrak{g}$ remains allowed (left-hand panel) / can be excluded (right-hand panel), after a confident detection of a lensed emission region $\text{LB}^{n\geq1}_\text{VLBI}$ (circular crescent; green regions contained in the lensing band; orange regions not contained in the lensing band).
    }
    \label{fig:scheme-example}
\end{figure}
From here on, we assume that a lensed emission region (i) has been observed and (ii) must lie, for any given spacetime $\mathfrak{g}$, within its $n=1$ lensing band, $\text{LB}^{n=1}(\mathfrak{g})$. Hence, if the observed lensed emission region $\text{LB}^{n\geq1}_\text{VLBI}$ does not match, or ``fit into,'' the lensing-band $\text{LB}^{n=1}(\mathfrak{g})$, then we can rule out the respective underlying spacetime geometry $\mathfrak{g}$, see Fig.~\ref{fig:scheme-example} for an schematic representation of this matching procedure.
The level of confidence with which the underlying spacetime geometry can be ruled out depends on (i) the level of confidence in the detection of the lensed emission region, and (ii) the level of confidence in the theoretical assumption that it must occur within the lensing band. 

We allow for an arbitrary centroid shift by an image-plane vector $(x_0,y_0)$ when performing the matching. The centroid shift amounts to the lack of a reference point. Potential observations of the inner shadow~\cite{Dokuchaev:2018kzk,Dokuchaev:2020wqk,Chael:2021rjo} or the base of the jet in M87*~\cite{Papoutsis:2022kzp} could provide additional priors for the centroid shift. Examples of such reference points will be shown in Sec.~\ref{sec:constraints}, but we will not include them in the statistical analysis presented in this work. Such additional priors will only strengthen evidence against the underlying spacetime $\mathfrak{g}$.

In the following, we present an explicit algorithmic scheme with which we decide -- given a lensed emission region $\text{LB}^{m\geq n}_\text{VLBI}$, as defined in~Sec.~\ref{sec:observing-lensed-emission-region} -- whether or not a given spacetime $\mathfrak{g}$ is excluded. The readers not interested in numerical technicalities may skip this section and proceed directly to the results (Sec.~\ref{sec:Kerr}). 

The scheme splits into three parts: first, an algorithm to determine the lensing band $\text{LB}^{n}(\mathfrak{g})$; second, a minimization of the subregion of $\text{LB}^{m\geq n}_\text{VLBI}$ which cannot be covered by $\text{LB}^{n}(\mathfrak{g})$; and third, an iteration of the former two steps as part of a Markov-chain Monte Carlo (MCMC) sampling algorithm. 

In the present work, we focus on the $n=1$ lensing band which is, at present, observationally most relevant. However, the scheme generalizes to arbitrary $n$ and we plan to determine the exclusion potential of a confident (space-based) $n=2$ detection in future work, cf. Ref.~\cite{Staelens:2023jgr}.

\subsubsection{Free-floating bisection to obtain the lensing-band region}

The numerical approximation of a lensing-band $\text{LB}^{n}(\mathfrak{g})$ requires us to determine\footnote{Strictly speaking, this split only holds as long as the topology of the lensing-band region remains unchanged. Here, we will assume that this is the case.}
\begin{enumerate}
    \item[(i)] its outer boundary $\partial^\text{(outer)}\text{LB}^{n}(\mathfrak{g})$ and
    \item[(ii)] its inner boundary $\partial^\text{(inner)}\text{LB}^{n}(\mathfrak{g})$.
\end{enumerate}
Each equatorial lensing-band $\text{LB}^{n}(\mathfrak{g})$ is defined by the number of times $N$ that the respective incident light ray has crossed the equatorial plane, i.e., $n = N - 1$. Equivalently, one can define the following binary conditions 
\begin{align}
    \mathcal{C}^\text{(outer)}_n &= 
    \begin{cases}
        \text{TRUE}&\text{if}\;N \geqslant n + 1
        \\
        \text{FALSE}&\text{if}\;N < n + 1
    \end{cases}\;,
    \\
    \mathcal{C}^\text{(inner)}_n &= 
    \begin{cases}
        \text{TRUE}&\text{if}\;N < n + 1
        \\
        \text{FALSE}&\text{if}\;N \geqslant n + 1
    \end{cases}\;.
\end{align} 
These conditions evaluate to TRUE whenever the respective image point is to the left of the respective lensing-band boundary $\partial^\text{(outer)}\text{LB}^{n}(\mathfrak{g})$ [or $\partial^\text{(inner)}\text{LB}^{n}(\mathfrak{g})$], where left and right are defined following the boundary counterclockwise. The only difference between determining the inner and outer boundary is a flip in the condition (or equivalently a flip in clockwise/counterclockwise direction). Other conventions may be chosen and are equivalent to the above. 

The task of numerically approximating a lensing band is, thereby, reduced to numerically approximating a closed boundary curve $\partial\mathcal{R}$ twice.  This curve demarcates a subregion $\mathcal{R}\subset\mathbb{R}^2$ in the 2D image plane with respect to a specified boundary bisection condition $\mathcal{C}$, i.e., with respect to the above conditions $\mathcal{C}^\text{(outer)}_n$ and $\mathcal{C}^\text{(inner)}_n$.

Except for a modification in the boundary bisection conditions, the individual bisections are fully equivalent to the well-known numerical bisection of the critical curve, see Ref.~\cite{Younsi:2016azx} for an example of such a bisection code. In particular, the numerical effort is equivalent. 

The abstract task of bisecting a closed curve can be thought of as a discretization of said curve. One may define an orthogonal ray on each such discrete boundary piece. If one can ensure that on each ray, two initial points are given---one on either side of the boundary---then one can perform a simple iterative bisection with respect to the defining boundary bisection condition $\mathcal{C}$.

Our key technical advancement lies in the choice of bisection rays. Previous numerical codes, see, e.g., Ref.~\cite{Younsi:2016azx} (for the critical curve) and Ref.~\cite{Cardenas-Avendano:2022csp} (for lensing bands), perform an `angular bisection' with respect to a (suitably) chosen central point. Instead, here we develop a ``free-floating'' bisection algorithm. Once two points on the boundary are known, the next bisection can then be chosen only in reference to these two points. We find that this has two major advantages:
\begin{itemize}
    \item 
    The first significant advantage is \emph{robustness with respect to non-star-convex boundaries.}
    When using an angular bisection with respect to a central reference point, bisection rays may intersect the boundary more than once. By definition, this necessarily occurs whenever the respective region is not star-convex. Whenever such a double intersection occurs, the boundary is not identified correctly by means of angular bisection. The free-floating bisection algorithm avoids these issues. In particular, it robustly identifies also non-star-convex boundaries\footnote{The only remaining ambiguity concerns the initial guess and whether it converges to the associated boundary. Here, we assume (and verify) that the screen point $(\alpha,\beta)=(0,0)$ lies within the inner shadow and that a random guess at radial distance of $100\,M$ lies within the asymptotic $n=0$ lensing-band region. We find that these assumptions are sufficient to identify all lensing-band boundaries investigated here.}. 
    \item
    The second major advantage is the \emph{adaptive step size.} The free-floating bisection initializes each subsequent bisection in reference to (at least two) previous points on the boundary. This makes it possible to dynamically adapt the step size and precision which, in turn, allows us to optimize efficiency while ensuring that the boundary is not lost.
    
\end{itemize}
We assume that the free-floating bisection algorithm can be adapted to any other (binary) boundary bisection condition in a 2D plane. Given the generality of this problem and given the above two advantages, the algorithm might be of more general use. We provide further details in App.~\ref{app:bisection}. 

Given this free-floating bisection algorithm, the problem is reduced to standard numerical ray tracing (in curved spacetime $\mathfrak{g}$) at each bisection step. In order to keep the code-development minimal, we use an adapted version of \texttt{Mathematica}'s \texttt{NDSolve} routine for the ray tracing\footnote{This is certainly not the most efficient ray tracing and we are inclined to make use of existing optimized ray tracing codes (such as the ones presented in Refs.~\cite{Vincent:2011wz,Bambi:2016sac,Sharma:2023nbk,Bozzola:2023daz}, for example) in future applications.}. The ray tracing is performed in Boyer--Lindquist type coordinates. The translation from screen coordinates to ray tracing coordinates is detailed in App.~\ref{app:asymptotics} (see also Ref.~\cite{Younsi:2016azx}) and, in fact, only requires that the coordinates converge to oblate spheroidal coordinates at $r\rightarrow\infty$. The observers screen is then placed at $r=10^5\,M$ which approximates the $r\rightarrow\infty$ limit sufficiently well\footnote{We have tested that the precision of the lensing-band boundary in the Kerr space time at high inclination (see Fig.~\ref{fig:LBGR} for some visual examples) remains at least at the subpercent level with the analytical results using the code described in Ref.~\cite{Cardenas-Avendano:2022csp}.
}

\subsubsection{Minimizing the uncovered emission region}

Once the lensing band is determined, the second task is to vary the centroid shift $(x_0,y_0)$ such that the subregion of $\text{LB}^{m\geq n}_\text{VLBI}$ that cannot be covered by $\text{LB}^{n}(\mathfrak{g})$ is minimized\footnote{In Ref.~\cite{Tahura:2023qqt}, a mismatch was defined for the regions within the critical curve, after co-centering them at the origin of the camara's reference frame.}.
In the mathematical language of set theory, the region which we want to minimize is the relative complement ${\text{LB}^{n}(\mathfrak{g})}^c$ of $\text{LB}^{n}(\mathfrak{g})$ in $\text{LB}^{m\geq n}_\text{VLBI}$, i.e., ${\text{LB}^{n}(\mathfrak{g})}^c\cap\text{LB}^{m\geq n}_\text{VLBI}\equiv \text{LB}^{m\geq n}_\text{VLBI}\;\backslash\;\text{LB}^{n}(\mathfrak{g})$. Finally, we normalize the area to the total area of $\text{LB}^{m\geq n}_\text{VLBI}$. We can thus summarize the abstract task as determining the relative covered area $\mathcal{A}$ defined by
\begin{align}
\label{eq:OverlappinArea}
    \mathcal{A}(\mathfrak{g}) \equiv  \frac{
        \text{Min}_{(x_0,y_0)}\left(\text{LB}^{m\geq n}_\text{VLBI}\;\backslash\;\text{LB}^{n}(\mathfrak{g})\right)
    }{
        \text{LB}^{m\geq n}_\text{VLBI}
    }\;.
\end{align}
The normalization ensures that $0\leqslant\mathcal{A}(\mathfrak{g})\leqslant1$ which is useful for numerical purposes. Once $\mathcal{A}$ is determined, we decide
\begin{itemize}
    \item if  $\mathcal{A}(\mathfrak{g})=0$, the geometry $\mathfrak{g}$ remains allowed;
    \item if $\mathcal{A}(\mathfrak{g})>0$, the geometry $\mathfrak{g}$ is excluded.
\end{itemize}
For the present purposes, we use \texttt{Mathematica}'s predefined region and minimization routine to perform this second step\footnote{To be explicit, we use \texttt{NMinimize} and specify the method to \texttt{RandomSearch}.}. While VLBI image reconstructions do not have an absolute reference frame, we enforce a constraint on the centroid shift, i.e.,
\begin{align}
    x_0^2 + y_0^2 < \text{rad}\left(\text{LB}^{n\geq1}_\text{VLBI}\right)^2\;,
\end{align}
where $\text{rad}(\text{LB}^{n\geq1}_\text{VLBI})$ denotes the radius of the observed/assumed lensed emission region. This constraint effectively enforces that the central brightness depression region does not move to unphysical regions. In practice, this constraint is required to avoid artefacts when the lensing band develops very broad regions, i.e., regions which exceed the diameter of the observed/assumed lensed emission region, see lower panels of Fig.~\ref{fig:LBGR} for examples.

\subsubsection{Bayesian inference}
\label{sec:MCMC-setup}

Given a confident measurement of lensed emission as discussed above, we estimate the parameters of the spacetime $\mathfrak{g}(M,A,p_{1},\ldots,p_{k})$, where $M$ and $A$ denote the asymptotic mass and spin, respectively, and $p_{i}$ denotes extra parameters (e.g., coupling constants or dimensionless ``bumpy'' parameters~\cite{Collins:2004ex}), for which the $n=1$ lensing band of an observer at infinity, and at inclination $i$ can cover the considered ring. In other words, we estimate the set of parameters $\vec{\lambda}=\left\{i,M,A,p_{1},\ldots,p_{k}\right \}$ such that its $n=1$ lensing band covers the observed VLBI feature,
as Fig.~\ref{fig:scheme-example} illustrates.

We perform the parameter estimation using the affine invariant MCMC sampler \texttt{emcee}~\cite{emcee}, to explore the posterior surface, i.e., the likelihood function $\mathfrak{L}$
\begin{align}
\log \left[ \mathfrak{L}\left(\theta_{\rm ring}=21.7\mu\text{as}|\vec{\lambda}\right)\right]=-\frac{1}{2}\left(\frac{\mathcal{A}}{\sigma}\right)^{2},
\end{align}
times the parameter priors. Here $\mathcal{A} \in [0,1]$ (Eq.~\ref{eq:OverlappinArea}) denotes the overlapping area between the VLBI feature and the lensing band, and $\sigma$ is the standard deviation of the distribution, which we assume it to be $\sigma=0.05$, as a proxy for the observational error of the VLBI feature.
We choose flat priors on all the parameters. For the parameters $A,i$ and $M$ the ranges are $-1\leq A/M\leq 1$,
$0\leq i \,[\rm{deg}]\leq 85$ and  $3\leq M/(10^9M_{\odot})\leq 8$, respectively. For the parameters $p_{i}$, however, the only requirement are for them to satisfy (i) the absence of further (Killing) horizons; (ii)
the absence of closed timelike curves; and (iii) no signature changes in the metric. In App.~\ref{app:paramRanges}, we present the explicit expressions for their range of validity.

The \texttt{emcee} ensemble of walkers is initialized by sampling from the prior distributions. For all the cases presented in this work, we run until $\sim10^5$ samples are obtained, and burn-in the initial $100$ samples. In general, parameter regions in which the lensing band is wide are easier to find within the posterior likelihood distribution. Recovering the theoretically expected symmetry of all results and correlations under the exchange of $A\leftrightarrow -A $ (as well as $a\leftrightarrow -a $) seems to be another good marker for convergence of the MCMC sampler. The presented results suggest that most (but not all) correlations are fully converged. Given the exploratory character of our work, we do not deem it justified to spend computation time on further convergence.

\section{Benchmark: Projected constraints in the Kerr spacetime}
\label{sec:Kerr}
%
\begin{figure}
    \centering
    \includegraphics[width=\linewidth]{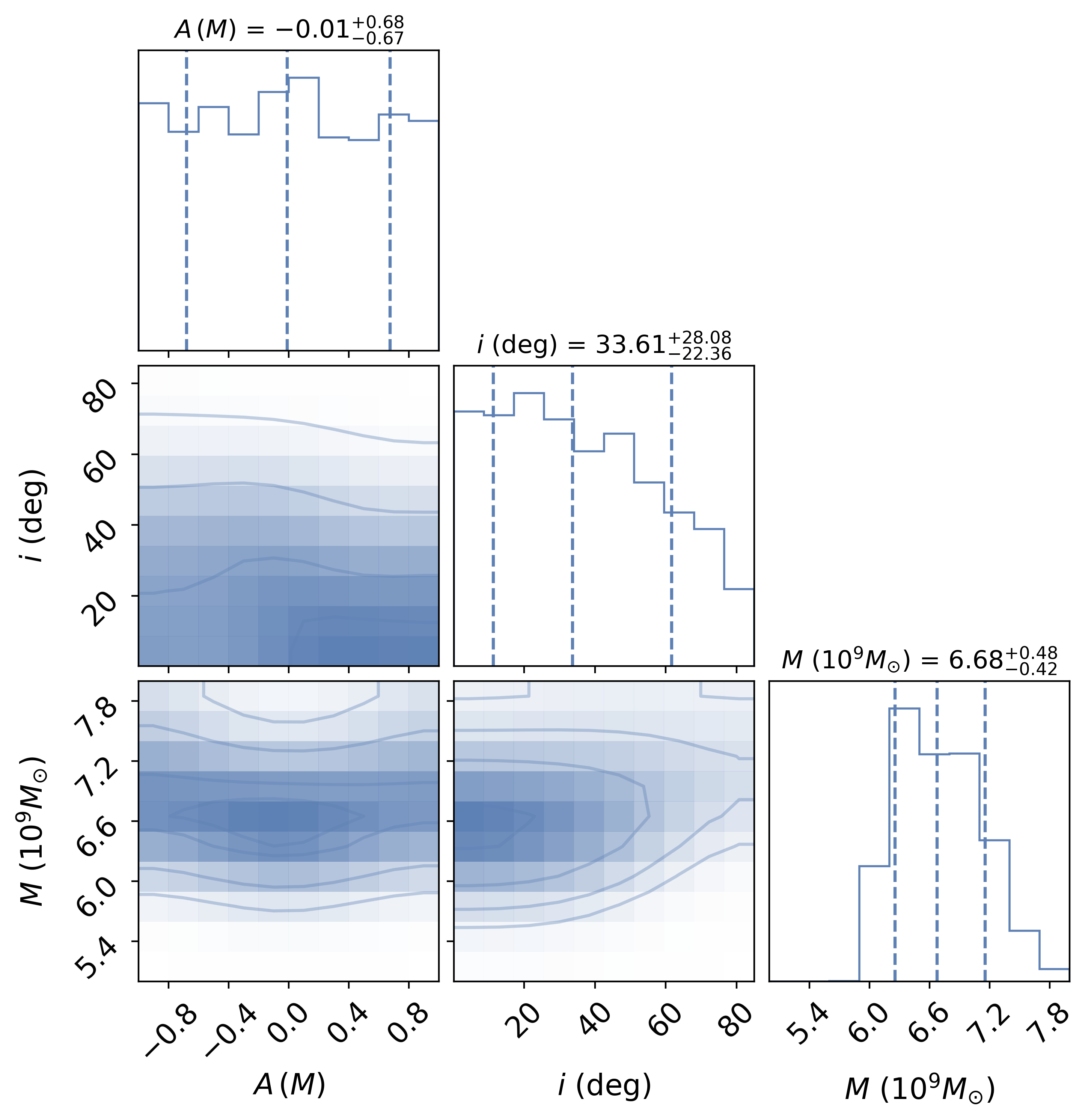}
    \caption{
    \emph{Projected constraints} obtained using the lensing-band framework applied to the Kerr spacetime. These results exemplify how, upon measuring a VLBI feature, as mentioned in~\cref{sec:observing-lensed-emission-region}, one can constrain parameters of the spacetime. Given that the lensed image assumed for this example is a circular ring, it is possible to constrain the inclination. The mass, as expected, is also constrained from the overall scale.}
    \label{fig:GRMCMC}
\end{figure}

\begin{figure}
    \centering
    \includegraphics[width=0.48\linewidth]{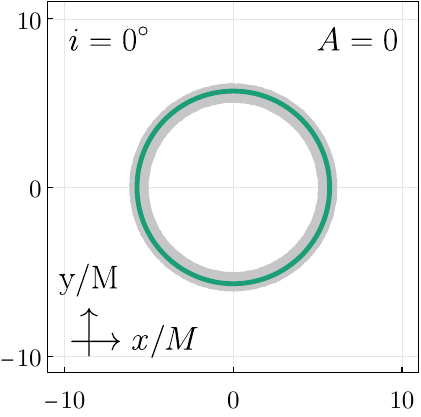}\hfill
    \includegraphics[width=0.48\linewidth]{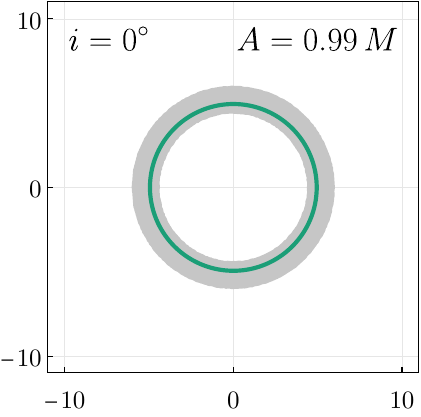}
    \\
    \includegraphics[width=0.48\linewidth]{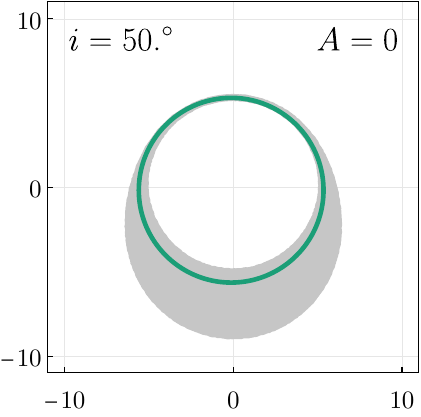}\hfill
    \includegraphics[width=0.48\linewidth]{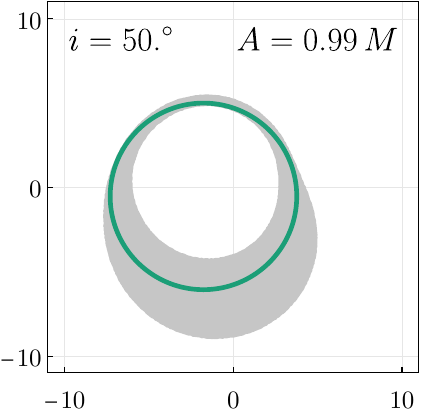}
    \\
    \includegraphics[width=0.48\linewidth]{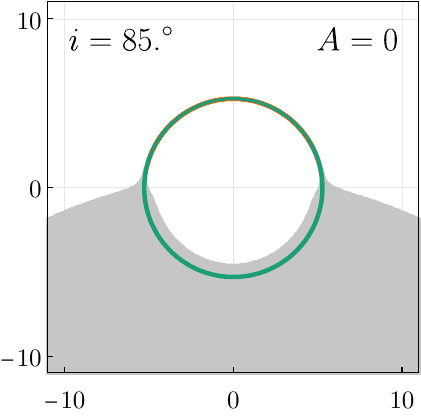}\hfill
    \includegraphics[width=0.48\linewidth]{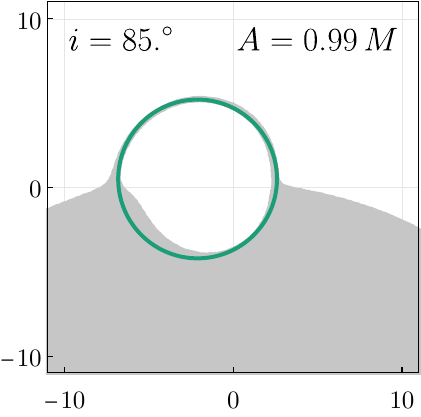}
    \caption{Examples of lensing bands (light gray regions) and the fitted lensed emission regions $\text{LB}^{n\geq1}_\text{VLBI}$ (green regions contained in the lensing band; orange regions not contained in the lensing band) in the Kerr spacetime. The three rows (top to bottom) distinguish different inclinations: $i=0^{\circ},\,50^{\circ}$, and $85^{\circ}$, respectively. The two columns distinguish different spins: $A=0$ (left) and $A=0.99\,M$ (right).
    }
    \label{fig:LBGR}
\end{figure}

As a benchmark application of the lensing-band scheme, we ask what can be inferred about the parameters of the Kerr spacetime itself. In Boyer--Lindquist coordinates the free parameters of the Kerr geometry are:
the inclination $i$;
the Kerr mass parameter $M$;
and the Kerr spin parameter $A/M$.
The results of the MCMC sampling are shown in Fig.~\ref{fig:GRMCMC}. All the results should be understood as \emph{projected constraints} which can be inferred once a confident detection of a lensing-band region, as discussed in Sec.~\ref{sec:observing-lensed-emission-region}, is available.
We further emphasize this interpretation, i.e., that these are not actual constraints obtained from data, by using the ``geometrically-equal-to'' sign $\doteqdot$ whenever referring to \emph{projected constraints}.

For the implemented example, it is reassuring that the mass posterior results in a \emph{projected constraint} for the mass of \mightyseven, i.e., $M \doteqdot 6.68^{+0.48}_{-0.42}\,10^9\,M_\odot$, that is in agreement with previous results~\cite{EventHorizonTelescope:2019dse}. We find no independent \emph{projected constraint} on the spin. We remind the reader that we ignore any external mass measurement and make only minimal assumptions on the astrophysics, cf.~Sec.~\ref{sec:framework}. Hence, we interpret this result as the statement that additional knowledge about, for example, the accretion physics, an external mass measurement, or higher-order lensing bands is needed to infer the spin.

We do, however, find an interesting correlation, albeit small, between mass and spin: the higher the absolute value of the spin, the larger the inferred mass, see the covariance of $A$ and $M$ in Fig.\ref{fig:GRMCMC}. 
The underlying reason for this correlation is the shrinking of the $n=1$ lensing band with increasing spin, see left vs right columns in Fig.~\ref{fig:LBGR}. With very different inference techniques, it has been previously found in Refs.~\cite{Broderick:2021ohx,Broderick_2022} that such correlation between mass and spin is more significant in the $n=1$ image than in the critical curve (i.e., in the $n=\infty$ image). It is reassuring that we recover this correlation.

As expected, given that the modeled VLBI feature is a circle, the results disfavor high inclinations. To be specific, the posterior for the \emph{projected constraint} on inclination results $i \doteqdot {34^{\circ}}^{+28^{\circ}}_{-22^{\circ}}\;$. Indeed, the $n=1$ lensing band becomes very thin as the inclination approaches $i\sim 90^{\circ}$, as shown by the different rows in Fig.~\ref{fig:LBGR}. We highlight that formally our analysis assumes, and thus all constraints rely on, a confident detection of lensed emission throughout the specified region $\text{LB}^{n\geq1}_\text{VLBI}$, see, for instance, the darker and colored regions in Figs.~\ref{fig:LBGR} and~\ref{fig:scheme-example} and the discussion in Subsec.~\ref{sec:observing-lensed-emission-region}. In particular, the quantitative \emph{projected constraints} (such as the one on the inclination) which arise from a thinning of the lensing band, implicitly rely on lensed emission being observed not just \emph{somewhere} but \emph{everywhere} within the observed region.

The above example highlights that our \emph{projected constraints} should be interpreted as an estimate for the constraining power of a given VLBI observation or as a means to uncover quantitative physical correlations among the underlying parameters of the geometry.

Dedicated future studies to determine how confident VLBI observations can constrain the thickness of a lensed emission region are vital to solidify constraints such as the above ones. We also emphasize that the definition of lensing bands (defined with respect to the equatorial plane) is ill defined at exactly edge-on inclination $i=90^{\circ}$. Respectively, when deforming, e.g., the $n=1$ lensing band with increasing inclination across $i=90^{\circ}$, one encounters a discontinuity, see e.g.,~\cite[Fig.~7]{Paugnat:2022qzy}. The quantitative impact of this remains to be scrutinized in future studies and, hence, we are cautious to interpret our results at close to edge-on inclination. 

\section{Application: Projected constraints on black hole non-uniqueness}
\label{sec:parametric}

In this section, we apply the developed lensing-band framework to parametrized deviations from the Kerr spacetime. This test case demonstrates that the framework is an efficient tool for first explorations of large parameter spaces of possible deviations from GR in the context of black-hole shadow observations. In particular, we demonstrate that the formalism enables us to
\begin{itemize}
    \item efficiently explore degeneracies and correlations in large parameter spaces,
    \item quantify the constraining power of a confident observation of a lensed emission region,
    \item identify specific detectable deviations which may then be tested for against the Kerr spacetime in explicit astrophysical models.
\end{itemize} 
As in the previous section, we present the results in the form of ``\emph{projected constraints}'' on the underlying spacetime parameters. In Subsec.~\ref{sec:circular-params}, we briefly review parametrized deviations in stationary and axisymmetric spacetimes beyond Kerr, and present a seven-parameter spacetime which we will use as a prototype test case beyond Kerr. In Subsec.~\ref{sec:constraints}, we then apply the lensing-band framework presented in Subsec.~\ref{sec:framework} and discuss the resulting \emph{projected} constraints.

\subsection{Parametrizing deviations to circular spacetimes}
\label{sec:circular-params}

There are various ways in which black-hole candidates may differ from a black hole solution in GR~\cite{Berti:2015itd}. In recent years, a ``theory-agnostic" approach to testing GR has been heavily pursued. This approach involves parametrizing deviations from GR solutions (such as the metric~\cite{Collins:2004ex} or gravitational potentials~\cite{Will:2014kxa}) or observables (such as the gravitational waveform~\cite{Yunes:2009ke} or quasinormal modes frequencies~\cite{Baibhav:2023clw}). As a result, a particular theory is neither employed nor tested, but the hope is that the parameters involved, in such an agnostic way, can then be mapped to physical constraints. 

The exploration of deviations at the level of the metric requires to (i) parametrize a given class of spacetimes in terms of free metric functions, and then (ii) expand these metric functions. In the first step, the free functions should ideally cover the entire class of spacetimes without any redundancy. While this may seem trivial, a satisfactory answer with regards to all possible stationary and axisymmetric spacetimes is---to the best of our knowledge---not known. In Subsec.~\ref{sec:classes-of-stat-axi-spacetimes}, we review what is known in subclasses of stationary and axisymmetric spacetimes with additional symmetries.
For the second step, a relation to physical quantities, such as multipole moments at asymptotic infinity, seems desirable. We briefly review such expansions in Subsec.~\ref{sec:expansions}.

\subsubsection{Subclasses of stationary and axisymmetric spacetimes with a redundancy-free metric ansatz}
\label{sec:classes-of-stat-axi-spacetimes}

Stationary and axisymmetric spacetimes are defined by the existence of a spacelike rotational Killing vector $\eta^\mu$ and an asymptotically timelike Killing vector $\xi^\mu$. A theorem by Carter ensures that the two Killing vectors always commute~\cite{Carter:1970ea}. 
Throughout the following discussion, we restrict to the physical case of four-dimensional Lorentzian spacetimes, but we do not assume vacuum GR. 
Moreover, we seek a coordinate system valid throughout the exterior black-hole region.

Due to the Killing symmetries, the metric can always be written in terms of ten free functions which depend only on two coordinates. Assuming a global $((2+1)+1)$ foliation of the spacetime, eight independent metric functions are sufficient~\cite{1993PhRvD..48.2635G}. When restricting to a coordinate patch in the vicinity of future null infinity, these can further be reduced~\cite{Bondi:1962px, Sachs:1962wk}. However, to the best of our knowledge, a fully satisfactory understanding regarding completeness and redundancy remains outstanding, see also~\cite{Ayon-Beato:2015xsz,Delaporte:2022acp}.

In contrast, the class of stationary and axisymmetric spacetimes contains several subclasses in which further progress has been made. Each of these subclasses can be understood in terms of some enhanced (either explicit or hidden) symmetry. Some of the known subclasses are successively contained within each other.

First, one may restrict to circular spacetimes~\cite{Papapetrou:1966zz,Kundt:1966zz,Wald:1984rg}. Physically, circular spacetimes allow for a \emph{global} foliation into two orthogonal 2-surfaces. This also implies a global symmetry under simultaneous inversion $\phi\leftrightarrow -\phi$ and $t\leftrightarrow -t$ of the two Killing coordinates (associated with $\eta^\mu$ and $\xi^\mu$). All vacuum solutions of GR are circular\footnote{Mathematically, circularity can be understood as a property of the Ricci-tensor $R_{\mu\nu}$ in relation to the two Killing vectors $\eta^\mu$ and $\xi^\mu$, i.e., that $\eta^{\mu} R_{\mu}^{\,\, [\nu}\xi^{\kappa}\eta^{\lambda]} = 0$ and $\xi^{\mu} R_{\mu}^{\,\, [\nu}\eta^{\kappa}\xi^{\lambda]} = 0$ everywhere (while there exists a point in the spacetime for which $\eta^{[\mu} \xi^{\nu}\nabla^{\kappa}\eta^{\lambda]} = 0$ and $\xi^{[\mu} \eta^{\nu}\nabla^{\kappa}\xi^{\lambda]} = 0$), see e.g., Ref.~\cite[Sec.~7.1]{Wald:1984rg}. It is now obvious that any Ricci-flat spacetime, hence, any vacuum solution of GR is circular.}.
Irrespective of vacuum GR solutions, any circular spacetime can be written as
\begin{align}
	ds^2_\text{circ}=&
	- g_{tt}(r, \theta)\,dt^2 - g_{t\phi}(r, \theta)(dt\,d\phi + d\phi\,dt) 
	\notag\\&
	+ g_{rr}(r,\theta)\left[dr^2 + \sigma(r,\theta) d\theta^2 \right] + g_{\phi\phi}(r, \theta)\,d\phi^2\;,
\label{eq:circularParam}
\end{align}
where the choice of $\sigma(r,\theta) = r^2+a^2\cos(\theta)^2$ specifies Boyer--Lindquist coordinates $(t,r,\theta,\phi)$.
The circular class requires five non-vanishing metric elements, described in terms of four independent functions\footnote{Equation.~\eqref{eq:circularParam} requires only four free functions but five non-vanishing metric elements. Hence, coordinate transformations can be used to, for instance, alter the relation between $dr^2$ and $d\theta^2$. This coordinate freedom can be used to reduce from five to four free functions, see, e.g., Ref.~\cite{Papapetrou:1966zz} for the well-known Papapetrou form. The latter choice of coordinates, however, does not contain Kerr in Boyer--Lindquist coordinates~\cite[App.~B]{Delaporte:2022acp}.}. It is both minimal and exhaustive in the number of required free functions. Exhaustiveness of Eq.~\ref{eq:circularParam} relies on circularity and has been established in Refs.~\cite{Papapetrou:1966zz,Kundt:1966zz,Wald:1984rg}.

One may further enhance the symmetry from stationarity to staticity. Any static and axisymmetric spacetime is a member of the Weyl class~\cite{Weyl:1917gp} with four non-vanishing metric elements written in terms of two independent functions.

Alternatively, one may restrict to spacetimes which, in addition to the two Killing vectors $\eta^\mu$ and $\xi^\mu$, admit for an independent rank-2 Killing tensor. The Killing tensor then guarantees a hidden constant of motion and thus separability and integrability of geodesic motion~\cite{Carter:1968rr,Benenti:1979erw}. The minimal and exhaustive form has been derived in Ref.~\cite{Benenti:1979erw}, and we will thus refer to this subclass as the Benenti--Francaviglia class. Both, the Weyl class and the Benenti--Francaviglia class are contained in the circular class.

Finally, in any of the above classes, one may or may not chose to break reflection symmetry about the equatorial plane, see e.g., Ref.~\cite{Chen:2020aix} for a study within the Benenti--Francaviglia class. Of course, one may also further restrict to spherical symmetry, where stationarity implies staticity.

Given that the circular class is the largest known subclass in which a redundancy-free metric ansatz has been established, we focus on this symmetry subclass of stationary and axisymmetric spacetimes in our analysis. Before we do so, we briefly summarize different expansion approaches.

\subsubsection{Expansion approaches to parametrized deviations}
\label{sec:expansions}

Multipole expansions of the Weyl class have been explored in Refs.~\cite{Collins:2004ex,Vigeland:2010xe}. They have been extended to deviations from Kerr spacetime by applying the Newman--Janis algorithm~\cite{Newman:1965tw} to the Weyl class~\cite{Vigeland:2009pr,Vigeland:2010xe,Vigeland:2011ji,Johannsen:2011dh}. The resulting spacetimes are no longer within the Weyl class. In Ref.~\cite{Vigeland:2011ji} it was shown that the resulting spacetimes do not necessarily admit for a rank-2 Killing tensor and thus need not remain within the Benenti--Francaviglia class.

Different Taylor expansions around asymptotic infinity within the Benenti--Francaviglia class have been explored in Refs.~\cite{Johannsen:2013szh,Carson:2020dez,Yagi:2023eap}. A continued fraction expansion around the horizon at $r=r_0$ of the circular class (first developed in spherical symmetry in Ref.~\cite{Rezzolla:2014mua}) has been set up by Konoplya, Rezzolla, and Zhidenko (KRZ)~\cite{Konoplya:2016jvv}, which we briefly review in App~\ref{app:KRZ}. To the best of our knowledge, the KRZ expansion covers all of the above expansions, while the reciprocal is not true. Hence, we focus on this continued-fraction expansion below.

\subsubsection{A seven-parameter family obtained from the leading-order terms in the KRZ expansion}

As detailed above, we focus on the KRZ expansion (see Subsec.~\ref{sec:expansions}) of circular spacetimes (see Subsec.~\ref{sec:classes-of-stat-axi-spacetimes}). Furthermore, we make sure to maintain oblate spheroidal coordinates at radial asymptotic infinity and match the leading order asymptotics to the mass~$M$, the angular momentum~$J$, and the PPN parameters~$\beta$ and~$\gamma$. The respective explicit derivation within the KRZ expansion is detailed in App.~\ref{app:KRZ}. These choices result in a seven-parameter family of spacetimes for which the metric functions in Eq.~\ref{eq:circularParam} can be written in the following simple form
\begingroup
\allowdisplaybreaks
\begin{align}
    g_{tt} &= - \frac{
        \left(
        		\Delta_{\beta\gamma} 
        		+ \frac{r_0^3 (r-r_0)}{r^2}\,a_{01}
        	\right)\sin^2\theta
        - g_{t\phi}^2
    }{
        g_{\phi\phi}
    }\;,
    \notag\\[0.5em]
    g_{rr} &= 
        \frac{
            \Sigma \left(1 - \frac{(1-\gamma)\,M}{r}\right)^2
        }{
            \left(
                \Delta_{\beta\gamma}
                + \frac{r_0^3 (r-r_0)}{r^2}\,a_{01}
            \right)
        }
    \;,
    \notag\\[0.5em]
    g_{\theta\theta} &= \Sigma\;,
    \notag\\[1em]
    g_{\phi\phi} &= \left[
        a^2+r^2
        +\frac{2 M r a^2}{\Sigma}\left(\frac{A}{a} - \frac{(a^2 + r_0^2)\cos^2\theta}{2\,M\,r_0}\right)
    \right]\sin^2\theta\;,
    \notag\\[0.5em]
	g_{t\phi} &= -\frac{2 M r A \sin ^2\theta}{\Sigma}\;.
    \label{eq:our-KRZ-truncation}
\end{align}
\endgroup
We use the common shorthand $A=J/M$, $\Sigma = r^2 + A^2\cos^2\theta$, and $\Delta=r^2 - 2Mr + A^2$ which are also often used to denote Kerr spacetime. In addition, we use the shorthand
\begin{align}	
    \Delta_{\beta\gamma} &= \frac{(r-r_0) \left[
        \Delta 
        -2 M^2 (\beta -\gamma + \frac{r_0}{M})
        +r_0 (r+r_0)
    \right]}{r}\;,
\end{align}
which reduces to $\Delta_{\beta\gamma} \rightarrow \Delta$ in the Kerr limit (see below).
The seven free parameters
\begin{align}
    \label{eq:7-params}
    \Xi = (r_0, M,J,\beta,\gamma,a,a_{01})\;,
\end{align}
have the following physical meaning:
\begin{itemize}
    \item The surface $r=r_0$ denotes the black-hole horizon. In App.~\ref{app:paramRanges} we present the theoretical bound on the other parameters to ensure that $r=r_0$ is the outermost (Killing) horizon.
    \item The parameters $M$ and $J$ correspond to asymptotic mass and angular momentum, respectively. Instead of $J$, we sometimes work with $A=J/M$.
    \item The parameters $\beta$ and $\gamma$ correspond to the leading-order PPN parameters in the point-mass limit.
    \item The parameter $a$ accounts for independent deviations of the horizon area\footnote{In the context of the KRZ expansion, see App.~\ref{app:KRZ}, $a$ is identified with the spin parameter of the background Kerr spacetime.} away from the horizon area of a Kerr black hole with equatorial horizon radius $r_0$ and asymptotic spin $A$.
    \item In the asymptotics, the parameter $a_{01}$ corresponds to $(r_0/r)^3$ corrections. We include this parameter to relate to previous studies~\cite{EventHorizonTelescope:2022urf,Ayzenberg:2022twz}. 
\end{itemize}
The Kerr spacetime in Boyer--Lindquist coordinates is recovered when $r_0\rightarrow M + \sqrt{M^2 - a^2}$, $J\rightarrow a\,M$, $\beta\rightarrow 1$, $\gamma\rightarrow 1$, and $a_{01}\rightarrow 0$.

\subsection{From size constraints to shape constraints}
\label{sec:constraints}

%
\begin{figure*}[htb]
    \includegraphics[width=\linewidth]{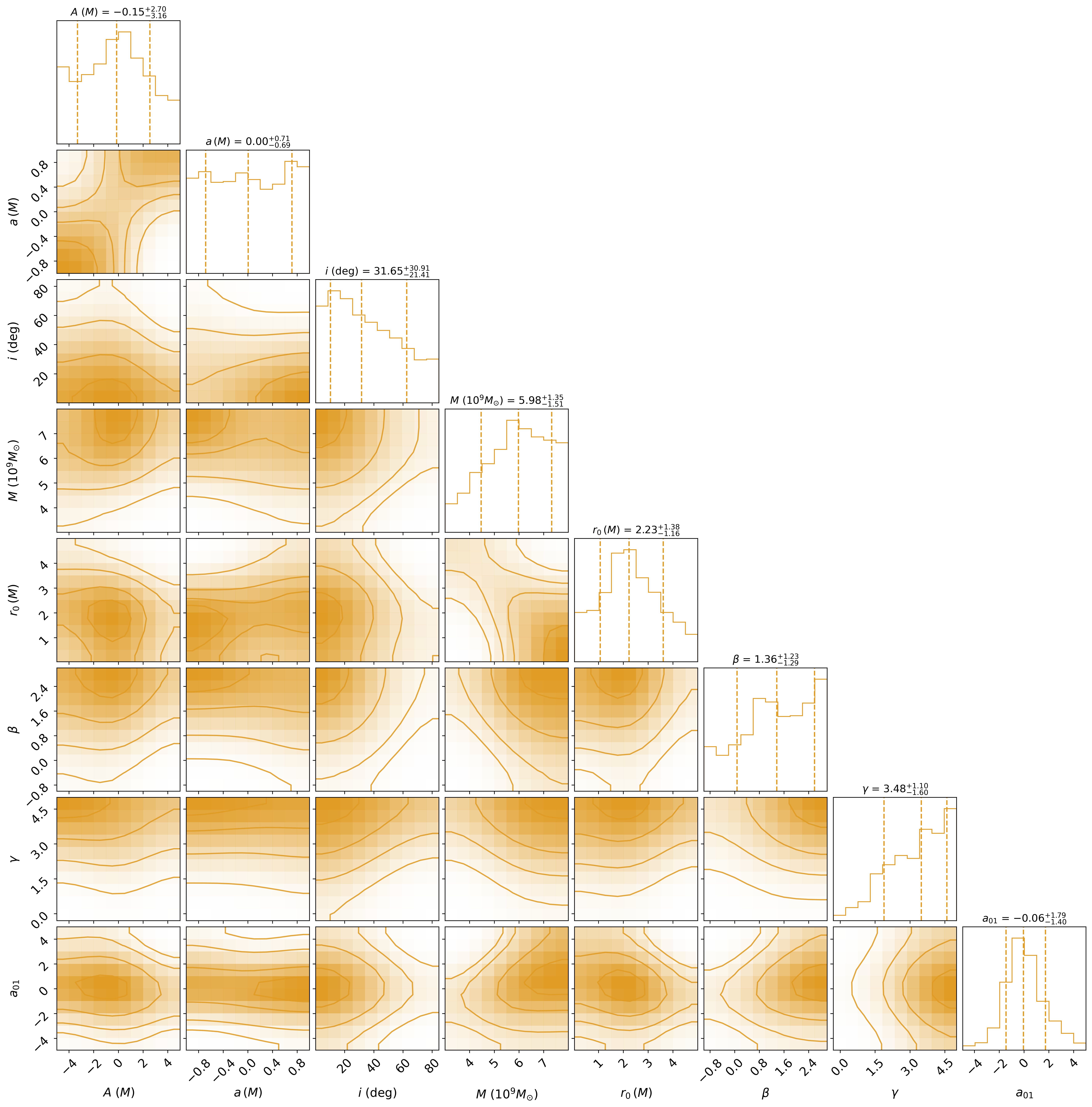}
    \caption{An application of the lensing-band framework to the seven free parameters of the beyond-GR spacetime given in Eq.~\ref{eq:our-KRZ-truncation} and the observer's inclination. Assuming the observation of a VLBI feature, as discussed in Subsec.~\ref{sec:observing-lensed-emission-region}, these \emph{projected constraints}, to one standard deviation, result by demanding it to be completely within the first lensing band of the spacetime considered.}
    \label{fig:NonGRAllMCMC}
\end{figure*}

Given the spacetime in \cref{eq:our-KRZ-truncation} and the lensed-emission region in Subsec.~\ref{sec:observing-lensed-emission-region}, we employ the lensing-band framework to obtain \emph{projected constraints} on the underlying free parameters of the circular beyond-Kerr geometry given in Eq.~\ref{eq:7-params}. 

The full results of the Bayesian parameter estimation, including deviations of all seven free parameters $\Xi = (r_0, M, A,\beta,\gamma,a,a_{01})$ of the underlying spacetime, is presented as corner plot in Fig.~\ref{fig:NonGRAllMCMC}. As in the previous section, we choose flat priors on all the parameters with the following bounds: $0.01\leq r_0\leq 5$, $-5\leq A/M\leq 5$, $-1\leq \beta \leq 3$, $-1\leq \gamma \leq 5$, $-1\leq a \leq 1$, and $-5\leq a_{01}\leq 5$, and check that the resulting metric obeys the theoretical bounds presented in App.~\ref{app:paramRanges}. As we will see, these flat priors are deformed into preferred regions, for most of these parameters, demonstrating the constraining power of a confident observation of a lensed emission region such as the one discussed in Subsec.~\ref{sec:observing-lensed-emission-region}.

We discuss several aspects of these results in more detail in the following section. To aid the discussion, we present explicit lensing bands of respective metric deformations and refer to the marginalized posteriors and covariances in Fig.~\ref{fig:NonGRAllMCMC}.

\subsubsection{Disentangling the asymptotic mass and the radial horizon location}

Previous constraints derived from data obtained in EHT observations of M87$^*$ and SgrA$^*$ assume that the asymptotic mass $M$ is tied to the horizon location $r_0$~\cite{EventHorizonTelescope:2019dse,Younsi:2021dxe,Ayzenberg:2022twz,EventHorizonTelescope:2022urf}. This can be understood as an expression of the uniqueness theorems of GR. As our parametrized spacetime disentangles $M$ and $r_0$, we are able to probe this particular uniqueness assumption. Indeed, our results indicate  independent \emph{projected constraints} for both parameters, with inferred values of $M \doteqdot 5.98^{+1.35}_{-1.51}\,10^9\,M_{\odot}$, and $r_0 \doteqdot 2.23^{+1.38}_{-1.16}\,M\,$. The outcome is, therefore, consistent with mass estimates obtained from the image of M87$^*$~\cite{EventHorizonTelescope:2019dse, Broderick_2022}. These values are also consistent with a horizon location in the full range of black holes admissible in GR, i.e., from Schwarzschild spacetime with $r_0=2\,M$ all the way to extremal Kerr spacetime, i.e., with $r_0=1\,M$. 

\subsubsection{The effects of the inclination}

As discussed in Sec.~\ref{sec:Kerr}, for the Kerr spacetime, the definition of equatorial lensing bands is ill-defined at edge-on inclination (i.e., at $i=90^\circ$), see, e.g.,~Fig.~7 in Ref.~\cite{Paugnat:2022qzy}. Hence, we are cautious to interpret results at close to edge-on inclination.

With the above caveat, and similar to the Kerr case in Sec.~\ref{sec:Kerr}, our results suggest a preference for low, i.e., close to face-on, inclination. Specifically, we find $i \doteqdot {32^{\circ}}^{+31^{\circ}}_{-21^{\circ}}\;$ which remains very similar to the posterior for the inclination obtained for the Kerr benchmark test in Sec.~\ref{sec:Kerr}.
Once more, the lensing band becomes very thin at high, i.e., close to edge-on, inclination. Hence, it becomes increasingly delicate (or at some point impossible) to fit the observed lensed emission region into the theoretical lensing band. We reiterate that such \emph{projected constraints} formally rely on a confident observation of lensed emission \emph{everywhere} within the associated image region.

With regards to the specific astrophysical source of M87$^*$--used here to quantitatively exemplify our lensing-band formalism--an observation of the jet can provide a strong independent prior on the inclination~\cite{mertens2016kinematics,2018ApJ...855..128W}, i.e., $i = 17.2^{\circ}\pm3.3^{\circ}$. Said prior is very far from edge-on such that the above caveat is likely not very relevant--at least for this particular source.
In the future, it would be interesting to perform a Bayesian analysis that uses this external prior on the inclination.

\subsubsection{Lifting (part of) the degeneracy among spherically symmetric PPN corrections~$(\beta,\gamma)$}

\begin{figure}
    \centering
    \includegraphics[width=\linewidth]{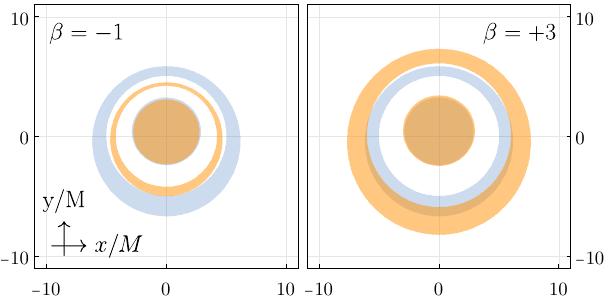}
    \\
    \includegraphics[width=\linewidth]{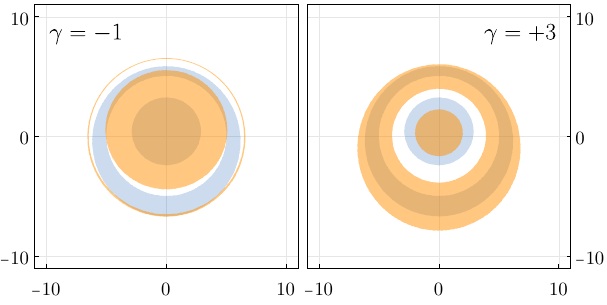}
    \caption{
    The inner shadow and the first ($n=1$) lensing band viewed at an inclination of $i=17^\circ$. The orange regions denote the lensing bands arising from individually deforming the PPN parameters $\beta$ (top panels) and $\gamma$ (bottom panels), while keeping all of other parameters fixed at their Schwarzschild values. As a reference, we also display the lensing band of the Schwarzschild spacetime in blue in all these panels.}
    \label{fig:LB_beta_gamma}
\end{figure}

The PPN corrections ($\beta$ and $\gamma$) enter as spherically-symmetric deformations. Both PPN corrections remain consistent with the values preferred by GR ($\beta_\text{GR}=1$ and $\gamma_\text{GR}=1$). Moreover, they are correlated which suggests that the strongest \emph{projected constraint} can actually be placed on the combination $(\beta-\gamma)$, see Fig.~\ref{fig:NonGRAllMCMC}. This combination enters in the considered spacetime metric via $\Delta_{\beta\gamma}$ and, therefore, affects the $g_{tt}$ component of the metric, see Eq.~\ref{eq:our-KRZ-truncation}. Several other works have quantified constraints directly in terms of constraints on the functional form of $g_{tt}$~\cite{EventHorizonTelescope:2020qrl,Volkel:2020xlc,Broderick:2023jfl,Salehi:2023eqy}. However, away from spherical symmetry or the critical curve, the lensing behaviour is not only controlled by the $g_{tt}$ component of the metric.

Beyond the combination $(\beta-\gamma)$, we find independent \emph{projected constraints} towards lower values of $\beta$ and $\gamma$, while we find that our priors are saturated towards larger values. From this we conclude that the presented lensing-band framework can only place lower bounds on the leading PPN parameters--at least not without external input from independent measurements.

As detailed below, an explanation for these one-sided \emph{projected constraints} can be understood in terms of (i) a thinning/broadening behaviour of the lensing-band deformation associated to these parameters and (ii) a degeneracy of several spherically symmetric deviation parameters.

Looking at single-parameter deformations away from the Schwarzschild spacetime (see Fig.~\ref{fig:LB_beta_gamma}), we identify that increasing $\beta$ or $\gamma$ broadens the lensing band. If there are no other deformations (or if all other deformations can be compensated for by other parameters such as the mass), then the observed emission region can still be accommodated.
Vice versa, decreasing $\beta$ or $\gamma$ thins the lensing band, thus eventually obstructs a complete overlap of the assumed VLBI feature, and, hence, places constraints on the respective parameter regions.

In addition to this broadening/thinning effect, deformations of $\beta$ or $\gamma$, also lead to an increase/decrease of the overall size of the lensing band. Such a deformation in the overall size occurs generically for many spherically-symmetric deformation parameters and has previously been discussed in Ref.~\cite{Volkel:2020xlc}. In fact, most constraints placed on deviations from GR to date reduce the geometrical information to the apparent size of the photon ring~\cite{Younsi:2021dxe,EventHorizonTelescope:2022urf} and are thus, as such, highly degenerate~\cite{Volkel:2020xlc}. 
The degeneracy with the mass $M$, can, of course, be lifted by means of an external mass measurement~\cite{EventHorizonTelescope:2019dse}. However, without resolving geometric information beyond the overall apparent shadow diameter, degeneracies among various possible spherically-symmetric deformation parameters are impossible to avoid~\cite{Cardenas-Avendano:2019zxd}. Our results demonstrate that the first lensing band provides an independent way to break (at least some of these) these degeneracies. In particular, the lensing-band framework captures more information than just one overall diameter. Even in spherical symmetry, it captures continuous information encoded in the width of the lensing band. As a result, these degeneracies can (at least partially) be lifted. For instance, the independent lower bounds on $\beta$ and $\gamma$ indicates such a breaking of degeneracy.

Several other ways to lift the degeneracy among various spherically symmetric parameters have been suggested in the literature. For example, using higher-order lensed images~\cite{Broderick:2005jj,2020ApJ...892..132T,Eichhorn:2021etc,Eichhorn:2021iwq,Broderick:2021ohx,Eichhorn:2022oma,Ayzenberg:2022twz}, or the inner shadow~\cite{Dokuchaev:2018kzk,Dokuchaev:2020wqk,Chael:2021rjo}. The latter is also demonstrated in Fig.~\ref{fig:LB_beta_gamma}, where we also show the inner shadow, where a comparison of the overall diameter of the first lensing band with the diameter of the inner shadow provides another powerful way of lifting the degeneracy, without relying on higher-order lensed images.

\subsubsection{Beyond PPN corrections: strong-field modifications}
\begin{figure}
    \centering
    \includegraphics[width=\linewidth]{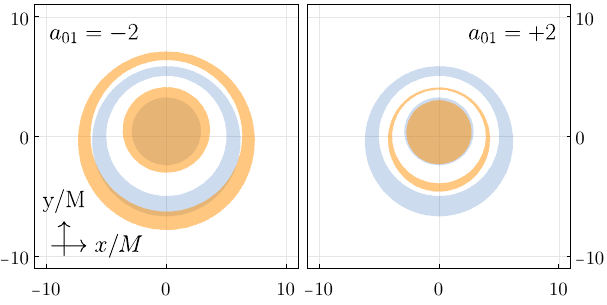}
    \caption{
    As in Fig.~\ref{fig:LB_beta_gamma}, but for single-parameter deviations in $a_{01}$.
    }
    \label{fig:LB_a01}
\end{figure}

We have chosen to include the parameter $a_{01}$ in order to compare with results in the literature~\cite{Younsi:2021dxe,Ayzenberg:2022twz,EventHorizonTelescope:2022urf}. These studies assume black-hole uniqueness and therefore focus on $a_{01}$ and other higher-order deviation parameters. In an expansion around radial asymptotic infinity, $a_{01}$ contributes to the next-to-leading order spherically-symmetric deviations once the leading-order PPN parameters (i.e., $\beta$ and $\gamma$) have been fixed to their GR values. The latter are well-constrained at Solar System scales, while the former is not. In this sense, if one assumes black-hole uniqueness, then $a_{01}$ captures the leading deviations which have not already been constrained by other observations.

We find that the lensing-band framework places an independent \emph{projected constraint} on $a_{01}$, both from above and from below, e.g, for the considered case
$a_{01} \doteqdot -0.06^{+1.79}_{-1.40}$. Together with the theoretical bounds (see App.~\ref{app:paramRanges}), the lensing-band shape itself seems to be sufficient to break the degeneracy with all other deviation parameters investigated here. 
On the one hand, a theoretical constraint on the existence of a Killing horizon (cf. App.~\ref{app:paramRanges}) bounds the value of $a_{01}$ from below~\cite{Kocherlakota:2022jnz}. 
On the other hand, larger values of $a_{01}$ lead to a thinner lensing band, see Fig.~\ref{fig:LB_a01}, hence, eventually to an upper bound. Contrary to previous results~\cite{Younsi:2021dxe,Ayzenberg:2022twz,EventHorizonTelescope:2022urf}, the obtained \emph{projected constraint} does not rely on an external mass measurement. The \emph{projected constraint} on $a_{01}$ has the same order of magnitude than the results inferred from the EHT observation of SgrA$^*$~\cite{EventHorizonTelescope:2022xqj}.

\subsubsection{A shape constraint on the asymptotic angular momentum~$J$}
\begin{figure}
    \centering
    \includegraphics[width=\linewidth]{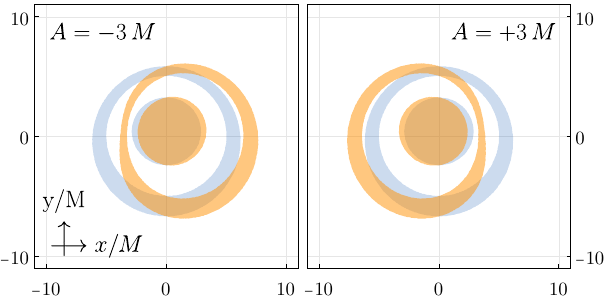}
    \caption{As in Fig.~\ref{fig:LB_beta_gamma}, but for single-parameter deviations in the asymptotic spin $A=J/M$.}
    \label{fig:LB_J}
\end{figure}

The lensing-band formalism also provides access to constraints which rely on the shape and not just on the overall size or width of the lensing band. In the following, we refer to these as ``shape constraints.'' Corresponding variations in the shape of the critical curve and in the second lensing band have been previously considered in Refs.~\cite{Bardeen:1973tla,Gralla:2020srx}. In particular, deformations of their approximately elliptical shape have lead to a proposed null test of GR~\cite{Gralla:2020srx,Cardenas-Avendano:2023dzo}, see also Ref.~\cite{Staelens:2023jgr} for beyond-GR deformations.

An explicit example of such a shape constraint is provided by deformations in the asymptotic angular momentum $J$, see Fig.~\ref{fig:LB_J}. At large angular momentum, exceeding the viable values in Kerr spacetime, the lensing band shape is deformed so far that the lensed emission region modelled in Subsec.~\ref{sec:observing-lensed-emission-region} can no longer be covered by the lensing band. Thus, the underlying geometry is excluded.

The asymptotic angular momentum, $J$, is the only parameter in our parametrized ansatz which explicitly breaks spherical symmetry. One may thus expect that, in fact, similar ``shape constraints'' can also be obtained for further non-spherically-symmetric deformation parameters. In this context, we also note that one can expect much tighter constraints when our formalism is applied to the $n=2$ lensing band and a potential space-based VLBI detection of the associated emission region. We will extend our analysis to the $n=2$ lensing band in future work.

\section{Discussion}
\label{sec:discussion}

We have introduced a framework for leveraging the observation of a lensed image of astrophysical emission surrounding a supermassive black hole to constrain the underlying spacetime geometry systematically. The approach is based on lensing bands: annular regions around the critical curve on the observer's screen, shaped exclusively by the spacetime geometry (and the observer's location).
We have specifically focused on an equatorial definition of the first lensing band, defined as the area in the image space in which incident light rays that have intersected the equatorial plane at least twice.

The framework relies only on three inputs: (i) a precise (geometric) definition of the lensing bands; (ii) a detection of a persistent VLBI feature; and (iii) the assumption that (i) and (ii) can be identified. Although the first requirement may sound simple at first glance, defining what a lensed image is can be challenging, as there are various ways to do so. The choice of geometric definition and the status of a detection of a persistent VLBI feature are discussed in Subsecs.~\ref{sec:lensing-bands} and~\ref{sec:observing-lensed-emission-region}, respectively.

As an explicit demonstrations of the framework, we have: (i) specified to a lensing-band definition which counts how many times an incident light ray has crossed the equatorial plane; and (ii) assumed a detection of an associated VLBI feature for which we model a thin ring. Given these choices, the precise exclusion statement which we quantitatively explore can be stated as follows:
\begin{quote}\textit{
    Upon a detection of a VLBI feature, we exclude spacetimes for which this feature cannot arise from geodesics that traversed the equatorial plane more than once. 
}\end{quote}
If one assumes a definition of the ``photon ring'' as the lensed image associated with the above lensing-band definition, and assumes that a VLBI measurement has detected said photon ring, then the precise exclusion statement can be rephrased as follows:
\begin{quote}\textit{
    We identify spacetimes which can be excluded by a confident detection of the photon ring.
}\end{quote}
Since a confident detection of the photon ring of \mightyseven~remains subject of current research~\cite{Broderick_2022,Lockhart:2022rui,Tiede:2022grp,Cardenas-Avendano:2023dzo,Palumbo:2023auc}, the application we have presented should be understood as a way to quantify the constraining capability of current and future VLBI observations. That is why we have referred to our results as \emph{projected constraints}, and not actual constraints.

In Subsec.~\ref{sec:scheme}, we have detailed the explicit implementation of the lensing-band framework. Our implementation encompasses (i) numerical ray tracing in stationary, axisymmetric, and asymptotically flat but otherwise arbitrary spacetimes, (ii) numerical bisection of potentially non-star-convex lensing-band boundaries, (iii) numerical minimization of the overlap of the resulting geometric lensing bands with an observational prior, and (iv) Bayesian parameter estimation, see also Fig.~\ref{fig:scheme-example} for a visual synthesis. In this context, we also develop a novel free-floating bisection algorithm, the details of which are delegated to App.~\ref{app:bisection}. 

Alongside the conceptual description in this paper, we provide links to the associated computational toolbox. The developed algorithm can also be used to generate adaptive pixel grids to compute high-resolution black-hole images by putting more pixels within the lensing bands, as done in, for example, Ref.~\cite{Cardenas-Avendano:2022csp}. We highlight that, while the present application is entirely focused on the $n=1$ lensing band, our framework is entirely general and can straightforwardly be applied to higher-order lensing bands, see Ref.~\cite{Staelens:2023jgr} for related work on the second photon ring. Constraints from higher-order photon rings will be much tighter because the thickness of the lensing bands (exponentially) decreases with their order. At the same time, higher-order photon rings are more challenging to resolve in observations.

Our work is complementary to previous work based on lensing bands. This includes the null tests of GR proposed for the first Ref.~\cite{Cardenas-Avendano:2023dzo} and second Ref.~\cite{Gralla:2020srx,Paugnat:2022qzy} photon rings. 
Beyond GR, our work complements previous studies which have focused on spherically-symmetric deviations~\cite{Wielgus:2021peu} and on deviations of the largest diameter of the second lensing band \cite{Staelens:2023jgr}.
In particular, the width and the angular shape of the lensing band have the potential to impose independent constraints. We find that these can be used to lift degeneracies, for instance, between(i) the black-hole mass and its horizon location; (ii) the asymptotic mass and angular momentum; and (iii) PPN parameters and horizon-scale modifications.
Nevertheless, the projected constraints obtained from the first lensing band are conservative, i.e., we make no assumption about the astrophysics. This has advantages and disadvantages. On the one hand, such conservative constraints cannot be impacted by incorrect assumptions on the astrophysics. On the other hand, this means that any correct assumption on the astrophysics is expected to improve the constraints. For instance, a fast radial falloff of the emissivity would significantly tighten constraints as the outer edge of the lensing band can be replaced with a new boundary corresponding to a finite radial location at which the intensity is effectively negligible.

Turning the projected constraints obtained in this exploratory study into actual constraints will require a concerted effort. As we have highlighted above, all our quantitative statements crucially rest on whether an observed VLBI feature can be identified and confidently associated with a precise definition of lensed emission.
We expect that such confidence is best achieved when inference methods based on hybrid imaging~\cite{Broderick_2022}, ringing in visibility space~\cite{Cardenas-Avendano:2023dzo}, and polarization~\cite{Palumbo:2023auc} will be combined.
It is likely that some systematic uncertainty in our assumptions will remain and we thus view the above only as a first step. In a second step, it is imperative to verify any resulting constraints by means of joint inference on parameters in the geometry and the astrophysics.
Any such joint-inference study would amount to a test of our assumptions. We emphasize that the latter statement does not rely on a perfect astrophysical modelling within the joint-inference study. It will thus be most valuable to compare to joint-inference studies beyond-GR based on disk models~\cite{Kocherlakota:2022jnz,Ayzenberg:2022twz}, semi-analytic models~\cite{Saurabh:2022oho}, and GRMHD~\cite{Nampalliwar:2022smp}.
Moreover, the second step of joint inference remains crucial, since any added insight on the astrophysics of the plasma surrounding the black hole is expected to tighten the obtained constraints.

With observational earth-based VLBI data becoming more precise and space-based VLBI data becoming available in the future, it is imperative to develop generic theoretical tools that go beyond the mere calculation of the critical curve. In this work, we propose a concrete framework to quantitatively constrain arbitrary stationary and axisymmetric spacetimes beyond GR with the first indirect image. This contributes, both, to a larger effort to quantify a geometric map between four-dimensional near-horizon spacetimes and their two-dimensional optical image on the celestial sphere, and to a larger effort to simultaneously infer geometry and astrophysics.

\begin{acknowledgments}
We thank C. Bambi, A. Chael, A. Eichhorn, D. Gates, R. Gold, P. Kocherlakota, A. Lupsasca, D. Mayerson, F. Pretorius, F. Vincent, and G. Wong for their comments. We also thank an anonymous referee for their constructive comments. A.H. thanks the Princeton Gravity Initiative for their hospitality during substantial parts of this work. A.H. acknowledges support from the PRIME programme of the German Academic Exchange Service (DAAD) with funds from the German Federal Ministry of Education and Research (BMBF). A.H. also acknowledges support by the Deutsche Forschungsgemeinschaft (DFG) under Grant No 406116891 within the Research Training Group RTG 2522/1. A.C.-A. acknowledges support from Will and Kacie Snellings and the Simons Foundation. Some of the simulations presented in this work were performed on computational resources managed and supported by Princeton Research Computing, a consortium of groups including the Princeton Institute for Computational Science and Engineering (PICSciE) and the Office of Information Technology's High Performance Computing Center and Visualization Laboratory at Princeton University. Map colors were based on www.ColorBrewer.org, by Cynthia A. Brewer, Penn State. In preparing this manuscript we made use of the \texttt{python} packages \texttt{numpy}~\cite{Harris:2020xlr}, \texttt{scipy}~\cite{Virtanen:2019joe}, and \texttt{matplotlib}~\cite{Hunter:2007}.

\end{acknowledgments}

\appendix

\section{Free-floating bisection algorithm} 
\label{app:bisection}

Here, we provide the details of an algorithm to solve the following numerical problem: we assume a 2D vector space $\mathbb{R}^2$ with a suitable norm $|\cdot|$, and a compact subregion $\mathcal{R}\subset\mathbb{R}^2$ defined by a binary condition $\mathcal{C}$ such that $\mathcal{C}(\vec{x})=\text{TRUE}$ if $\vec{x}\in\mathcal{R}$ and $\mathcal{C}(\vec{x})=\text{FALSE}$ otherwise. The goal of the algorithm is to numerically approximate one closed piece of the boundary $\partial\mathcal{R}$ of this region. If the region is topologically nontrivial, then there may be several such closed pieces which will then have to be obtained individually\footnote{For the present use-case of a lensing-band region, there are two distinct closed pieces of boundary---one inner and one outer boundary of the lensing band.}.

We assume some prior implementation of a standard bisection algorithm. Given any two points $\vec{x}_1,\,\vec{x}_2\in\mathbb{R}^2$, such standard bisection will approximate the intersection of the boundary $\partial\mathcal{R}$ with the straight line connecting $\vec{x}_1$ and $\vec{x}_2$. If there is no such intersection the bisection will converge to one of the boundary points. Depending on the specific binary bisection condition $\mathcal{C}$, such a standard bisection algorithm will require to pass on suitable information. We refer to this information collectively as \textit{hyper parameters}. For the lensing-band case investigated in the main text, the hyper parameters include all the necessary information to perform ray tracing which then serves to evaluate the bisection condition $\mathcal{C}$.

The abstract algorithm detailed below subdivides the goal of determining the closed boundary. The algorithm is realized in three steps, the last of which is iterated until the a full closed boundary piece has been obtained:
\begin{enumerate}
    \item \textit{Determine an initial section of the boundary:}
    First, we need to obtain two initial points $\vec{x}_1,\,\vec{x}_2\in\mathbb{R}^2$ which approximate two distinct, but sufficiently close, boundary points. With the phrase ``distinct, but sufficiently close'', we refer to $0<|\vec{x}_2 - \vec{x}_1|<\epsilon$ such that $\epsilon$ is smaller than the desired (user-defined) initial numerical precision. This step either requires some prior knowledge about where (at least a piece of) the boundary is located, or needs to be iterated by drawing random pairs of $\vec{x}_1,\,\vec{x}_2\in\mathbb{R}^2$ until an initial boundary section has been successfully identified. We detail in the main text how this initial step can be robustly implemented for the specific application to lensing bands.
    \item \textit{Determine the initial guardrail points:}
    Given $\vec{x}_1$ and $\vec{x}_2$, their distance may serve as the initial precision $p=|\vec{x}_2 - \vec{x}_1|$, their tangent $\vec{t} = \vec{x}_2 - \vec{x}_1$ may provide an initial bisection direction, and, following their normal by a distance $p$ on either side, we define so-called inner and outer ``guard-rail points'' $\vec{i}$ and $\vec{o}$. The latter are defined such that initially $\vec{n} = \vec{o} - \vec{i}$ is perpendicular to $\vec{t}$. By construction, we choose $\vec{x}_2$, $\vec{i}$, and $\vec{o}$ to lie on one straight line.
    \item
    \textit{Take a free-floating bisection step (iterated):}
    Given the initial section of the boundary (defined by $\vec{x}_1$ and $\vec{x}_2$), the precision $p$, and the initial inner and outer guard-rail points $\vec{i}$ and $\vec{o}$, we can now perform the key iterative step of the algorithm. This step is further subdivided into the following tasks:
    \begin{enumerate}
        \item We normalize the tangent by the precision $p$. 
        \item We move forward either $\vec{i}$ or $\vec{o}$ along the tangent $\vec{t}$. In the first iteration this choice is arbitrary but in each subsequent step we determine whether to move forward $\vec{i}$ and $\vec{o}$ depending on whether the scalar product $\vec{t}\cdot\vec{n}$ is positive or negative. This ensures that each bisection step will (approximately) follow the curvature of the boundary.
        \item We perform a standard boundary bisection between the updated $\vec{i}'$ and $\vec{o}'$. If this bisection converges (up to precision $p$) to either of the guardrail points, then we can no longer be sure that the boundary remains between the guardrail points. If the latter occurs, then we increase the precision $p' = a_\text{tighten}\,p$ by a factor of $a_\text{tighten}<1$ and repeat the above three tasks. This increase in the precision eventually ensures that the boundary remains between the guardrail points.
    \end{enumerate}
    Finally, once successful, we relax the precision $p$ by a factor $a_\text{relax}>1$ in favour of efficiency, update $\vec{i}=\vec{i}'$ and $\vec{o}=\vec{o}'$, and append the newly obtained boundary point to the boundary section.
\end{enumerate}
Step (3) can now be iterated until the last point on the current boundary section is only a distance $p$ from the first point, i.e., the boundary section has (approximately) closed.

We implement the above algorithm in the current version of \texttt{LBeyondGR}. We find that the choice of $a_\text{tighten}=1/2$ and $a_\text{relax}=4$ seems to work well but we have not yet performed a dedicated runtime optimization study for these parameters. In general, we expect that the optimal choice of $a_\text{tighten}$ and $a_\text{relax}$ will depend on the specific application at hand.

The two key advantages of such a free-floating bisection algorithm are (i)  that it guarantees a reliable bisection of a closed but otherwise arbitrary (in particular, not necessarily star-convex) boundary and (ii) it incorporates an adaptive step size control. 

\section{The KRZ expansion of circular spacetimes} 
\label{app:KRZ}

In this Appendix we briefly review the expansion of circular spacetimes provided in Ref.~\cite{Konoplya:2016jvv}. For specific details, we refer the reader to Refs.~\cite{Rezzolla:2014mua,Konoplya:2016jvv}. This particular parametrization uses both polynomial and continued-fraction expansions to represent the metric in Eq.~\ref{eq:circularParam} by five free functions $B$, $K$, $N$, $W$, and $\Sigma$, i.e.,
\begin{align}
    \label{eq:gtt-KRZ-functions}
	g_{tt} &= \frac{N(r,\theta)^2 - W(r,\theta)^2\,\sin(\theta)^2}{K(r,\theta)^2}\;,
	\\
    \label{eq:gtphi-KRZ-functions}
	g_{t\phi} & = 2\,W(r,\theta)\,r\,\sin(\theta)^2\;,
	\\
    \label{eq:gphiphi-KRZ-functions}
	g_{\phi\phi} & = K(r,\theta)^2\,r^2\,\sin(\theta)^2\;,
	\\
    \label{eq:grr-KRZ-functions}
	g_{rr} & = \Sigma(r,\theta)\frac{B(r,\theta)^2}{N(r,\theta)^2}\;,
	\\
    \label{eq:grtheta-KRZ-functions}
	\sigma(r,\theta)\,g_{rr} \equiv g_{\theta\theta} & = \Sigma(r,\theta)\,r^2\;.
\end{align}
The functional form in which the free metric functions appear is arbitrary as long as the functions are exact. If, however, the functions are expanded, the specific choice of the functional form may lead to differences at finite order in a respective expansion. It appears that the specific functional form in Eqs.~\ref{eq:gtt-KRZ-functions} and~\ref{eq:grtheta-KRZ-functions} (see Ref.~\cite{Konoplya:2016jvv} for the specifics) is chosen such that (i) the leading asymptotics are fixed by single coefficients and (ii) the Kerr metric is represented by comparatively simple polynomial functions.

One of the free functions is related to coordinate freedom, so the authors in Ref.~\cite{Konoplya:2016jvv} choose to fix
\begin{align}
	\Sigma(r,\theta) = 1 + \frac{A_\text{KRZ}^2}{r^2}\cos(\theta)^2\;,
\end{align}
which is compatible with a representation of Kerr spacetime in Boyer--Lindquist coordinates, cf. $\sigma(r,\theta)$ below Eq.~\ref{eq:circularParam}. 
In the latter case, $A_\text{KRZ}$ matches the spin parameter of Kerr spacetime. More generally, the parameter $A_\text{KRZ}$ is related to the asymptotic choice of coordinates, i.e., to the focal parameter of the oblate spheroidal coordinates in which the asymptotically flat limit of the spacetime is represented--see App.~\ref{app:asymptotics} for further discussion.

By introducing a polar polynomial and a radial coordinate~\cite{Konoplya:2016jvv} 
\begin{align}
    y=\cos(\theta)\;,
    \hspace{0.1\linewidth}
    x = 1-\frac{r_{0}}{r}\;,
\end{align}
respectively, the functions are expanded in a mixture of a low-order polynomial and a continued-fraction expansion. This choice guarantees an asymptotically flat limit and that the lowest-order expansion coefficients map to the first few asymptotic corrections, i.e., map to the PPN parameters. This is not the only choice and fixing more/less of the asymptotic behaviour is, a priori, equally (un)justified. More explicitly, in Ref.~\cite{Konoplya:2016jvv} the polynomial expansion in the angular coordinate $y$ chosen is
\begin{align}
	N(x,y)^2 &= x\,A_0(x) + \sum_{i=1}^\infty A_i(x)\,y^i\;,
	\\
	B(x,y) &= 1 + \sum_{i=0}^\infty B_i(x)\,y^i\;,
	\\
	W(x,y) &= \frac{1}{\Sigma}\sum_{i=1}^\infty W_i(x)\,y^i\;,
	\\
	K^2 - \frac{A_\text{KRZ}}{r}\,W &= 1 + \frac{1}{\Sigma}\sum_{i=1}^\infty K_i(x)\,y^i\;.
\end{align}
The lowest polynomial orders in $(1-x)\equiv r_{0}/r$ are treated separately, i.e.,
\begin{align}
	A_0(x) &= 1 - \epsilon_0(1-x) + (a_{00} - \epsilon_0 + k_{00})(1-x)^2 
	\notag\\
	&\quad
	+ \tilde{A}_0(x)(1-x)^3\;,
	\\
	A_{i>0}(x) &= K_i + \epsilon_i(1-x)^2 + a_{i0}(1-x)^3 
	\notag\\
	&\quad
	+ \tilde{A}_i(x)(1-x)^4\;,
	\\
	B_i(x) &= b_{i0}(1-x) + \tilde{B}_i(x)(1-x)^2\;,
	\\
	K_i(x) &= w_{i0}(1-x)^2 + \tilde{K}_i(x)(1-x)^3\;,
	\\
	W_i(x) &= w_{i0}(1-x)^2 + \tilde{W}_i(x)(1-x)^3\;,
\end{align}
while higher (polynomial) orders are part of a continued-fraction expansion, i.e.,
\begin{align}
	\tilde{A}_i(x) &= 
	\frac{a_{i1}}{1 + \frac{a_{i2}\,x}{1 + \dots}}\;,
	\\
	\tilde{B}_i(x) &= 
	\frac{b_{i1}}{1 + \frac{b_{i2}\,x}{1 + \dots}}\;,
	\\
	\tilde{K}_i(x) &= 
	\frac{k_{i1}}{1 + \frac{k_{i2}\,x}{1 + \dots}}\;,
	\\
	\tilde{W}_i(x) &= 
	\frac{w_{i1}}{1 + \frac{w_{i2}\,x}{1 + \dots}}\;.
\end{align}
We summarize the explicit matching to the Kerr solution in Tab.~\ref{tab:KRZ} and in App.~\ref{app:KRZ}. Note that the matching to the Kerr metric presented in Ref.~\cite{Konoplya:2016jvv} needs additional terms that are not shown explicitly in their paper~\cite{Nampalliwar:2019iti}. The last term in~\cite[Eq.~(A6)]{Konoplya:2016jvv} should read 
\begin{equation}
    \frac{k_{21} r^3_0}{r^3 \left[ 1+ \frac{k_{22}\left(1 -\frac{r_0}{r} \right) }{1+k_{23}\left(1 -\frac{r_0}{r} \right)}\right]},
\end{equation} 
with $k_{22}$ and $k_{23}$ given as in Tab.~\ref{tab:KRZ}.
Nevertheless, the Kerr metric has an exact representation at finite order in the KRZ expansion.
The expansion (or ``bumpy'') parameters are thereby denoted by
$(
    \epsilon_i,\,
    a_{ij},\,
    b_{ij},\,
    w_{ij},\,
    k_{ij},
)$
where the index $i$ ($j$) denotes the order of the (continued-fraction) expansion in $y$ ($x$)~\cite{Konoplya:2016jvv}. 

Similarly, as in Ref.~\cite{Ni:2016uik}, here we use the KRZ metric to parametrize deviations from Kerr as follows. First, we linearly perturb the KRZ parameters $
\Xi=(
        \epsilon_i,\,
        a_{ij},\,
        b_{ij},\,
        w_{ij},\,
        k_{ij}
    )^T
$, around their respective Kerr value, i.e. around $\Xi_\text{Kerr}$. This effectively introduces deviations (``bumpy'' parameters), which we collectively label by $\delta\Xi$, to all of the KRZ parameters, i.e.,
\begin{align}
    \Xi = \Xi_\text{Kerr} + \delta\Xi = 
    \begin{pmatrix}
        \epsilon_i^\text{Kerr}\\
        a_{ij}^\text{Kerr}\\
        b_{ij}^\text{Kerr}\\
        \omega_{ij}^\text{Kerr}\\
        k_{ij}^\text{Kerr}
    \end{pmatrix}
    +
    \begin{pmatrix}
        \delta\epsilon_i\\
        \delta a_{ij}\\
        \delta b_{ij}\\
        \delta \omega_{ij}\\
        \delta k_{ij}
    \end{pmatrix}\;.
\end{align}
Since we seek a modification of the Kerr metric, only two parameters are required to describe the ``background'' KRZ parameters $\Xi_\text{Kerr}$. We chose to parametrize them as a function of the mass $m$ and spin parameter $a$, i.e., $\Xi_\text{Kerr}(m,a)$, and we take $r_0 = m+\sqrt{m^2+a^2}$ as in, for example, Ref.~\cite{EventHorizonTelescope:2022xqj} for the spherically symmetric case and Ref.~\cite{Ni:2016uik} for the axis symmetric case. Note, however, that under this description, $m$ and $a$ are \emph{not} necessarily the asymptotic mass and spin.

Once all expansion parameters, up to the order at which the Kerr metric is exactly represented, are included, the ``background'' KRZ parameters become redundant with a suitable deformation of the deviation parameters.

The coefficients $k_{i0}$ play a special role because they are not suppressed by $r_{0}/r$ and thus modify the oblate spheroidal coordinates to which the Boyer--Lindquist coordinated converge in the flat asymptotic limit, as shown in App.~\ref{app:asymptotics}. We choose to fix all the $k_{i0}$ as in Ref.~\cite{Konoplya:2016jvv}, i.e.,
\begin{align}
    k_{00} =\frac{A_\text{KRZ}^2}{r_{0}^2}\;,
    \quad
    k_{0i}|_{i>0} = 0\;,
\end{align}
such that the parameter $A_\text{KRZ}$ corresponds to the focal parameter of the asymptotically flat oblate spheroidal coordinates (cf. App.~\ref{app:asymptotics}). The coefficients $\epsilon_i$, $b_{i0}$, $\omega_{i0}$, and $k_{i0}$ contribute at order $\mathcal{O}(r_{0}/r)$, while $a_{i0}$ contributes at order $\mathcal{O}(r_{0}^2/r^2)$, cf.~also~ Tab.~\ref{tab:KRZ}.

Without the uniqueness theorems of GR, there is no a priori reason to expect that the Kerr parameters $(m,a)$ are related to the asymptotics of the spacetime. In fact, the asymptotic (Komar) mass and angular momentum are now (to leading order) given by 
\begin{align}
    M = m 
    + \frac{r_0}{2}\,\delta\epsilon_0\;,
    \quad\quad
    J = am
    + \frac{r_0^2}{2}\,\delta\omega_{00}\;.
    \label{eq:asymptotic-mass-and-spin}
\end{align}
When the deformation parameters $\delta\epsilon_0=0$ and $\delta\omega_{00}=0$, the Kerr parameters $(m,a)$ agree with the asymptotic mass and spin, i.e., $M=m$ and $J = aM$, respectively. Therefore, the relation between asymptotic $(M,J)$ and Kerr parameters $(m,a)$ can be viewed as an expression of the uniqueness theorems of GR.

Two of the other KRZ deformations parameters can be identified with the PPN parameters $\gamma$ (related to how much spatial curvature is produced by unit rest mass) and $\beta$ (related to how much ``nonlinearity'' is introduced in the superposition law for gravity) as
\begin{align}
    \gamma &= 1 + \frac{r_0\,\delta b_{00}}{M}\;,
    \quad\quad
    \beta = \gamma - \frac{r_0^2\,\delta a_{00}}{2\,M^2}\;.
    \label{eq:asymptotic-ppn}
\end{align}
Thus, with regards to the leasing asymptotics, the physical parameters of the metric in this parametrization can be identified by trading $(\delta\epsilon_0,\,\delta\omega_{00})$ for $(M,\,J)$ (using Eqs.~\ref{eq:asymptotic-mass-and-spin}) and $(\delta a_{00},\,\delta b_{00})$ for $(\beta,\,\gamma)$ [using Eqs.~\ref{eq:asymptotic-ppn}], to get
\begin{align}
    \delta\epsilon_0 &= \frac{2(M-m)}{r_0}\;,
    \quad
    &\delta\omega_{00} &= \frac{2(J-a\,m)}{r_0^2}\;,
    \\
    \delta a_{00} &= \frac{2\,M^2(\gamma-\beta)}{r_0^2}\;,
    \quad
    &\delta b_{00} &= \frac{M(\gamma-1)}{r_0}\;.
\end{align}
If one assumes uniqueness and Solar System constraints on the PPN parameters, $\gamma$ and $\beta$ are required to be close to one~\cite{Will:2014kxa}. Higher-order coefficients correspond to higher-order corrections to the behaviour at asymptotic infinity. The mapping to the PPN metric of a point particle is shown explicitly in App.~\ref{app:asymptotics}. 
\begin{table*}[]
    {\renewcommand{\arraystretch}{1.5}
    \begin{tabular}{c||c|c|c|c|c||c|c|c|c||c|c|c|c||c|c|c|c}
         & \multicolumn{5}{c||}{Polynomial coefficients}
         & \multicolumn{4}{c||}{\parbox[b]{3cm}{Leading\\continued fraction}}
         & \multicolumn{8}{c}{\parbox[b]{5cm}{Subleading\\continued fraction}}
         \\\hline\hline
         KRZ parameter &
         $\bm{\epsilon_0}$ &  
         $\bm{a_{00}}$ & $\bm{b_{00}}$ & $k_{00}$ & $\bm{\omega_{00}}$ &
         $\bm{a_{01}}$ & $b_{01}$ & $k_{01}$ & $\omega_{01}$ &
         $a_{02}$ & $b_{02}$ & $k_{02}$ & $\omega_{02}$ &
         $a_{03}$ & $b_{03}$ & $k_{03}$ & $\omega_{03}$
         \\
         Kerr &
         $\frac{a^2}{r_{0}^2}$ &
         $0$ & $0$ & $\frac{A_\text{KRZ}^2}{r_{0}^2}$  & 
         $\left(1+\frac{a^2}{r_{0}^2}\right)\frac{a}{r_{0}}$ &
         $0$ & $0$ & $0$ & $0$ & 
         --- & --- & --- & --- &
         --- & --- & --- & ---
         \\[0.5em]
         $\mathcal{O}(r^{-n})$ &
         $1$ &
         $2$ & $1$ & $0$ & $1$ &
         $3$ & $2$ & $1$ & $2$ &
         \multicolumn{4}{c||}{$\cdots$} &
         \multicolumn{4}{c}{$\cdots$}
         \\\hline \hline
         KRZ parameter &
         $\epsilon_1$ & 
         $a_{10}$ & $b_{10}$ & $k_{10}$ & $\omega_{10}$ &
         $a_{11}$ & $b_{11}$ & $k_{11}$ & $\omega_{11}$ &
         $a_{12}$ & $b_{12}$ & $k_{12}$ & $\omega_{12}$ &
         $a_{13}$ & $b_{13}$ & $k_{13}$ & $\omega_{13}$
         \\
         Kerr &
         $0$ & 
         $0$ & $0$ & $0$ & $0$ &
         $0$ & $0$ & $0$ & $0$ & 
         --- & --- & --- & --- &
         --- & --- & --- & ---
         \\[0.5em]
         $\mathcal{O}(r^{-n})$  &
         $2$ &
         $3$ & $1$ & $0$ & $1$ &
         $4$ & $2$ & $1$ & $2$ &
         \multicolumn{4}{c||}{$\cdots$} &
         \multicolumn{4}{c}{$\cdots$}
         \\\hline \hline
         KRZ parameter &
         $\epsilon_2$ & 
         $a_{20}$ & $b_{20}$ & $k_{20}$ & $\omega_{20}$ &
         $a_{21}$ & $b_{21}$ & $k_{21}$ & $\omega_{21}$ &
         $a_{22}$ & $b_{22}$ & $k_{22}$ & $\omega_{22}$ &
         $a_{23}$ & $b_{23}$ & $k_{23}$ & $\omega_{23}$
         \\
         Kerr &
         $0$ &
         $\left(1+\frac{a^2}{r_{0}^2}\right)\frac{a^2}{r_{0}^2}$ &
         $0$ & $0$ & $0$ &
         $-\frac{a^4}{r_{0}^4}$ & $0$ & $-\frac{a^2}{r_{0}^2}$ & $0$ & 
         $0$ & --- & $-\frac{a^2}{r_{0}^2}$ & --- &
         --- & --- & $\frac{a^2}{r_{0}^2}$ & ---
         \\[0.5em]
         $\mathcal{O}(r^{-n})$  &
         $2$ &
         $3$ & $1$ & $0$ & $1$ &
         $4$ & $2$ & $1$ & $2$ &
         \multicolumn{4}{c||}{$\cdots$} &
         \multicolumn{4}{c}{$\cdots$}
         \\\hline\hline
         & \multicolumn{5}{c||}{$\vdots$}
         & \multicolumn{4}{c||}{$\vdots$}
         & \multicolumn{8}{c}{$\vdots$}
    \end{tabular}
    }
    \caption{
    \label{tab:KRZ} The $51$ parameters $(\epsilon_i,\,a_{ij},\,b_{ij},\,w_{ij},\,k_{ij})$, up to $\mathcal{O}(x^3y^2)$, ( $i\leqslant2$ and $j\leqslant3$), required in the KRZ expansion to represent the Kerr spacetime exactly. The KRZ parameters $\epsilon_i$ and those with $j=0$ are chosen to obtain a specific form of asymptotics. In contrast, the KRZ parameters with $j\geqslant1$ are part of a continued-fraction expansion. Once any of the continued-fraction parameters vanishes, all corresponding higher-order parameters do not contribute. The KRZ parameters with odd $i$ break reflection symmetry about the equatorial plane, and thus all vanish for Kerr spacetime. 
    We do not consider such deviations in the present work. We list the respective values reproducing the Kerr spacetime, where $r_0 = m + \sqrt{m^2-a^2}$. 
    We also indicate the order in the $1/r$ expansion around $r=\infty$ at which the respective deviation parameter affects the asymptotic behaviour of the spacetime by $\mathcal{O}(r^{-n})$. All the coefficients for which $n=0$ affect the asymptotic definition of oblate spheroidal coordinates, cf. App.~\ref{app:asymptotics}.
    Finally, we highlight in bold the five deviation parameters which are investigated in the present work. Some of the leading and subleading continued-fraction parameters have also been investigated in Refs.~\cite{Ayzenberg:2022twz,Younsi:2021dxe}.
    }
\end{table*}
In summary, the resulting expansion around the Kerr spacetime within the KRZ parametrization thus exhibits the following parameters:
\begin{itemize}
    \item 
    the location of the horizon:
    $r_0$,
    \item 
    a horizon parameter of the background Kerr geometry parameter:
    $a$,
    \item
    leading asymptotics:
    $(M,\,J)$,
    \item
    PPN asymptotics:
    $(\beta,\,\gamma)$,
    \item
    beyond-PPN ($\theta$-dependent) asymptotics:\\
    $(\delta\epsilon_{i},\,\delta a_{i0},\,\delta b_{i0},\,\delta k_{i0},\,\delta\omega_{i0})\;\forall i\geqslant1$, which we will not modify here,
    \item
    continued-fraction coefficients:\\
    $(\delta a_{ij},\,\delta b_{ij},\,\delta k_{ij},\,\delta\omega_{ij})\;\forall j\geqslant1$, of which we modify only $\delta a_{01}$.
\end{itemize}
As a result, we obtain a spacetime with asymptotic mass and angular momentum $(M,\,J)$ and five deviation parameters $(r_0,\,s,\,\beta,\,\gamma,\,\delta a_{01})$. When $r_0=M+\sqrt{M^2 - J^2/M^2}$, $a=J/M$, $\beta=\gamma=1$, and $\delta a_{01}=0$, the spacetime reduces to Kerr spacetime in Boyer--Lindquist coordinates. More generally, its metric elements are compactly written as
\begin{align}
	g_{t\phi} &= -\frac{2 M r A \sin ^2\theta}{\Sigma}\;,
    \notag\\[0.5em]
    g_{rr} &= 
        \frac{
            \Sigma \left(1 - \frac{(1-\gamma)\,M}{r}\right)^2
        }{
            \left(
                \Delta_{\beta\gamma}
                + \frac{r_0^3 (r-r_0)}{r^2}\,a_{01}
            \right),
        }
    \notag\\[0.5em]
    g_{\theta\theta} &= \Sigma\;,
    \notag\\[1em]
    g_{\phi\phi} &= \left(
        a^2+r^2
        +\frac{2 M r a^2 \sin^2\theta}{\Sigma}\,\Delta_{aA_\text{KRZ}}
    \right)\sin^2\theta\;,
    \notag\\[0.5em]
    g_{tt} &= - \frac{
        \left(
        		\Delta_{\beta\gamma} 
        		+ \frac{r_0^3 (r-r_0)}{r^2}\,a_{01}
        	\right)\sin^2\theta
        - g_{t\phi}^2
    }{
        g_{\phi\phi}
    }\;.
    \label{eq:our-KRZ-truncation-before-identification}
\end{align}
We use the common shorthand $A=J/M$ as well as
\begin{align}
    \Sigma &= r^2 + A_\text{KRZ}^2\cos^2\theta\;,
    \\
    \Delta &= r^2 - 2Mr + A_\text{KRZ}^2.
\end{align}
It is $A_\text{KRZ}$, not $A=J/M$ or $a$, which appears in shorthand.
In addition, we use additional deviation shorthand, i.e.,
\begin{align}	
    \Delta_{\beta\gamma} &= \frac{(r-r_0) \left[
        \Delta 
        -2 M^2 (\beta -\gamma + \frac{r_0}{M})
        +r_0 (r+r_0)
    \right]}{r}\;,
    \\[0.5em]
    \Delta_{aA_\text{KRZ}} &= 
    \left(\frac{A\,A_\text{KRZ}}{a^2} - \frac{r(a^2 - A_\text{KRZ}^2)}{2\,M\,a^2}\right)\frac{1}{\sin^2\theta}
    \notag\\&\quad
    + \left(\frac{a^2 - A_\text{KRZ}^2}{2\,M\,r} - \frac{a^2 + r_0^2}{2\,M\,r_0}\right)\frac{\cos^2\theta}{\sin^2\theta}.
\end{align}
The latter reduce to $\Delta_{\beta\gamma} \rightarrow \Delta$ and $\Delta_{aA_\text{KRZ}} \rightarrow 1$ in the Kerr limit.

The focal parameter $A_\text{KRZ}$, the asymptotic angular momentum $A=J/M$, and the background spin parameter $a$ are, in principle, independent quantities. In the main text, we choose to identify the focal parameter of the asymptotic coordinates with the background spin parameter, i.e., $A_\text{KRZ}\equiv a$ but we keep the asymptotic spin parameter $A=J/M$ as an independent parameter.

\section{KRZ parameters and the asymptotic limit} 
\label{app:asymptotics}

\subsection{The far-field metric in GR in quasi-isotropic coordinates}
\label{sec:far-fieldQI}

Kerr black holes are unique in GR (see Ref.~\cite{Robinson:1975bv} for a proof and its mathematically precise statement). Thus, in quasi-isotropic coordinates~\cite{Brandt:1996si} $(t,r_{\rm I},\theta,\phi)$, the asymptotic behavior of any black hole in GR must match the point-mass PPN metric~\cite{Will:1972zz}
\begin{align}
    ds^2 =&
    -\left[1-2\frac{M}{r_{\rm I}} + \beta \frac{M^2}{r_{\rm I}^2} + \mathcal{O}\left( \frac{1}{r_{\rm I}^3}\right)\right]dt^2
    \notag\\&
    +\left[1+2\,\gamma\,\frac{M}{r_{\rm I}} + \mathcal{O}\left(\frac{1}{r_{\rm I}^2}\right) \right]dr_{\rm I}^2
    + \left[r_{\rm I}^2 + \mathcal{O}(1)\right]d\Omega
    \notag\\&
    -\left[4\,\sin(\theta)^2\frac{J}{M}\frac{M}{r_{\rm I}} + \mathcal{O}\left(\frac{1}{r_{\rm I}^2}\right)\right]dt\,d\phi\;,
\end{align}
where $M$ and $J$ can be identified as the Newtonian mass and angular momentum (in agreement with the mass and angular momentum of the full Kerr solution), and $d\Omega = d\theta^2 + \sin(\theta)^2\,d\phi^2$. The two PPN parameters in GR are $\beta_\text{GR}\equiv\gamma_\text{GR}\equiv 1$~\cite{Will:2014kxa}. Under the assumption that AGNs are black holes, this is sufficient to test the asymptotic behavior required by GR. 

Other uniqueness theorems extend the importance of the above asymptotic behavior to the vacuum exterior of generic spacetimes in two ways. First, in spherical symmetry, i.e., for $J\rightarrow0$, the Jebsen--Birkhoff theorem~\cite{2005GReGr..37.2253J, 1923rmp..book.....B} guarantees that the same asymptotic behavior applies to the vacuum exterior of any source~\cite{1923rmp..book.....B}. Second, the leading-order (i.e., up to $\mathcal{O}(1/r^2)$) behavior also applies to the vacuum exterior of any stationary and axisymmetric source~\cite{1995CQGra..12..149K}.

\subsection{The far-field metric in GR in oblate spheroidal coordinates}
\label{sec:far-fieldOS}

We rewrite the Kerr spacetime in Boyer--Lindquist coordinates as deviations from flat space (see,e.g., Ref.~\cite{Visser:2007fj}), i.e.,
\begin{align}
    ds^2 =& 
    -  dt^2 
    + \frac{r^2+ a^2\cos^2\theta}{r^2+a^2}\; dr^2
    \\\notag&
    + (r^2+ a^2\cos\theta^2) \; d\theta^2
    + (r^2+a^2)\; \sin^2\theta   \; d\phi^2
    \\\notag&
    + \frac{r_{0}}{r} \left[
        \frac{\left(d t - a \sin^2\theta\; d\phi\right)^2}{(1+ a^2\cos^2\theta/r^2)}
        + \frac{(1+a^2\cos^2\theta/r^2) \; d r^2}{1-r_{0}/r+a^2/r^2} 
    \right].
\end{align}
In this form, the first two lines denote Minkowski space in oblate spheroidal coordinates, while the last line denotes corrections starting at order $\mathcal{O}(r_{0}/r)$. 

When comparing this to the asymptotic limit (neglecting $\mathcal{O}(r_{0}/r)$) of the KRZ expansion, i.e., to 
\begin{align}
    ds^2 =& 
    - dt^2 
    + dr^2
    + (r^2+ A_\text{KRZ}^2\cos\theta^2) \; d\theta^2
    \\\notag&
    + \left[r^2 + r_{0}^2(k_{00} + k_{20}\cos^2\theta + \dots)\right]\; \sin^2\theta   \; d\phi^2
    \\\notag&
    +\mathcal{O}(r_{0}/r)\;,
\end{align}
we find that all the $k_{i0}$ contribute to the definition of the oblate spheroidal coordinates in which the flat limit of the KRZ spacetime is expressed. In particular, we find that the choices
\begin{align}
    k_{00} &= \frac{A_\text{KRZ}^2}{r_0^2}\;,
    \\
    k_{i0} &= 0\;,
\end{align}
are required such that the asymptotics match to those of of a spinning spacetime (in oblate spheroidal coordinates). Hence, the parameter $A_\text{KRZ}$ determines how to transform from Boyer--Lindquist to screen coordinates and thus how to set initial conditions for ray tracing geodesics backwards in time, see,  e.g., Ref.~\cite{Younsi:2016azx}.

\section{Theoretical constraints on the KRZ parameters} 
\label{app:paramRanges}

Any fixed set of non-vanishing KRZ parameters is constrained by theoretical consistency considerations in the external spacetime i.e., for $r>r_{0}$. In particular, this includes (i) the absence of further (Killing) horizons; (ii) the absence of closed timelike curves; and (iii) no signature changes in the metric.
\begin{enumerate}
    \item[(i)]
    Here, we assume that event horizons coincide with the Killing horizons. In GR, this equivalence is guaranteed by the Hawking rigidity theorem. We caution that, beyond GR, it is a non-trivial assumption. For stationary and axisymmetric spacetimes (with $t$ and $\phi$ the two Killing coordinates), Killing horizons occur whenever
    \begin{align}
        \label{eq:Killing-condition}
        g_{t\phi}^2 - g_{tt}g_{\phi\phi} = 0\;.
    \end{align}
    We thus need to make sure that $r=r_{0}$ remains the largest root of Eq.~\ref{eq:Killing-condition}.
    \item[(ii)]
    The sign of $g_{\phi\phi}$ determines whether changes in $\phi$ are spacelike or timelike. At the same time, asymptotic flatness requires that $\phi$ is a periodic coordinate in $\phi\in[0,2\pi]$. Thus, whenever $g_{\phi\phi}(r,\theta)<0$, it means that there exists a closed timelike curve. If we want to avoid such causality violations, we need to therefore require
    \begin{align}
        \label{eq:noCTC-condition}
        g_{\phi\phi}>0\;,
    \end{align}
    at least outside the horizon.
    \item[(iii)]
    We also check that the signature of the metric does not change, i.e., that
    \begin{align}
        \det(g)<0\;.
    \end{align}
    However, we find that this does not impose conditions on the investigated parameters. 
\end{enumerate}

In the following, we collect the implied theoretical bounds on the parameter sets which we investigate in the present work. We implement these theoretical constraints as priors in the Bayesian analysis presented in the main text, and rule out the respective spacetimes before applying any lensing-band constraints.

Let us now focus on the explicit bounds obtained when identifying $A_\text{KRZ}\equiv a$ (but keeping $A=J/M$ independent) as in the main text. From the form of the metric in Eq.~\ref{eq:our-KRZ-truncation} (or even in the more general form in Eq.~\ref{eq:our-KRZ-truncation-before-identification}) and the Killing-horizon condition in Eq.~\ref{eq:Killing-condition}, we find that Killing horizons occur whenever
\begin{align}
    0 &= \Delta_{\beta\gamma} + \frac{r_0^3 (r-r_0)}{r^2}\,a_{01}
    \\\notag &
    = \frac{(r-r_0)}{r}\!\left[\Delta -2 M (\beta\!-\!\gamma\!+\!r_0)+r_0 (r+r_0) + \frac{r_0^3}{r} \,a_{01}\right]
    \,.
\end{align}
Hence, the (Killing) horizon condition (i) is fulfilled whenever the largest real-valued root 
$r=r_{+}$ of the above equation remains inside of the horizon, i.e., $r_{+}\leqslant r_{0}$. (Note that $r=r_{0}$ itself always remains a root, i.e., a Killing horizon.)

Regarding closed timelike curves, we find that $g_{\phi\phi}>0$, is guaranteed whenever
\begin{align}
    2\,A\,s\,M + s^2\,r_0 + r_0^3
    > 0\;.
\end{align}
To obtain the explicit expression, we have divided out manifestly positive factors and specified to the equatorial plane, i.e., to $\chi\equiv\cos(\theta)=0$ (which can be shown to imply the tightest bound as it comes with a prefactor which is manifestly positive for all $r>r_{0}$). Similarly, the tightest bound arises when evaluating the condition at $r=r_{0}$.
 
For the considered metric parameters, condition (iii) does not imply further constraints.

\bibliography{bibliography}

\begin{thebibliography}{133}%
\makeatletter
\providecommand \@ifxundefined [1]{%
 \@ifx{#1\undefined}
}%
\providecommand \@ifnum [1]{%
 \ifnum #1\expandafter \@firstoftwo
 \else \expandafter \@secondoftwo
 \fi
}%
\providecommand \@ifx [1]{%
 \ifx #1\expandafter \@firstoftwo
 \else \expandafter \@secondoftwo
 \fi
}%
\providecommand \natexlab [1]{#1}%
\providecommand \enquote  [1]{``#1''}%
\providecommand \bibnamefont  [1]{#1}%
\providecommand \bibfnamefont [1]{#1}%
\providecommand \citenamefont [1]{#1}%
\providecommand \href@noop [0]{\@secondoftwo}%
\providecommand \href [0]{\begingroup \@sanitize@url \@href}%
\providecommand \@href[1]{\@@startlink{#1}\@@href}%
\providecommand \@@href[1]{\endgroup#1\@@endlink}%
\providecommand \@sanitize@url [0]{\catcode `\\12\catcode `\$12\catcode
  `\&12\catcode `\#12\catcode `\^12\catcode `\_12\catcode `\%12\relax}%
\providecommand \@@startlink[1]{}%
\providecommand \@@endlink[0]{}%
\providecommand \url  [0]{\begingroup\@sanitize@url \@url }%
\providecommand \@url [1]{\endgroup\@href {#1}{\urlprefix }}%
\providecommand \urlprefix  [0]{URL }%
\providecommand \Eprint [0]{\href }%
\providecommand \doibase [0]{https://doi.org/}%
\providecommand \selectlanguage [0]{\@gobble}%
\providecommand \bibinfo  [0]{\@secondoftwo}%
\providecommand \bibfield  [0]{\@secondoftwo}%
\providecommand \translation [1]{[#1]}%
\providecommand \BibitemOpen [0]{}%
\providecommand \bibitemStop [0]{}%
\providecommand \bibitemNoStop [0]{.\EOS\space}%
\providecommand \EOS [0]{\spacefactor3000\relax}%
\providecommand \BibitemShut  [1]{\csname bibitem#1\endcsname}%
\let\auto@bib@innerbib\@empty
\bibitem [{\citenamefont {{Jebsen}}(2005)}]{2005GReGr..37.2253J}%
  \BibitemOpen
  \bibfield  {author} {\bibinfo {author} {\bibfnamefont {J.~T.}\ \bibnamefont
  {{Jebsen}}},\ }\bibfield  {title} {\bibinfo {title} {{On the general
  spherically symmetric solutions of Einstein's gravitational equations in
  vacuo}},\ }\href {https://doi.org/10.1007/s10714-005-0168-y} {\bibfield
  {journal} {\bibinfo  {journal} {General Relativity and Gravitation}\ }\textbf
  {\bibinfo {volume} {37}},\ \bibinfo {pages} {2253} (\bibinfo {year}
  {2005})}\BibitemShut {NoStop}%
\bibitem [{\citenamefont {{Birkhoff}}\ and\ \citenamefont
  {{Langer}}(1923)}]{1923rmp..book.....B}%
  \BibitemOpen
  \bibfield  {author} {\bibinfo {author} {\bibfnamefont {G.~D.}\ \bibnamefont
  {{Birkhoff}}}\ and\ \bibinfo {author} {\bibfnamefont {R.~E.}\ \bibnamefont
  {{Langer}}},\ }\href@noop {} {\emph {\bibinfo {title} {{Relativity and modern
  physics}}}}\ (\bibinfo {year} {1923})\BibitemShut {NoStop}%
\bibitem [{\citenamefont {Robinson}(1975)}]{Robinson:1975bv}%
  \BibitemOpen
  \bibfield  {author} {\bibinfo {author} {\bibfnamefont {D.~C.}\ \bibnamefont
  {Robinson}},\ }\bibfield  {title} {\bibinfo {title} {{Uniqueness of the Kerr
  black hole}},\ }\href {https://doi.org/10.1103/PhysRevLett.34.905} {\bibfield
   {journal} {\bibinfo  {journal} {Phys. Rev. Lett.}\ }\textbf {\bibinfo
  {volume} {34}},\ \bibinfo {pages} {905} (\bibinfo {year} {1975})}\BibitemShut
  {NoStop}%
\bibitem [{\citenamefont {{Kennefick}}\ and\ \citenamefont {{{\'O}
  Murchadha}}(1995)}]{1995CQGra..12..149K}%
  \BibitemOpen
  \bibfield  {author} {\bibinfo {author} {\bibfnamefont {D.}~\bibnamefont
  {{Kennefick}}}\ and\ \bibinfo {author} {\bibfnamefont {N.}~\bibnamefont
  {{{\'O} Murchadha}}},\ }\bibfield  {title} {\bibinfo {title} {{Weakly
  decaying asymptotically flat static and stationary solutions to the Einstein
  equations}},\ }\href {https://doi.org/10.1088/0264-9381/12/1/013} {\bibfield
  {journal} {\bibinfo  {journal} {Classical and Quantum Gravity}\ }\textbf
  {\bibinfo {volume} {12}},\ \bibinfo {pages} {149} (\bibinfo {year} {1995})},\
  \Eprint {https://arxiv.org/abs/gr-qc/9311012} {arXiv:gr-qc/9311012 [gr-qc]}
  \BibitemShut {NoStop}%
\bibitem [{\citenamefont {Nordtvedt}(1968)}]{Nordtvedt:1968qs}%
  \BibitemOpen
  \bibfield  {author} {\bibinfo {author} {\bibfnamefont {K.}~\bibnamefont
  {Nordtvedt}},\ }\bibfield  {title} {\bibinfo {title} {{Equivalence Principle
  for Massive Bodies. 2. Theory}},\ }\href
  {https://doi.org/10.1103/PhysRev.169.1017} {\bibfield  {journal} {\bibinfo
  {journal} {Phys. Rev.}\ }\textbf {\bibinfo {volume} {169}},\ \bibinfo {pages}
  {1017} (\bibinfo {year} {1968})}\BibitemShut {NoStop}%
\bibitem [{\citenamefont {Will}(1971)}]{Will:1971zzb}%
  \BibitemOpen
  \bibfield  {author} {\bibinfo {author} {\bibfnamefont {C.~M.}\ \bibnamefont
  {Will}},\ }\bibfield  {title} {\bibinfo {title} {{Theoretical Frameworks for
  Testing Relativistic Gravity. 2. Parametrized Post-Newtonian Hydrodynamics,
  and the Nordtvedt Effect}},\ }\href {https://doi.org/10.1086/150804}
  {\bibfield  {journal} {\bibinfo  {journal} {Astrophys. J.}\ }\textbf
  {\bibinfo {volume} {163}},\ \bibinfo {pages} {611} (\bibinfo {year}
  {1971})}\BibitemShut {NoStop}%
\bibitem [{\citenamefont {{Will}}(1993)}]{1993tegp.book.....W}%
  \BibitemOpen
  \bibfield  {author} {\bibinfo {author} {\bibfnamefont {C.~M.}\ \bibnamefont
  {{Will}}},\ }\href@noop {} {\emph {\bibinfo {title} {{Theory and Experiment
  in Gravitational Physics}}}}\ (\bibinfo {year} {1993})\BibitemShut {NoStop}%
\bibitem [{\citenamefont {Will}(2014)}]{Will:2014kxa}%
  \BibitemOpen
  \bibfield  {author} {\bibinfo {author} {\bibfnamefont {C.~M.}\ \bibnamefont
  {Will}},\ }\bibfield  {title} {\bibinfo {title} {{The Confrontation between
  General Relativity and Experiment}},\ }\href
  {https://doi.org/10.12942/lrr-2014-4} {\bibfield  {journal} {\bibinfo
  {journal} {Living Rev. Rel.}\ }\textbf {\bibinfo {volume} {17}},\ \bibinfo
  {pages} {4} (\bibinfo {year} {2014})},\ \Eprint
  {https://arxiv.org/abs/1403.7377} {arXiv:1403.7377 [gr-qc]} \BibitemShut
  {NoStop}%
\bibitem [{\citenamefont {Baker}\ \emph {et~al.}(2015)\citenamefont {Baker},
  \citenamefont {Psaltis},\ and\ \citenamefont {Skordis}}]{Baker:2014zba}%
  \BibitemOpen
  \bibfield  {author} {\bibinfo {author} {\bibfnamefont {T.}~\bibnamefont
  {Baker}}, \bibinfo {author} {\bibfnamefont {D.}~\bibnamefont {Psaltis}},\
  and\ \bibinfo {author} {\bibfnamefont {C.}~\bibnamefont {Skordis}},\
  }\bibfield  {title} {\bibinfo {title} {{Linking Tests of Gravity On All
  Scales: from the Strong-Field Regime to Cosmology}},\ }\href
  {https://doi.org/10.1088/0004-637X/802/1/63} {\bibfield  {journal} {\bibinfo
  {journal} {Astrophys. J.}\ }\textbf {\bibinfo {volume} {802}},\ \bibinfo
  {pages} {63} (\bibinfo {year} {2015})},\ \Eprint
  {https://arxiv.org/abs/1412.3455} {arXiv:1412.3455 [astro-ph.CO]}
  \BibitemShut {NoStop}%
\bibitem [{\citenamefont {Arnowitt}\ \emph {et~al.}(1962)\citenamefont
  {Arnowitt}, \citenamefont {Deser},\ and\ \citenamefont
  {Misner}}]{arnowitt1962dynamics}%
  \BibitemOpen
  \bibfield  {author} {\bibinfo {author} {\bibfnamefont {R.}~\bibnamefont
  {Arnowitt}}, \bibinfo {author} {\bibfnamefont {S.}~\bibnamefont {Deser}},\
  and\ \bibinfo {author} {\bibfnamefont {C.~W.}\ \bibnamefont {Misner}},\
  }\bibfield  {title} {\bibinfo {title} {The dynamics of general relativity,
  gravitation: An introduction to current research},\ }\href@noop {} {\bibfield
   {journal} {\bibinfo  {journal} {Chap}\ }\textbf {\bibinfo {volume} {7}},\
  \bibinfo {pages} {227} (\bibinfo {year} {1962})}\BibitemShut {NoStop}%
\bibitem [{\citenamefont {Brown}\ and\ \citenamefont
  {York}(1993)}]{Brown:1992br}%
  \BibitemOpen
  \bibfield  {author} {\bibinfo {author} {\bibfnamefont {J.~D.}\ \bibnamefont
  {Brown}}\ and\ \bibinfo {author} {\bibfnamefont {J.~W.}\ \bibnamefont {York},
  \bibfnamefont {Jr.}},\ }\bibfield  {title} {\bibinfo {title} {{Quasilocal
  energy and conserved charges derived from the gravitational action}},\ }\href
  {https://doi.org/10.1103/PhysRevD.47.1407} {\bibfield  {journal} {\bibinfo
  {journal} {Phys. Rev. D}\ }\textbf {\bibinfo {volume} {47}},\ \bibinfo
  {pages} {1407} (\bibinfo {year} {1993})},\ \Eprint
  {https://arxiv.org/abs/gr-qc/9209012} {arXiv:gr-qc/9209012} \BibitemShut
  {NoStop}%
\bibitem [{\citenamefont {Geraci}\ \emph {et~al.}(2008)\citenamefont {Geraci},
  \citenamefont {Smullin}, \citenamefont {Weld}, \citenamefont {Chiaverini},\
  and\ \citenamefont {Kapitulnik}}]{Geraci:2008hb}%
  \BibitemOpen
  \bibfield  {author} {\bibinfo {author} {\bibfnamefont {A.~A.}\ \bibnamefont
  {Geraci}}, \bibinfo {author} {\bibfnamefont {S.~J.}\ \bibnamefont {Smullin}},
  \bibinfo {author} {\bibfnamefont {D.~M.}\ \bibnamefont {Weld}}, \bibinfo
  {author} {\bibfnamefont {J.}~\bibnamefont {Chiaverini}},\ and\ \bibinfo
  {author} {\bibfnamefont {A.}~\bibnamefont {Kapitulnik}},\ }\bibfield  {title}
  {\bibinfo {title} {{Improved constraints on non-Newtonian forces at 10
  microns}},\ }\href {https://doi.org/10.1103/PhysRevD.78.022002} {\bibfield
  {journal} {\bibinfo  {journal} {Phys. Rev. D}\ }\textbf {\bibinfo {volume}
  {78}},\ \bibinfo {pages} {022002} (\bibinfo {year} {2008})},\ \Eprint
  {https://arxiv.org/abs/0802.2350} {arXiv:0802.2350 [hep-ex]} \BibitemShut
  {NoStop}%
\bibitem [{\citenamefont {Westphal}\ \emph {et~al.}(2021)\citenamefont
  {Westphal}, \citenamefont {Hepach}, \citenamefont {Pfaff},\ and\
  \citenamefont {Aspelmeyer}}]{Westphal:2020okx}%
  \BibitemOpen
  \bibfield  {author} {\bibinfo {author} {\bibfnamefont {T.}~\bibnamefont
  {Westphal}}, \bibinfo {author} {\bibfnamefont {H.}~\bibnamefont {Hepach}},
  \bibinfo {author} {\bibfnamefont {J.}~\bibnamefont {Pfaff}},\ and\ \bibinfo
  {author} {\bibfnamefont {M.}~\bibnamefont {Aspelmeyer}},\ }\bibfield  {title}
  {\bibinfo {title} {{Measurement of gravitational coupling between
  millimetre-sized masses}},\ }\href
  {https://doi.org/10.1038/s41586-021-03250-7} {\bibfield  {journal} {\bibinfo
  {journal} {Nature}\ }\textbf {\bibinfo {volume} {591}},\ \bibinfo {pages}
  {225} (\bibinfo {year} {2021})},\ \Eprint {https://arxiv.org/abs/2009.09546}
  {arXiv:2009.09546 [gr-qc]} \BibitemShut {NoStop}%
\bibitem [{\citenamefont {Akiyama}\ \emph
  {et~al.}(2019{\natexlab{a}})\citenamefont {Akiyama} \emph
  {et~al.}}]{EventHorizonTelescope:2019dse}%
  \BibitemOpen
  \bibfield  {author} {\bibinfo {author} {\bibfnamefont {K.}~\bibnamefont
  {Akiyama}} \emph {et~al.} (\bibinfo {collaboration} {Event Horizon
  Telescope}),\ }\bibfield  {title} {\bibinfo {title} {{First M87 Event Horizon
  Telescope Results. I. The Shadow of the Supermassive Black Hole}},\ }\href
  {https://doi.org/10.3847/2041-8213/ab0ec7} {\bibfield  {journal} {\bibinfo
  {journal} {Astrophys. J. Lett.}\ }\textbf {\bibinfo {volume} {875}},\
  \bibinfo {pages} {L1} (\bibinfo {year} {2019}{\natexlab{a}})},\ \Eprint
  {https://arxiv.org/abs/1906.11238} {arXiv:1906.11238 [astro-ph.GA]}
  \BibitemShut {NoStop}%
\bibitem [{\citenamefont {Akiyama}\ \emph
  {et~al.}(2022{\natexlab{a}})\citenamefont {Akiyama} \emph
  {et~al.}}]{EventHorizonTelescope:2022xnr}%
  \BibitemOpen
  \bibfield  {author} {\bibinfo {author} {\bibfnamefont {K.}~\bibnamefont
  {Akiyama}} \emph {et~al.} (\bibinfo {collaboration} {Event Horizon
  Telescope}),\ }\bibfield  {title} {\bibinfo {title} {{First Sagittarius A*
  Event Horizon Telescope Results. I. The Shadow of the Supermassive Black Hole
  in the Center of the Milky Way}},\ }\href
  {https://doi.org/10.3847/2041-8213/ac6674} {\bibfield  {journal} {\bibinfo
  {journal} {Astrophys. J. Lett.}\ }\textbf {\bibinfo {volume} {930}},\
  \bibinfo {pages} {L12} (\bibinfo {year} {2022}{\natexlab{a}})}\BibitemShut
  {NoStop}%
\bibitem [{\citenamefont {Bambi}\ \emph {et~al.}(2017)\citenamefont {Bambi},
  \citenamefont {Cardenas-Avendano}, \citenamefont {Dauser}, \citenamefont
  {Garcia},\ and\ \citenamefont {Nampalliwar}}]{Bambi:2016sac}%
  \BibitemOpen
  \bibfield  {author} {\bibinfo {author} {\bibfnamefont {C.}~\bibnamefont
  {Bambi}}, \bibinfo {author} {\bibfnamefont {A.}~\bibnamefont
  {Cardenas-Avendano}}, \bibinfo {author} {\bibfnamefont {T.}~\bibnamefont
  {Dauser}}, \bibinfo {author} {\bibfnamefont {J.~A.}\ \bibnamefont {Garcia}},\
  and\ \bibinfo {author} {\bibfnamefont {S.}~\bibnamefont {Nampalliwar}},\
  }\bibfield  {title} {\bibinfo {title} {{Testing the Kerr black hole
  hypothesis using X-ray reflection spectroscopy}},\ }\href
  {https://doi.org/10.3847/1538-4357/aa74c0} {\bibfield  {journal} {\bibinfo
  {journal} {Astrophys. J.}\ }\textbf {\bibinfo {volume} {842}},\ \bibinfo
  {pages} {76} (\bibinfo {year} {2017})},\ \Eprint
  {https://arxiv.org/abs/1607.00596} {arXiv:1607.00596 [gr-qc]} \BibitemShut
  {NoStop}%
\bibitem [{\citenamefont {Bambi}(2022)}]{Bambi:2022dtw}%
  \BibitemOpen
  \bibfield  {author} {\bibinfo {author} {\bibfnamefont {C.}~\bibnamefont
  {Bambi}},\ }\bibfield  {title} {\bibinfo {title} {{Testing Gravity with Black
  Hole X-Ray Data}},\ }\href@noop {} {\  (\bibinfo {year} {2022})},\ \Eprint
  {https://arxiv.org/abs/2210.05322} {arXiv:2210.05322 [gr-qc]} \BibitemShut
  {NoStop}%
\bibitem [{\citenamefont {{GRAVITY Collaboration}}\ \emph
  {et~al.}(2018)\citenamefont {{GRAVITY Collaboration}}, \citenamefont
  {{Abuter}}, \citenamefont {{Amorim}}, \citenamefont {{Baub{\"o}ck}},
  \citenamefont {{Berger}}, \citenamefont {{Bonnet}}, \citenamefont
  {{Brandner}}, \citenamefont {{Cl{\'e}net}}, \citenamefont {{Coud{\'e} Du
  Foresto}}, \citenamefont {{de Zeeuw}}, \citenamefont {{Deen}}, \citenamefont
  {{Dexter}}, \citenamefont {{Duvert}}, \citenamefont {{Eckart}}, \citenamefont
  {{Eisenhauer}}, \citenamefont {{F{\"o}rster Schreiber}}, \citenamefont
  {{Garcia}}, \citenamefont {{Gao}}, \citenamefont {{Gendron}}, \citenamefont
  {{Genzel}}, \citenamefont {{Gillessen}}, \citenamefont {{Guajardo}},
  \citenamefont {{Habibi}}, \citenamefont {{Haubois}}, \citenamefont
  {{Henning}}, \citenamefont {{Hippler}}, \citenamefont {{Horrobin}},
  \citenamefont {{Huber}}, \citenamefont {{Jim{\'e}nez-Rosales}}, \citenamefont
  {{Jocou}}, \citenamefont {{Kervella}}, \citenamefont {{Lacour}},
  \citenamefont {{Lapeyr{\`e}re}}, \citenamefont {{Lazareff}}, \citenamefont
  {{Le Bouquin}}, \citenamefont {{L{\'e}na}}, \citenamefont {{Lippa}},
  \citenamefont {{Ott}}, \citenamefont {{Panduro}}, \citenamefont {{Paumard}},
  \citenamefont {{Perraut}}, \citenamefont {{Perrin}}, \citenamefont {{Pfuhl}},
  \citenamefont {{Plewa}}, \citenamefont {{Rabien}}, \citenamefont
  {{Rodr{\'\i}guez-Coira}}, \citenamefont {{Rousset}}, \citenamefont
  {{Sternberg}}, \citenamefont {{Straub}}, \citenamefont {{Straubmeier}},
  \citenamefont {{Sturm}}, \citenamefont {{Tacconi}}, \citenamefont
  {{Vincent}}, \citenamefont {{von Fellenberg}}, \citenamefont {{Waisberg}},
  \citenamefont {{Widmann}}, \citenamefont {{Wieprecht}}, \citenamefont
  {{Wiezorrek}}, \citenamefont {{Woillez}},\ and\ \citenamefont
  {{Yazici}}}]{Gravity1}%
  \BibitemOpen
  \bibfield  {author} {\bibinfo {author} {\bibnamefont {{GRAVITY
  Collaboration}}}, \bibinfo {author} {\bibfnamefont {R.}~\bibnamefont
  {{Abuter}}}, \bibinfo {author} {\bibfnamefont {A.}~\bibnamefont {{Amorim}}},
  \bibinfo {author} {\bibfnamefont {M.}~\bibnamefont {{Baub{\"o}ck}}}, \bibinfo
  {author} {\bibfnamefont {J.~P.}\ \bibnamefont {{Berger}}}, \bibinfo {author}
  {\bibfnamefont {H.}~\bibnamefont {{Bonnet}}}, \bibinfo {author}
  {\bibfnamefont {W.}~\bibnamefont {{Brandner}}}, \bibinfo {author}
  {\bibfnamefont {Y.}~\bibnamefont {{Cl{\'e}net}}}, \bibinfo {author}
  {\bibfnamefont {V.}~\bibnamefont {{Coud{\'e} Du Foresto}}}, \bibinfo {author}
  {\bibfnamefont {P.~T.}\ \bibnamefont {{de Zeeuw}}}, \bibinfo {author}
  {\bibfnamefont {C.}~\bibnamefont {{Deen}}}, \bibinfo {author} {\bibfnamefont
  {J.}~\bibnamefont {{Dexter}}}, \bibinfo {author} {\bibfnamefont
  {G.}~\bibnamefont {{Duvert}}}, \bibinfo {author} {\bibfnamefont
  {A.}~\bibnamefont {{Eckart}}}, \bibinfo {author} {\bibfnamefont
  {F.}~\bibnamefont {{Eisenhauer}}}, \bibinfo {author} {\bibfnamefont {N.~M.}\
  \bibnamefont {{F{\"o}rster Schreiber}}}, \bibinfo {author} {\bibfnamefont
  {P.}~\bibnamefont {{Garcia}}}, \bibinfo {author} {\bibfnamefont
  {F.}~\bibnamefont {{Gao}}}, \bibinfo {author} {\bibfnamefont
  {E.}~\bibnamefont {{Gendron}}}, \bibinfo {author} {\bibfnamefont
  {R.}~\bibnamefont {{Genzel}}}, \bibinfo {author} {\bibfnamefont
  {S.}~\bibnamefont {{Gillessen}}}, \bibinfo {author} {\bibfnamefont
  {P.}~\bibnamefont {{Guajardo}}}, \bibinfo {author} {\bibfnamefont
  {M.}~\bibnamefont {{Habibi}}}, \bibinfo {author} {\bibfnamefont
  {X.}~\bibnamefont {{Haubois}}}, \bibinfo {author} {\bibfnamefont
  {T.}~\bibnamefont {{Henning}}}, \bibinfo {author} {\bibfnamefont
  {S.}~\bibnamefont {{Hippler}}}, \bibinfo {author} {\bibfnamefont
  {M.}~\bibnamefont {{Horrobin}}}, \bibinfo {author} {\bibfnamefont
  {A.}~\bibnamefont {{Huber}}}, \bibinfo {author} {\bibfnamefont
  {A.}~\bibnamefont {{Jim{\'e}nez-Rosales}}}, \bibinfo {author} {\bibfnamefont
  {L.}~\bibnamefont {{Jocou}}}, \bibinfo {author} {\bibfnamefont
  {P.}~\bibnamefont {{Kervella}}}, \bibinfo {author} {\bibfnamefont
  {S.}~\bibnamefont {{Lacour}}}, \bibinfo {author} {\bibfnamefont
  {V.}~\bibnamefont {{Lapeyr{\`e}re}}}, \bibinfo {author} {\bibfnamefont
  {B.}~\bibnamefont {{Lazareff}}}, \bibinfo {author} {\bibfnamefont {J.~B.}\
  \bibnamefont {{Le Bouquin}}}, \bibinfo {author} {\bibfnamefont
  {P.}~\bibnamefont {{L{\'e}na}}}, \bibinfo {author} {\bibfnamefont
  {M.}~\bibnamefont {{Lippa}}}, \bibinfo {author} {\bibfnamefont
  {T.}~\bibnamefont {{Ott}}}, \bibinfo {author} {\bibfnamefont
  {J.}~\bibnamefont {{Panduro}}}, \bibinfo {author} {\bibfnamefont
  {T.}~\bibnamefont {{Paumard}}}, \bibinfo {author} {\bibfnamefont
  {K.}~\bibnamefont {{Perraut}}}, \bibinfo {author} {\bibfnamefont
  {G.}~\bibnamefont {{Perrin}}}, \bibinfo {author} {\bibfnamefont
  {O.}~\bibnamefont {{Pfuhl}}}, \bibinfo {author} {\bibfnamefont {P.~M.}\
  \bibnamefont {{Plewa}}}, \bibinfo {author} {\bibfnamefont {S.}~\bibnamefont
  {{Rabien}}}, \bibinfo {author} {\bibfnamefont {G.}~\bibnamefont
  {{Rodr{\'\i}guez-Coira}}}, \bibinfo {author} {\bibfnamefont {G.}~\bibnamefont
  {{Rousset}}}, \bibinfo {author} {\bibfnamefont {A.}~\bibnamefont
  {{Sternberg}}}, \bibinfo {author} {\bibfnamefont {O.}~\bibnamefont
  {{Straub}}}, \bibinfo {author} {\bibfnamefont {C.}~\bibnamefont
  {{Straubmeier}}}, \bibinfo {author} {\bibfnamefont {E.}~\bibnamefont
  {{Sturm}}}, \bibinfo {author} {\bibfnamefont {L.~J.}\ \bibnamefont
  {{Tacconi}}}, \bibinfo {author} {\bibfnamefont {F.}~\bibnamefont
  {{Vincent}}}, \bibinfo {author} {\bibfnamefont {S.}~\bibnamefont {{von
  Fellenberg}}}, \bibinfo {author} {\bibfnamefont {I.}~\bibnamefont
  {{Waisberg}}}, \bibinfo {author} {\bibfnamefont {F.}~\bibnamefont
  {{Widmann}}}, \bibinfo {author} {\bibfnamefont {E.}~\bibnamefont
  {{Wieprecht}}}, \bibinfo {author} {\bibfnamefont {E.}~\bibnamefont
  {{Wiezorrek}}}, \bibinfo {author} {\bibfnamefont {J.}~\bibnamefont
  {{Woillez}}},\ and\ \bibinfo {author} {\bibfnamefont {S.}~\bibnamefont
  {{Yazici}}},\ }\bibfield  {title} {\bibinfo {title} {{Detection of orbital
  motions near the last stable circular orbit of the massive black hole
  SgrA*}},\ }\href {https://doi.org/10.1051/0004-6361/201834294} {\bibfield
  {journal} {\bibinfo  {journal} {Astronomy \& Astrophysics}\ }\textbf
  {\bibinfo {volume} {618}},\ \bibinfo {eid} {L10} (\bibinfo {year} {2018})},\
  \Eprint {https://arxiv.org/abs/1810.12641} {arXiv:1810.12641 [astro-ph.GA]}
  \BibitemShut {NoStop}%
\bibitem [{\citenamefont {Foschi}\ \emph {et~al.}(2023)\citenamefont {Foschi}
  \emph {et~al.}}]{GRAVITY:2023cjt}%
  \BibitemOpen
  \bibfield  {author} {\bibinfo {author} {\bibfnamefont {A.}~\bibnamefont
  {Foschi}} \emph {et~al.} (\bibinfo {collaboration} {GRAVITY}),\ }\bibfield
  {title} {\bibinfo {title} {{Using the motion of S2 to constrain scalar clouds
  around Sgr A*}},\ }\href {https://doi.org/10.1093/mnras/stad1939} {\bibfield
  {journal} {\bibinfo  {journal} {Mon. Not. Roy. Astron. Soc.}\ }\textbf
  {\bibinfo {volume} {524}},\ \bibinfo {pages} {1075} (\bibinfo {year}
  {2023})},\ \Eprint {https://arxiv.org/abs/2306.17215} {arXiv:2306.17215
  [astro-ph.GA]} \BibitemShut {NoStop}%
\bibitem [{\citenamefont {Arun}\ \emph {et~al.}(2022)\citenamefont {Arun} \emph
  {et~al.}}]{LISA:2022kgy}%
  \BibitemOpen
  \bibfield  {author} {\bibinfo {author} {\bibfnamefont {K.~G.}\ \bibnamefont
  {Arun}} \emph {et~al.} (\bibinfo {collaboration} {LISA}),\ }\bibfield
  {title} {\bibinfo {title} {{New horizons for fundamental physics with
  LISA}},\ }\href {https://doi.org/10.1007/s41114-022-00036-9} {\bibfield
  {journal} {\bibinfo  {journal} {Living Rev. Rel.}\ }\textbf {\bibinfo
  {volume} {25}},\ \bibinfo {pages} {4} (\bibinfo {year} {2022})},\ \Eprint
  {https://arxiv.org/abs/2205.01597} {arXiv:2205.01597 [gr-qc]} \BibitemShut
  {NoStop}%
\bibitem [{\citenamefont {Perlick}\ and\ \citenamefont
  {Tsupko}(2022)}]{Perlick:2021aok}%
  \BibitemOpen
  \bibfield  {author} {\bibinfo {author} {\bibfnamefont {V.}~\bibnamefont
  {Perlick}}\ and\ \bibinfo {author} {\bibfnamefont {O.~Y.}\ \bibnamefont
  {Tsupko}},\ }\bibfield  {title} {\bibinfo {title} {{Calculating black hole
  shadows: Review of analytical studies}},\ }\href
  {https://doi.org/10.1016/j.physrep.2021.10.004} {\bibfield  {journal}
  {\bibinfo  {journal} {Phys. Rept.}\ }\textbf {\bibinfo {volume} {947}},\
  \bibinfo {pages} {1} (\bibinfo {year} {2022})},\ \Eprint
  {https://arxiv.org/abs/2105.07101} {arXiv:2105.07101 [gr-qc]} \BibitemShut
  {NoStop}%
\bibitem [{\citenamefont {{Teo}}(2003)}]{2003GReGr..35.1909T}%
  \BibitemOpen
  \bibfield  {author} {\bibinfo {author} {\bibfnamefont {E.}~\bibnamefont
  {{Teo}}},\ }\bibfield  {title} {\bibinfo {title} {{Spherical Photon Orbits
  Around a Kerr Black Hole}},\ }\href {https://doi.org/10.1023/A:1026286607562}
  {\bibfield  {journal} {\bibinfo  {journal} {General Relativity and
  Gravitation}\ }\textbf {\bibinfo {volume} {35}},\ \bibinfo {pages} {1909}
  (\bibinfo {year} {2003})}\BibitemShut {NoStop}%
\bibitem [{\citenamefont {{Perlick}}(2004)}]{Perlick2004}%
  \BibitemOpen
  \bibfield  {author} {\bibinfo {author} {\bibfnamefont {V.}~\bibnamefont
  {{Perlick}}},\ }\bibfield  {title} {\bibinfo {title} {{Gravitational Lensing
  from a Spacetime Perspective}},\ }\href {https://doi.org/10.12942/lrr-2004-9}
  {\bibfield  {journal} {\bibinfo  {journal} {Living Reviews in Relativity}\
  }\textbf {\bibinfo {volume} {7}},\ \bibinfo {eid} {9} (\bibinfo {year}
  {2004})}\BibitemShut {NoStop}%
\bibitem [{\citenamefont {Johnson}\ \emph {et~al.}(2020)\citenamefont {Johnson}
  \emph {et~al.}}]{Johnson:2019ljv}%
  \BibitemOpen
  \bibfield  {author} {\bibinfo {author} {\bibfnamefont {M.~D.}\ \bibnamefont
  {Johnson}} \emph {et~al.},\ }\bibfield  {title} {\bibinfo {title} {{Universal
  interferometric signatures of a black hole\textquoteright{}s photon ring}},\
  }\href {https://doi.org/10.1126/sciadv.aaz1310} {\bibfield  {journal}
  {\bibinfo  {journal} {Sci. Adv.}\ }\textbf {\bibinfo {volume} {6}},\ \bibinfo
  {pages} {eaaz1310} (\bibinfo {year} {2020})},\ \Eprint
  {https://arxiv.org/abs/1907.04329} {arXiv:1907.04329 [astro-ph.IM]}
  \BibitemShut {NoStop}%
\bibitem [{\citenamefont {Bardeen}(1973)}]{Bardeen:1973tla}%
  \BibitemOpen
  \bibfield  {author} {\bibinfo {author} {\bibfnamefont {J.~M.}\ \bibnamefont
  {Bardeen}},\ }\bibfield  {title} {\bibinfo {title} {{Timelike and null
  geodesics in the Kerr metric}},\ }in\ \href@noop {} {\emph {\bibinfo
  {booktitle} {{Les Houches Summer School of Theoretical Physics}: {Black
  Holes}}}}\ (\bibinfo {year} {1973})\ pp.\ \bibinfo {pages}
  {215--240}\BibitemShut {NoStop}%
\bibitem [{\citenamefont {Gralla}\ \emph {et~al.}(2019)\citenamefont {Gralla},
  \citenamefont {Holz},\ and\ \citenamefont {Wald}}]{Gralla:2019xty}%
  \BibitemOpen
  \bibfield  {author} {\bibinfo {author} {\bibfnamefont {S.~E.}\ \bibnamefont
  {Gralla}}, \bibinfo {author} {\bibfnamefont {D.~E.}\ \bibnamefont {Holz}},\
  and\ \bibinfo {author} {\bibfnamefont {R.~M.}\ \bibnamefont {Wald}},\
  }\bibfield  {title} {\bibinfo {title} {{Black Hole Shadows, Photon Rings, and
  Lensing Rings}},\ }\href {https://doi.org/10.1103/PhysRevD.100.024018}
  {\bibfield  {journal} {\bibinfo  {journal} {Phys. Rev. D}\ }\textbf {\bibinfo
  {volume} {100}},\ \bibinfo {pages} {024018} (\bibinfo {year} {2019})},\
  \Eprint {https://arxiv.org/abs/1906.00873} {arXiv:1906.00873 [astro-ph.HE]}
  \BibitemShut {NoStop}%
\bibitem [{\citenamefont {{Darwin}}(1959)}]{Darwin1959}%
  \BibitemOpen
  \bibfield  {author} {\bibinfo {author} {\bibfnamefont {C.}~\bibnamefont
  {{Darwin}}},\ }\bibfield  {title} {\bibinfo {title} {{The Gravity Field of a
  Particle}},\ }\href {https://doi.org/10.1098/rspa.1959.0015} {\bibfield
  {journal} {\bibinfo  {journal} {Proceedings of the Royal Society of London
  Series A}\ }\textbf {\bibinfo {volume} {249}},\ \bibinfo {pages} {180}
  (\bibinfo {year} {1959})}\BibitemShut {NoStop}%
\bibitem [{\citenamefont {Luminet}(1979)}]{Luminet:1979nyg}%
  \BibitemOpen
  \bibfield  {author} {\bibinfo {author} {\bibfnamefont {J.~P.}\ \bibnamefont
  {Luminet}},\ }\bibfield  {title} {\bibinfo {title} {{Image of a spherical
  black hole with thin accretion disk}},\ }\href@noop {} {\bibfield  {journal}
  {\bibinfo  {journal} {Astron. Astrophys.}\ }\textbf {\bibinfo {volume}
  {75}},\ \bibinfo {pages} {228} (\bibinfo {year} {1979})}\BibitemShut
  {NoStop}%
\bibitem [{\citenamefont {Ohanian}(1987)}]{ohanian1987black}%
  \BibitemOpen
  \bibfield  {author} {\bibinfo {author} {\bibfnamefont {H.~C.}\ \bibnamefont
  {Ohanian}},\ }\bibfield  {title} {\bibinfo {title} {The black hole as a
  gravitational ‘‘lens’’},\ }\href@noop {} {\bibfield  {journal}
  {\bibinfo  {journal} {American Journal of Physics}\ }\textbf {\bibinfo
  {volume} {55}},\ \bibinfo {pages} {428} (\bibinfo {year} {1987})}\BibitemShut
  {NoStop}%
\bibitem [{\citenamefont {Dokuchaev}\ and\ \citenamefont
  {Nazarova}(2019)}]{Dokuchaev:2018kzk}%
  \BibitemOpen
  \bibfield  {author} {\bibinfo {author} {\bibfnamefont {V.~I.}\ \bibnamefont
  {Dokuchaev}}\ and\ \bibinfo {author} {\bibfnamefont {N.~O.}\ \bibnamefont
  {Nazarova}},\ }\bibfield  {title} {\bibinfo {title} {{Event horizon image
  within black hole shadow}},\ }\href
  {https://doi.org/10.1134/S1063776119030026} {\bibfield  {journal} {\bibinfo
  {journal} {J. Exp. Theor. Phys.}\ }\textbf {\bibinfo {volume} {128}},\
  \bibinfo {pages} {578} (\bibinfo {year} {2019})},\ \Eprint
  {https://arxiv.org/abs/1804.08030} {arXiv:1804.08030 [astro-ph.HE]}
  \BibitemShut {NoStop}%
\bibitem [{\citenamefont {Dokuchaev}\ and\ \citenamefont
  {Nazarova}(2020)}]{Dokuchaev:2020wqk}%
  \BibitemOpen
  \bibfield  {author} {\bibinfo {author} {\bibfnamefont {V.~I.}\ \bibnamefont
  {Dokuchaev}}\ and\ \bibinfo {author} {\bibfnamefont {N.~O.}\ \bibnamefont
  {Nazarova}},\ }\bibfield  {title} {\bibinfo {title} {{Visible shapes of black
  holes M87* and SgrA*}},\ }\href {https://doi.org/10.3390/universe6090154}
  {\bibfield  {journal} {\bibinfo  {journal} {Universe}\ }\textbf {\bibinfo
  {volume} {6}},\ \bibinfo {pages} {154} (\bibinfo {year} {2020})},\ \Eprint
  {https://arxiv.org/abs/2007.14121} {arXiv:2007.14121 [astro-ph.HE]}
  \BibitemShut {NoStop}%
\bibitem [{\citenamefont {Chael}\ \emph {et~al.}(2021)\citenamefont {Chael},
  \citenamefont {Johnson},\ and\ \citenamefont {Lupsasca}}]{Chael:2021rjo}%
  \BibitemOpen
  \bibfield  {author} {\bibinfo {author} {\bibfnamefont {A.}~\bibnamefont
  {Chael}}, \bibinfo {author} {\bibfnamefont {M.~D.}\ \bibnamefont {Johnson}},\
  and\ \bibinfo {author} {\bibfnamefont {A.}~\bibnamefont {Lupsasca}},\
  }\bibfield  {title} {\bibinfo {title} {{Observing the Inner Shadow of a Black
  Hole: A Direct View of the Event Horizon}},\ }\href
  {https://doi.org/10.3847/1538-4357/ac09ee} {\bibfield  {journal} {\bibinfo
  {journal} {Astrophys. J.}\ }\textbf {\bibinfo {volume} {918}},\ \bibinfo
  {pages} {6} (\bibinfo {year} {2021})},\ \Eprint
  {https://arxiv.org/abs/2106.00683} {arXiv:2106.00683 [astro-ph.HE]}
  \BibitemShut {NoStop}%
\bibitem [{\citenamefont {Vincent}\ \emph {et~al.}(2021)\citenamefont
  {Vincent}, \citenamefont {Wielgus}, \citenamefont {Abramowicz}, \citenamefont
  {Gourgoulhon}, \citenamefont {Lasota}, \citenamefont {Paumard},\ and\
  \citenamefont {Perrin}}]{Vincent:2020dij}%
  \BibitemOpen
  \bibfield  {author} {\bibinfo {author} {\bibfnamefont {F.~H.}\ \bibnamefont
  {Vincent}}, \bibinfo {author} {\bibfnamefont {M.}~\bibnamefont {Wielgus}},
  \bibinfo {author} {\bibfnamefont {M.~A.}\ \bibnamefont {Abramowicz}},
  \bibinfo {author} {\bibfnamefont {E.}~\bibnamefont {Gourgoulhon}}, \bibinfo
  {author} {\bibfnamefont {J.~P.}\ \bibnamefont {Lasota}}, \bibinfo {author}
  {\bibfnamefont {T.}~\bibnamefont {Paumard}},\ and\ \bibinfo {author}
  {\bibfnamefont {G.}~\bibnamefont {Perrin}},\ }\bibfield  {title} {\bibinfo
  {title} {{Geometric modeling of M87* as a Kerr black hole or a non-Kerr
  compact object}},\ }\href {https://doi.org/10.1051/0004-6361/202037787}
  {\bibfield  {journal} {\bibinfo  {journal} {Astron. Astrophys.}\ }\textbf
  {\bibinfo {volume} {646}},\ \bibinfo {pages} {A37} (\bibinfo {year}
  {2021})},\ \Eprint {https://arxiv.org/abs/2002.09226} {arXiv:2002.09226
  [gr-qc]} \BibitemShut {NoStop}%
\bibitem [{\citenamefont {Akiyama}\ \emph
  {et~al.}(2022{\natexlab{b}})\citenamefont {Akiyama} \emph
  {et~al.}}]{EventHorizonTelescope:2022xqj}%
  \BibitemOpen
  \bibfield  {author} {\bibinfo {author} {\bibfnamefont {K.}~\bibnamefont
  {Akiyama}} \emph {et~al.} (\bibinfo {collaboration} {Event Horizon
  Telescope}),\ }\bibfield  {title} {\bibinfo {title} {{First Sagittarius A*
  Event Horizon Telescope Results. VI. Testing the Black Hole Metric}},\ }\href
  {https://doi.org/10.3847/2041-8213/ac6756} {\bibfield  {journal} {\bibinfo
  {journal} {Astrophys. J. Lett.}\ }\textbf {\bibinfo {volume} {930}},\
  \bibinfo {pages} {L17} (\bibinfo {year} {2022}{\natexlab{b}})}\BibitemShut
  {NoStop}%
\bibitem [{\citenamefont {Eichhorn}\ \emph
  {et~al.}(2023{\natexlab{a}})\citenamefont {Eichhorn}, \citenamefont {Gold},\
  and\ \citenamefont {Held}}]{Eichhorn:2022fcl}%
  \BibitemOpen
  \bibfield  {author} {\bibinfo {author} {\bibfnamefont {A.}~\bibnamefont
  {Eichhorn}}, \bibinfo {author} {\bibfnamefont {R.}~\bibnamefont {Gold}},\
  and\ \bibinfo {author} {\bibfnamefont {A.}~\bibnamefont {Held}},\ }\bibfield
  {title} {\bibinfo {title} {{Horizonless Spacetimes As Seen by Present and
  Next-generation Event Horizon Telescope Arrays}},\ }\href
  {https://doi.org/10.3847/1538-4357/accced} {\bibfield  {journal} {\bibinfo
  {journal} {Astrophys. J.}\ }\textbf {\bibinfo {volume} {950}},\ \bibinfo
  {pages} {117} (\bibinfo {year} {2023}{\natexlab{a}})},\ \Eprint
  {https://arxiv.org/abs/2205.14883} {arXiv:2205.14883 [astro-ph.HE]}
  \BibitemShut {NoStop}%
\bibitem [{\citenamefont {Akiyama}\ \emph
  {et~al.}(2022{\natexlab{c}})\citenamefont {Akiyama} \emph
  {et~al.}}]{EventHorizonTelescope:2022wkp}%
  \BibitemOpen
  \bibfield  {author} {\bibinfo {author} {\bibfnamefont {K.}~\bibnamefont
  {Akiyama}} \emph {et~al.} (\bibinfo {collaboration} {Event Horizon
  Telescope}),\ }\bibfield  {title} {\bibinfo {title} {{First Sagittarius A*
  Event Horizon Telescope Results. I. The Shadow of the Supermassive Black Hole
  in the Center of the Milky Way}},\ }\href
  {https://doi.org/10.3847/2041-8213/ac6674} {\bibfield  {journal} {\bibinfo
  {journal} {Astrophys. J. Lett.}\ }\textbf {\bibinfo {volume} {930}},\
  \bibinfo {pages} {L12} (\bibinfo {year} {2022}{\natexlab{c}})},\ \Eprint
  {https://arxiv.org/abs/2311.08680} {arXiv:2311.08680 [astro-ph.HE]}
  \BibitemShut {NoStop}%
\bibitem [{\citenamefont {{Gralla}}(2021)}]{Gralla2021}%
  \BibitemOpen
  \bibfield  {author} {\bibinfo {author} {\bibfnamefont {S.~E.}\ \bibnamefont
  {{Gralla}}},\ }\bibfield  {title} {\bibinfo {title} {{Can the EHT M87 results
  be used to test general relativity?}},\ }\href
  {https://doi.org/10.1103/PhysRevD.103.024023} {\bibfield  {journal} {\bibinfo
   {journal} {\prd}\ }\textbf {\bibinfo {volume} {103}},\ \bibinfo {eid}
  {024023} (\bibinfo {year} {2021})},\ \Eprint
  {https://arxiv.org/abs/2010.08557} {arXiv:2010.08557 [astro-ph.HE]}
  \BibitemShut {NoStop}%
\bibitem [{\citenamefont {V\"olkel}\ \emph {et~al.}(2021)\citenamefont
  {V\"olkel}, \citenamefont {Barausse}, \citenamefont {Franchini},\ and\
  \citenamefont {Broderick}}]{Volkel:2020xlc}%
  \BibitemOpen
  \bibfield  {author} {\bibinfo {author} {\bibfnamefont {S.~H.}\ \bibnamefont
  {V\"olkel}}, \bibinfo {author} {\bibfnamefont {E.}~\bibnamefont {Barausse}},
  \bibinfo {author} {\bibfnamefont {N.}~\bibnamefont {Franchini}},\ and\
  \bibinfo {author} {\bibfnamefont {A.~E.}\ \bibnamefont {Broderick}},\
  }\bibfield  {title} {\bibinfo {title} {{EHT tests of the strong-field regime
  of general relativity}},\ }\href {https://doi.org/10.1088/1361-6382/ac27ed}
  {\bibfield  {journal} {\bibinfo  {journal} {Class. Quant. Grav.}\ }\textbf
  {\bibinfo {volume} {38}},\ \bibinfo {pages} {21LT01} (\bibinfo {year}
  {2021})},\ \Eprint {https://arxiv.org/abs/2011.06812} {arXiv:2011.06812
  [gr-qc]} \BibitemShut {NoStop}%
\bibitem [{\citenamefont {Bauer}\ \emph {et~al.}(2022)\citenamefont {Bauer},
  \citenamefont {C\'ardenas-Avenda\~no}, \citenamefont {Gammie},\ and\
  \citenamefont {Yunes}}]{Bauer:2021atk}%
  \BibitemOpen
  \bibfield  {author} {\bibinfo {author} {\bibfnamefont {A.~M.}\ \bibnamefont
  {Bauer}}, \bibinfo {author} {\bibfnamefont {A.}~\bibnamefont
  {C\'ardenas-Avenda\~no}}, \bibinfo {author} {\bibfnamefont {C.~F.}\
  \bibnamefont {Gammie}},\ and\ \bibinfo {author} {\bibfnamefont
  {N.}~\bibnamefont {Yunes}},\ }\bibfield  {title} {\bibinfo {title}
  {{Spherical Accretion in Alternative Theories of Gravity}},\ }\href
  {https://doi.org/10.3847/1538-4357/ac3a03} {\bibfield  {journal} {\bibinfo
  {journal} {Astrophys. J.}\ }\textbf {\bibinfo {volume} {925}},\ \bibinfo
  {pages} {119} (\bibinfo {year} {2022})},\ \Eprint
  {https://arxiv.org/abs/2111.02178} {arXiv:2111.02178 [gr-qc]} \BibitemShut
  {NoStop}%
\bibitem [{\citenamefont {Broderick}\ \emph
  {et~al.}(2022{\natexlab{a}})\citenamefont {Broderick} \emph
  {et~al.}}]{Broderick_2022}%
  \BibitemOpen
  \bibfield  {author} {\bibinfo {author} {\bibfnamefont {A.~E.}\ \bibnamefont
  {Broderick}} \emph {et~al.},\ }\bibfield  {title} {\bibinfo {title} {{The
  Photon Ring in M87*}},\ }\href {https://doi.org/10.3847/1538-4357/ac7c1d}
  {\bibfield  {journal} {\bibinfo  {journal} {Astrophys. J.}\ }\textbf
  {\bibinfo {volume} {935}},\ \bibinfo {pages} {61} (\bibinfo {year}
  {2022}{\natexlab{a}})},\ \Eprint {https://arxiv.org/abs/2208.09004}
  {arXiv:2208.09004 [astro-ph.HE]} \BibitemShut {NoStop}%
\bibitem [{\citenamefont {Lockhart}\ and\ \citenamefont
  {Gralla}(2022)}]{Lockhart:2022rui}%
  \BibitemOpen
  \bibfield  {author} {\bibinfo {author} {\bibfnamefont {W.}~\bibnamefont
  {Lockhart}}\ and\ \bibinfo {author} {\bibfnamefont {S.~E.}\ \bibnamefont
  {Gralla}},\ }\bibfield  {title} {\bibinfo {title} {{How narrow is the M87*
  ring \textendash{} II. A new geometric model}},\ }\href
  {https://doi.org/10.1093/mnras/stac2743} {\bibfield  {journal} {\bibinfo
  {journal} {Mon. Not. Roy. Astron. Soc.}\ }\textbf {\bibinfo {volume} {517}},\
  \bibinfo {pages} {2462} (\bibinfo {year} {2022})},\ \Eprint
  {https://arxiv.org/abs/2208.09989} {arXiv:2208.09989 [astro-ph.HE]}
  \BibitemShut {NoStop}%
\bibitem [{\citenamefont {Gralla}\ \emph {et~al.}(2020)\citenamefont {Gralla},
  \citenamefont {Lupsasca},\ and\ \citenamefont {Marrone}}]{Gralla:2020srx}%
  \BibitemOpen
  \bibfield  {author} {\bibinfo {author} {\bibfnamefont {S.~E.}\ \bibnamefont
  {Gralla}}, \bibinfo {author} {\bibfnamefont {A.}~\bibnamefont {Lupsasca}},\
  and\ \bibinfo {author} {\bibfnamefont {D.~P.}\ \bibnamefont {Marrone}},\
  }\bibfield  {title} {\bibinfo {title} {{The shape of the black hole photon
  ring: A precise test of strong-field general relativity}},\ }\href
  {https://doi.org/10.1103/PhysRevD.102.124004} {\bibfield  {journal} {\bibinfo
   {journal} {Phys. Rev. D}\ }\textbf {\bibinfo {volume} {102}},\ \bibinfo
  {pages} {124004} (\bibinfo {year} {2020})},\ \Eprint
  {https://arxiv.org/abs/2008.03879} {arXiv:2008.03879 [gr-qc]} \BibitemShut
  {NoStop}%
\bibitem [{\citenamefont {Gralla}\ and\ \citenamefont
  {Lupsasca}(2020{\natexlab{a}})}]{Gralla:2019ceu}%
  \BibitemOpen
  \bibfield  {author} {\bibinfo {author} {\bibfnamefont {S.~E.}\ \bibnamefont
  {Gralla}}\ and\ \bibinfo {author} {\bibfnamefont {A.}~\bibnamefont
  {Lupsasca}},\ }\bibfield  {title} {\bibinfo {title} {{Null geodesics of the
  Kerr exterior}},\ }\href {https://doi.org/10.1103/PhysRevD.101.044032}
  {\bibfield  {journal} {\bibinfo  {journal} {Phys. Rev. D}\ }\textbf {\bibinfo
  {volume} {101}},\ \bibinfo {pages} {044032} (\bibinfo {year}
  {2020}{\natexlab{a}})},\ \Eprint {https://arxiv.org/abs/1910.12881}
  {arXiv:1910.12881 [gr-qc]} \BibitemShut {NoStop}%
\bibitem [{\citenamefont {Gralla}\ and\ \citenamefont
  {Lupsasca}(2020{\natexlab{b}})}]{Gralla:2020yvo}%
  \BibitemOpen
  \bibfield  {author} {\bibinfo {author} {\bibfnamefont {S.~E.}\ \bibnamefont
  {Gralla}}\ and\ \bibinfo {author} {\bibfnamefont {A.}~\bibnamefont
  {Lupsasca}},\ }\bibfield  {title} {\bibinfo {title} {{Observable shape of
  black hole photon rings}},\ }\href
  {https://doi.org/10.1103/PhysRevD.102.124003} {\bibfield  {journal} {\bibinfo
   {journal} {Phys. Rev. D}\ }\textbf {\bibinfo {volume} {102}},\ \bibinfo
  {pages} {124003} (\bibinfo {year} {2020}{\natexlab{b}})},\ \Eprint
  {https://arxiv.org/abs/2007.10336} {arXiv:2007.10336 [gr-qc]} \BibitemShut
  {NoStop}%
\bibitem [{\citenamefont {Paugnat}\ \emph {et~al.}(2022)\citenamefont
  {Paugnat}, \citenamefont {Lupsasca}, \citenamefont {Vincent},\ and\
  \citenamefont {Wielgus}}]{Paugnat:2022qzy}%
  \BibitemOpen
  \bibfield  {author} {\bibinfo {author} {\bibfnamefont {H.}~\bibnamefont
  {Paugnat}}, \bibinfo {author} {\bibfnamefont {A.}~\bibnamefont {Lupsasca}},
  \bibinfo {author} {\bibfnamefont {F.}~\bibnamefont {Vincent}},\ and\ \bibinfo
  {author} {\bibfnamefont {M.}~\bibnamefont {Wielgus}},\ }\bibfield  {title}
  {\bibinfo {title} {{Photon ring test of the Kerr hypothesis: variation in the
  ring shape}},\ }\href@noop {} {\  (\bibinfo {year} {2022})},\ \Eprint
  {https://arxiv.org/abs/2206.02781} {arXiv:2206.02781 [astro-ph.HE]}
  \BibitemShut {NoStop}%
\bibitem [{\citenamefont {C\'ardenas-Avenda\~no}\ and\ \citenamefont
  {Lupsasca}(2023)}]{Cardenas-Avendano:2023dzo}%
  \BibitemOpen
  \bibfield  {author} {\bibinfo {author} {\bibfnamefont {A.}~\bibnamefont
  {C\'ardenas-Avenda\~no}}\ and\ \bibinfo {author} {\bibfnamefont
  {A.}~\bibnamefont {Lupsasca}},\ }\bibfield  {title} {\bibinfo {title}
  {{Prediction for the interferometric shape of the first black hole photon
  ring}},\ }\href {https://doi.org/10.1103/PhysRevD.108.064043} {\bibfield
  {journal} {\bibinfo  {journal} {Phys. Rev. D}\ }\textbf {\bibinfo {volume}
  {108}},\ \bibinfo {pages} {064043} (\bibinfo {year} {2023})},\ \Eprint
  {https://arxiv.org/abs/2305.12956} {arXiv:2305.12956 [gr-qc]} \BibitemShut
  {NoStop}%
\bibitem [{\citenamefont {Psaltis}\ \emph {et~al.}(2020)\citenamefont {Psaltis}
  \emph {et~al.}}]{EventHorizonTelescope:2020qrl}%
  \BibitemOpen
  \bibfield  {author} {\bibinfo {author} {\bibfnamefont {D.}~\bibnamefont
  {Psaltis}} \emph {et~al.} (\bibinfo {collaboration} {Event Horizon
  Telescope}),\ }\bibfield  {title} {\bibinfo {title} {{Gravitational Test
  Beyond the First Post-Newtonian Order with the Shadow of the M87 Black
  Hole}},\ }\href {https://doi.org/10.1103/PhysRevLett.125.141104} {\bibfield
  {journal} {\bibinfo  {journal} {Phys. Rev. Lett.}\ }\textbf {\bibinfo
  {volume} {125}},\ \bibinfo {pages} {141104} (\bibinfo {year} {2020})},\
  \Eprint {https://arxiv.org/abs/2010.01055} {arXiv:2010.01055 [gr-qc]}
  \BibitemShut {NoStop}%
\bibitem [{\citenamefont {Younsi}\ \emph {et~al.}(2021)\citenamefont {Younsi},
  \citenamefont {Psaltis},\ and\ \citenamefont {\"Ozel}}]{Younsi:2021dxe}%
  \BibitemOpen
  \bibfield  {author} {\bibinfo {author} {\bibfnamefont {Z.}~\bibnamefont
  {Younsi}}, \bibinfo {author} {\bibfnamefont {D.}~\bibnamefont {Psaltis}},\
  and\ \bibinfo {author} {\bibfnamefont {F.}~\bibnamefont {\"Ozel}},\
  }\bibfield  {title} {\bibinfo {title} {{Black Hole Images as Tests of General
  Relativity: Effects of Spacetime Geometry}},\ }\href@noop {} {\  (\bibinfo
  {year} {2021})},\ \Eprint {https://arxiv.org/abs/2111.01752}
  {arXiv:2111.01752 [astro-ph.HE]} \BibitemShut {NoStop}%
\bibitem [{\citenamefont {Ayzenberg}(2022)}]{Ayzenberg:2022twz}%
  \BibitemOpen
  \bibfield  {author} {\bibinfo {author} {\bibfnamefont {D.}~\bibnamefont
  {Ayzenberg}},\ }\bibfield  {title} {\bibinfo {title} {{Testing gravity with
  black hole shadow subrings}},\ }\href
  {https://doi.org/10.1088/1361-6382/ac655d} {\bibfield  {journal} {\bibinfo
  {journal} {Class. Quant. Grav.}\ }\textbf {\bibinfo {volume} {39}},\ \bibinfo
  {pages} {105009} (\bibinfo {year} {2022})},\ \Eprint
  {https://arxiv.org/abs/2202.02355} {arXiv:2202.02355 [gr-qc]} \BibitemShut
  {NoStop}%
\bibitem [{\citenamefont {Staelens}\ \emph {et~al.}(2023)\citenamefont
  {Staelens}, \citenamefont {Mayerson}, \citenamefont {Bacchini}, \citenamefont
  {Ripperda},\ and\ \citenamefont {K\"uchler}}]{Staelens:2023jgr}%
  \BibitemOpen
  \bibfield  {author} {\bibinfo {author} {\bibfnamefont {S.}~\bibnamefont
  {Staelens}}, \bibinfo {author} {\bibfnamefont {D.~R.}\ \bibnamefont
  {Mayerson}}, \bibinfo {author} {\bibfnamefont {F.}~\bibnamefont {Bacchini}},
  \bibinfo {author} {\bibfnamefont {B.}~\bibnamefont {Ripperda}},\ and\
  \bibinfo {author} {\bibfnamefont {L.}~\bibnamefont {K\"uchler}},\ }\bibfield
  {title} {\bibinfo {title} {{Black Hole Photon Rings Beyond General
  Relativity}},\ }\href@noop {} {\  (\bibinfo {year} {2023})},\ \Eprint
  {https://arxiv.org/abs/2303.02111} {arXiv:2303.02111 [gr-qc]} \BibitemShut
  {NoStop}%
\bibitem [{\citenamefont {Babichev}\ and\ \citenamefont
  {Deffayet}(2013)}]{Babichev:2013usa}%
  \BibitemOpen
  \bibfield  {author} {\bibinfo {author} {\bibfnamefont {E.}~\bibnamefont
  {Babichev}}\ and\ \bibinfo {author} {\bibfnamefont {C.}~\bibnamefont
  {Deffayet}},\ }\bibfield  {title} {\bibinfo {title} {{An introduction to the
  Vainshtein mechanism}},\ }\href
  {https://doi.org/10.1088/0264-9381/30/18/184001} {\bibfield  {journal}
  {\bibinfo  {journal} {Class. Quant. Grav.}\ }\textbf {\bibinfo {volume}
  {30}},\ \bibinfo {pages} {184001} (\bibinfo {year} {2013})},\ \Eprint
  {https://arxiv.org/abs/1304.7240} {arXiv:1304.7240 [gr-qc]} \BibitemShut
  {NoStop}%
\bibitem [{\citenamefont {Doneva}\ and\ \citenamefont
  {Yazadjiev}(2018)}]{Doneva:2017bvd}%
  \BibitemOpen
  \bibfield  {author} {\bibinfo {author} {\bibfnamefont {D.~D.}\ \bibnamefont
  {Doneva}}\ and\ \bibinfo {author} {\bibfnamefont {S.~S.}\ \bibnamefont
  {Yazadjiev}},\ }\bibfield  {title} {\bibinfo {title} {{New Gauss-Bonnet Black
  Holes with Curvature-Induced Scalarization in Extended Scalar-Tensor
  Theories}},\ }\href {https://doi.org/10.1103/PhysRevLett.120.131103}
  {\bibfield  {journal} {\bibinfo  {journal} {Phys. Rev. Lett.}\ }\textbf
  {\bibinfo {volume} {120}},\ \bibinfo {pages} {131103} (\bibinfo {year}
  {2018})},\ \Eprint {https://arxiv.org/abs/1711.01187} {arXiv:1711.01187
  [gr-qc]} \BibitemShut {NoStop}%
\bibitem [{\citenamefont {Held}\ and\ \citenamefont
  {Zhang}(2023)}]{Held:2022abx}%
  \BibitemOpen
  \bibfield  {author} {\bibinfo {author} {\bibfnamefont {A.}~\bibnamefont
  {Held}}\ and\ \bibinfo {author} {\bibfnamefont {J.}~\bibnamefont {Zhang}},\
  }\bibfield  {title} {\bibinfo {title} {{Instability of spherically symmetric
  black holes in quadratic gravity}},\ }\href
  {https://doi.org/10.1103/PhysRevD.107.064060} {\bibfield  {journal} {\bibinfo
   {journal} {Phys. Rev. D}\ }\textbf {\bibinfo {volume} {107}},\ \bibinfo
  {pages} {064060} (\bibinfo {year} {2023})},\ \Eprint
  {https://arxiv.org/abs/2209.01867} {arXiv:2209.01867 [gr-qc]} \BibitemShut
  {NoStop}%
\bibitem [{\citenamefont {Fabian}(2012)}]{Fabian:2012xr}%
  \BibitemOpen
  \bibfield  {author} {\bibinfo {author} {\bibfnamefont {A.~C.}\ \bibnamefont
  {Fabian}},\ }\bibfield  {title} {\bibinfo {title} {{Observational Evidence of
  AGN Feedback}},\ }\href {https://doi.org/10.1146/annurev-astro-081811-125521}
  {\bibfield  {journal} {\bibinfo  {journal} {Ann. Rev. Astron. Astrophys.}\
  }\textbf {\bibinfo {volume} {50}},\ \bibinfo {pages} {455} (\bibinfo {year}
  {2012})},\ \Eprint {https://arxiv.org/abs/1204.4114} {arXiv:1204.4114
  [astro-ph.CO]} \BibitemShut {NoStop}%
\bibitem [{\citenamefont {Volonteri}\ \emph {et~al.}(2021)\citenamefont
  {Volonteri}, \citenamefont {Habouzit},\ and\ \citenamefont
  {Colpi}}]{Volonteri:2021sfo}%
  \BibitemOpen
  \bibfield  {author} {\bibinfo {author} {\bibfnamefont {M.}~\bibnamefont
  {Volonteri}}, \bibinfo {author} {\bibfnamefont {M.}~\bibnamefont
  {Habouzit}},\ and\ \bibinfo {author} {\bibfnamefont {M.}~\bibnamefont
  {Colpi}},\ }\bibfield  {title} {\bibinfo {title} {{The origins of massive
  black holes}},\ }\href {https://doi.org/10.1038/s42254-021-00364-9}
  {\bibfield  {journal} {\bibinfo  {journal} {Nature Rev. Phys.}\ }\textbf
  {\bibinfo {volume} {3}},\ \bibinfo {pages} {732} (\bibinfo {year} {2021})},\
  \Eprint {https://arxiv.org/abs/2110.10175} {arXiv:2110.10175 [astro-ph.GA]}
  \BibitemShut {NoStop}%
\bibitem [{\citenamefont {Akiyama}\ \emph
  {et~al.}(2022{\natexlab{d}})\citenamefont {Akiyama} \emph
  {et~al.}}]{EventHorizonTelescope:2022urf}%
  \BibitemOpen
  \bibfield  {author} {\bibinfo {author} {\bibfnamefont {K.}~\bibnamefont
  {Akiyama}} \emph {et~al.} (\bibinfo {collaboration} {Event Horizon
  Telescope}),\ }\bibfield  {title} {\bibinfo {title} {{First Sagittarius A*
  Event Horizon Telescope Results. V. Testing Astrophysical Models of the
  Galactic Center Black Hole}},\ }\href
  {https://doi.org/10.3847/2041-8213/ac6672} {\bibfield  {journal} {\bibinfo
  {journal} {Astrophys. J. Lett.}\ }\textbf {\bibinfo {volume} {930}},\
  \bibinfo {pages} {L16} (\bibinfo {year} {2022}{\natexlab{d}})}\BibitemShut
  {NoStop}%
\bibitem [{\citenamefont {Lara}\ \emph {et~al.}(2021)\citenamefont {Lara},
  \citenamefont {V\"olkel},\ and\ \citenamefont {Barausse}}]{Lara:2021zth}%
  \BibitemOpen
  \bibfield  {author} {\bibinfo {author} {\bibfnamefont {G.}~\bibnamefont
  {Lara}}, \bibinfo {author} {\bibfnamefont {S.~H.}\ \bibnamefont {V\"olkel}},\
  and\ \bibinfo {author} {\bibfnamefont {E.}~\bibnamefont {Barausse}},\
  }\bibfield  {title} {\bibinfo {title} {{Separating astrophysics and geometry
  in black hole images}},\ }\href {https://doi.org/10.1103/PhysRevD.104.124041}
  {\bibfield  {journal} {\bibinfo  {journal} {Phys. Rev. D}\ }\textbf {\bibinfo
  {volume} {104}},\ \bibinfo {pages} {124041} (\bibinfo {year} {2021})},\
  \Eprint {https://arxiv.org/abs/2110.00026} {arXiv:2110.00026 [gr-qc]}
  \BibitemShut {NoStop}%
\bibitem [{\citenamefont {Nampalliwar}\ and\ \citenamefont
  {K}(2021)}]{Nampalliwar:2021oqr}%
  \BibitemOpen
  \bibfield  {author} {\bibinfo {author} {\bibfnamefont {S.}~\bibnamefont
  {Nampalliwar}}\ and\ \bibinfo {author} {\bibfnamefont {S.}~\bibnamefont
  {K}},\ }\bibfield  {title} {\bibinfo {title} {{Theory-agnostic tests of
  gravity with black hole shadows}},\ }\href@noop {} {\  (\bibinfo {year}
  {2021})},\ \Eprint {https://arxiv.org/abs/2108.01190} {arXiv:2108.01190
  [gr-qc]} \BibitemShut {NoStop}%
\bibitem [{\citenamefont {Kocherlakota}\ and\ \citenamefont
  {Rezzolla}(2022)}]{Kocherlakota:2022jnz}%
  \BibitemOpen
  \bibfield  {author} {\bibinfo {author} {\bibfnamefont {P.}~\bibnamefont
  {Kocherlakota}}\ and\ \bibinfo {author} {\bibfnamefont {L.}~\bibnamefont
  {Rezzolla}},\ }\bibfield  {title} {\bibinfo {title} {{Distinguishing
  gravitational and emission physics in black hole imaging: spherical
  symmetry}},\ }\href {https://doi.org/10.1093/mnras/stac891} {\bibfield
  {journal} {\bibinfo  {journal} {Mon. Not. Roy. Astron. Soc.}\ }\textbf
  {\bibinfo {volume} {513}},\ \bibinfo {pages} {1229} (\bibinfo {year}
  {2022})},\ \Eprint {https://arxiv.org/abs/2201.05641} {arXiv:2201.05641
  [gr-qc]} \BibitemShut {NoStop}%
\bibitem [{\citenamefont {Nampalliwar}\ \emph {et~al.}(2022)\citenamefont
  {Nampalliwar}, \citenamefont {Yfantis},\ and\ \citenamefont
  {Kokkotas}}]{Nampalliwar:2022smp}%
  \BibitemOpen
  \bibfield  {author} {\bibinfo {author} {\bibfnamefont {S.}~\bibnamefont
  {Nampalliwar}}, \bibinfo {author} {\bibfnamefont {A.~I.}\ \bibnamefont
  {Yfantis}},\ and\ \bibinfo {author} {\bibfnamefont {K.~D.}\ \bibnamefont
  {Kokkotas}},\ }\bibfield  {title} {\bibinfo {title} {{Extending GRMHD for
  thin disks to non-Kerr spacetimes}},\ }\href
  {https://doi.org/10.1103/PhysRevD.106.063009} {\bibfield  {journal} {\bibinfo
   {journal} {Phys. Rev. D}\ }\textbf {\bibinfo {volume} {106}},\ \bibinfo
  {pages} {063009} (\bibinfo {year} {2022})}\BibitemShut {NoStop}%
\bibitem [{\citenamefont {Kocherlakota}\ \emph {et~al.}(2021)\citenamefont
  {Kocherlakota} \emph {et~al.}}]{EventHorizonTelescope:2021dqv}%
  \BibitemOpen
  \bibfield  {author} {\bibinfo {author} {\bibfnamefont {P.}~\bibnamefont
  {Kocherlakota}} \emph {et~al.} (\bibinfo {collaboration} {Event Horizon
  Telescope}),\ }\bibfield  {title} {\bibinfo {title} {{Constraints on
  black-hole charges with the 2017 EHT observations of M87*}},\ }\href
  {https://doi.org/10.1103/PhysRevD.103.104047} {\bibfield  {journal} {\bibinfo
   {journal} {Phys. Rev. D}\ }\textbf {\bibinfo {volume} {103}},\ \bibinfo
  {pages} {104047} (\bibinfo {year} {2021})},\ \Eprint
  {https://arxiv.org/abs/2105.09343} {arXiv:2105.09343 [gr-qc]} \BibitemShut
  {NoStop}%
\bibitem [{\citenamefont {Vagnozzi}\ \emph {et~al.}(2023)\citenamefont
  {Vagnozzi} \emph {et~al.}}]{Vagnozzi:2022moj}%
  \BibitemOpen
  \bibfield  {author} {\bibinfo {author} {\bibfnamefont {S.}~\bibnamefont
  {Vagnozzi}} \emph {et~al.},\ }\bibfield  {title} {\bibinfo {title}
  {{Horizon-scale tests of gravity theories and fundamental physics from the
  Event Horizon Telescope image of Sagittarius A}},\ }\href
  {https://doi.org/10.1088/1361-6382/acd97b} {\bibfield  {journal} {\bibinfo
  {journal} {Class. Quant. Grav.}\ }\textbf {\bibinfo {volume} {40}},\ \bibinfo
  {pages} {165007} (\bibinfo {year} {2023})},\ \Eprint
  {https://arxiv.org/abs/2205.07787} {arXiv:2205.07787 [gr-qc]} \BibitemShut
  {NoStop}%
\bibitem [{\citenamefont {Salehi}\ \emph {et~al.}(2023)\citenamefont {Salehi},
  \citenamefont {Broderick},\ and\ \citenamefont {Georgiev}}]{Salehi:2023eqy}%
  \BibitemOpen
  \bibfield  {author} {\bibinfo {author} {\bibfnamefont {K.}~\bibnamefont
  {Salehi}}, \bibinfo {author} {\bibfnamefont {A.}~\bibnamefont {Broderick}},\
  and\ \bibinfo {author} {\bibfnamefont {B.}~\bibnamefont {Georgiev}},\
  }\bibfield  {title} {\bibinfo {title} {{Photon Rings and Shadow Size for
  General Integrable Spacetimes}},\ }\href@noop {} {\  (\bibinfo {year}
  {2023})},\ \Eprint {https://arxiv.org/abs/2311.01495} {arXiv:2311.01495
  [gr-qc]} \BibitemShut {NoStop}%
\bibitem [{\citenamefont {C\'ardenas-Avenda\~no}\ \emph
  {et~al.}(2023)\citenamefont {C\'ardenas-Avenda\~no}, \citenamefont
  {Lupsasca},\ and\ \citenamefont {Zhu}}]{Cardenas-Avendano:2022csp}%
  \BibitemOpen
  \bibfield  {author} {\bibinfo {author} {\bibfnamefont {A.}~\bibnamefont
  {C\'ardenas-Avenda\~no}}, \bibinfo {author} {\bibfnamefont {A.}~\bibnamefont
  {Lupsasca}},\ and\ \bibinfo {author} {\bibfnamefont {H.}~\bibnamefont
  {Zhu}},\ }\bibfield  {title} {\bibinfo {title} {{Adaptive analytical ray
  tracing of black hole photon rings}},\ }\href
  {https://doi.org/10.1103/PhysRevD.107.043030} {\bibfield  {journal} {\bibinfo
   {journal} {Phys. Rev. D}\ }\textbf {\bibinfo {volume} {107}},\ \bibinfo
  {pages} {043030} (\bibinfo {year} {2023})},\ \Eprint
  {https://arxiv.org/abs/2211.07469} {arXiv:2211.07469 [gr-qc]} \BibitemShut
  {NoStop}%
\bibitem [{\citenamefont {Wielgus}(2021)}]{Wielgus:2021peu}%
  \BibitemOpen
  \bibfield  {author} {\bibinfo {author} {\bibfnamefont {M.}~\bibnamefont
  {Wielgus}},\ }\bibfield  {title} {\bibinfo {title} {{Photon rings of
  spherically symmetric black holes and robust tests of non-Kerr metrics}},\
  }\href {https://doi.org/10.1103/PhysRevD.104.124058} {\bibfield  {journal}
  {\bibinfo  {journal} {Phys. Rev. D}\ }\textbf {\bibinfo {volume} {104}},\
  \bibinfo {pages} {124058} (\bibinfo {year} {2021})},\ \Eprint
  {https://arxiv.org/abs/2109.10840} {arXiv:2109.10840 [gr-qc]} \BibitemShut
  {NoStop}%
\bibitem [{\citenamefont {Iyer}\ and\ \citenamefont
  {Hansen}(2009)}]{Iyer:2009wa}%
  \BibitemOpen
  \bibfield  {author} {\bibinfo {author} {\bibfnamefont {S.~V.}\ \bibnamefont
  {Iyer}}\ and\ \bibinfo {author} {\bibfnamefont {E.~C.}\ \bibnamefont
  {Hansen}},\ }\bibfield  {title} {\bibinfo {title} {{Light's Bending Angle in
  the Equatorial Plane of a Kerr Black Hole}},\ }\href
  {https://doi.org/10.1103/PhysRevD.80.124023} {\bibfield  {journal} {\bibinfo
  {journal} {Phys. Rev. D}\ }\textbf {\bibinfo {volume} {80}},\ \bibinfo
  {pages} {124023} (\bibinfo {year} {2009})},\ \Eprint
  {https://arxiv.org/abs/0907.5352} {arXiv:0907.5352 [gr-qc]} \BibitemShut
  {NoStop}%
\bibitem [{\citenamefont {Gralla}\ and\ \citenamefont
  {Lupsasca}(2020{\natexlab{c}})}]{Gralla:2019drh}%
  \BibitemOpen
  \bibfield  {author} {\bibinfo {author} {\bibfnamefont {S.~E.}\ \bibnamefont
  {Gralla}}\ and\ \bibinfo {author} {\bibfnamefont {A.}~\bibnamefont
  {Lupsasca}},\ }\bibfield  {title} {\bibinfo {title} {{Lensing by Kerr Black
  Holes}},\ }\href {https://doi.org/10.1103/PhysRevD.101.044031} {\bibfield
  {journal} {\bibinfo  {journal} {Phys. Rev. D}\ }\textbf {\bibinfo {volume}
  {101}},\ \bibinfo {pages} {044031} (\bibinfo {year} {2020}{\natexlab{c}})},\
  \Eprint {https://arxiv.org/abs/1910.12873} {arXiv:1910.12873 [gr-qc]}
  \BibitemShut {NoStop}%
\bibitem [{\citenamefont {Eichhorn}\ and\ \citenamefont
  {Held}(2023)}]{Eichhorn:2022bbn}%
  \BibitemOpen
  \bibfield  {author} {\bibinfo {author} {\bibfnamefont {A.}~\bibnamefont
  {Eichhorn}}\ and\ \bibinfo {author} {\bibfnamefont {A.}~\bibnamefont
  {Held}},\ }\bibfield  {title} {\bibinfo {title} {{Quantum gravity lights up
  spinning black holes}},\ }\href
  {https://doi.org/10.1088/1475-7516/2023/01/032} {\bibfield  {journal}
  {\bibinfo  {journal} {JCAP}\ }\textbf {\bibinfo {volume} {01}},\ \bibinfo
  {pages} {032}},\ \Eprint {https://arxiv.org/abs/2206.11152} {arXiv:2206.11152
  [gr-qc]} \BibitemShut {NoStop}%
\bibitem [{\citenamefont {Eichhorn}\ and\ \citenamefont
  {Held}(2021{\natexlab{a}})}]{Eichhorn:2021iwq}%
  \BibitemOpen
  \bibfield  {author} {\bibinfo {author} {\bibfnamefont {A.}~\bibnamefont
  {Eichhorn}}\ and\ \bibinfo {author} {\bibfnamefont {A.}~\bibnamefont
  {Held}},\ }\bibfield  {title} {\bibinfo {title} {{From a locality-principle
  for new physics to image features of regular spinning black holes with
  disks}},\ }\href {https://doi.org/10.1088/1475-7516/2021/05/073} {\bibfield
  {journal} {\bibinfo  {journal} {JCAP}\ }\textbf {\bibinfo {volume} {05}},\
  \bibinfo {pages} {073}},\ \Eprint {https://arxiv.org/abs/2103.13163}
  {arXiv:2103.13163 [gr-qc]} \BibitemShut {NoStop}%
\bibitem [{\citenamefont {Broderick}\ \emph
  {et~al.}(2020{\natexlab{a}})\citenamefont {Broderick} \emph
  {et~al.}}]{EventHorizonTelescope:2020eky}%
  \BibitemOpen
  \bibfield  {author} {\bibinfo {author} {\bibfnamefont {A.~E.}\ \bibnamefont
  {Broderick}} \emph {et~al.} (\bibinfo {collaboration} {Event Horizon
  Telescope}),\ }\bibfield  {title} {\bibinfo {title} {{THEMIS: A Parameter
  Estimation Framework for the Event Horizon Telescope}},\ }\href
  {https://doi.org/10.3847/1538-4357/ab91a4} {\bibfield  {journal} {\bibinfo
  {journal} {Astrophys. J.}\ }\textbf {\bibinfo {volume} {897}},\ \bibinfo
  {pages} {139} (\bibinfo {year} {2020}{\natexlab{a}})}\BibitemShut {NoStop}%
\bibitem [{\citenamefont {Broderick}\ \emph
  {et~al.}(2020{\natexlab{b}})\citenamefont {Broderick}, \citenamefont {Pesce},
  \citenamefont {Tiede}, \citenamefont {Pu},\ and\ \citenamefont
  {Gold}}]{Broderick:2020wda}%
  \BibitemOpen
  \bibfield  {author} {\bibinfo {author} {\bibfnamefont {A.~E.}\ \bibnamefont
  {Broderick}}, \bibinfo {author} {\bibfnamefont {D.~W.}\ \bibnamefont
  {Pesce}}, \bibinfo {author} {\bibfnamefont {P.}~\bibnamefont {Tiede}},
  \bibinfo {author} {\bibfnamefont {H.-Y.}\ \bibnamefont {Pu}},\ and\ \bibinfo
  {author} {\bibfnamefont {R.}~\bibnamefont {Gold}},\ }\bibfield  {title}
  {\bibinfo {title} {{Hybrid Very Long Baseline Interferometry Imaging and
  Modeling with THEMIS}},\ }\href {https://doi.org/10.3847/1538-4357/ab9c1f}
  {\bibfield  {journal} {\bibinfo  {journal} {Astrophys. J.}\ }\textbf
  {\bibinfo {volume} {898}},\ \bibinfo {pages} {9} (\bibinfo {year}
  {2020}{\natexlab{b}})},\ \Eprint {https://arxiv.org/abs/2208.09003}
  {arXiv:2208.09003 [astro-ph.IM]} \BibitemShut {NoStop}%
\bibitem [{\citenamefont {Tiede}\ \emph {et~al.}(2022)\citenamefont {Tiede},
  \citenamefont {Johnson}, \citenamefont {Pesce}, \citenamefont {Palumbo},
  \citenamefont {Chang},\ and\ \citenamefont {Galison}}]{Tiede:2022grp}%
  \BibitemOpen
  \bibfield  {author} {\bibinfo {author} {\bibfnamefont {P.}~\bibnamefont
  {Tiede}}, \bibinfo {author} {\bibfnamefont {M.~D.}\ \bibnamefont {Johnson}},
  \bibinfo {author} {\bibfnamefont {D.~W.}\ \bibnamefont {Pesce}}, \bibinfo
  {author} {\bibfnamefont {D.~C.~M.}\ \bibnamefont {Palumbo}}, \bibinfo
  {author} {\bibfnamefont {D.~O.}\ \bibnamefont {Chang}},\ and\ \bibinfo
  {author} {\bibfnamefont {P.}~\bibnamefont {Galison}},\ }\bibfield  {title}
  {\bibinfo {title} {{Measuring Photon Rings with the ngEHT}},\ }\href
  {https://doi.org/10.3390/galaxies10060111} {\bibfield  {journal} {\bibinfo
  {journal} {Galaxies}\ }\textbf {\bibinfo {volume} {10}},\ \bibinfo {pages}
  {111} (\bibinfo {year} {2022})},\ \Eprint {https://arxiv.org/abs/2210.13498}
  {arXiv:2210.13498 [astro-ph.HE]} \BibitemShut {NoStop}%
\bibitem [{\citenamefont {Himwich}\ \emph {et~al.}(2020)\citenamefont
  {Himwich}, \citenamefont {Johnson}, \citenamefont {Lupsasca},\ and\
  \citenamefont {Strominger}}]{Himwich:2020msm}%
  \BibitemOpen
  \bibfield  {author} {\bibinfo {author} {\bibfnamefont {E.}~\bibnamefont
  {Himwich}}, \bibinfo {author} {\bibfnamefont {M.~D.}\ \bibnamefont
  {Johnson}}, \bibinfo {author} {\bibfnamefont {A.}~\bibnamefont {Lupsasca}},\
  and\ \bibinfo {author} {\bibfnamefont {A.}~\bibnamefont {Strominger}},\
  }\bibfield  {title} {\bibinfo {title} {{Universal polarimetric signatures of
  the black hole photon ring}},\ }\href
  {https://doi.org/10.1103/PhysRevD.101.084020} {\bibfield  {journal} {\bibinfo
   {journal} {Phys. Rev. D}\ }\textbf {\bibinfo {volume} {101}},\ \bibinfo
  {pages} {084020} (\bibinfo {year} {2020})},\ \Eprint
  {https://arxiv.org/abs/2001.08750} {arXiv:2001.08750 [gr-qc]} \BibitemShut
  {NoStop}%
\bibitem [{\citenamefont {Palumbo}\ \emph {et~al.}(2023)\citenamefont
  {Palumbo}, \citenamefont {Wong}, \citenamefont {Chael},\ and\ \citenamefont
  {Johnson}}]{Palumbo:2023auc}%
  \BibitemOpen
  \bibfield  {author} {\bibinfo {author} {\bibfnamefont {D.~C.~M.}\
  \bibnamefont {Palumbo}}, \bibinfo {author} {\bibfnamefont {G.~N.}\
  \bibnamefont {Wong}}, \bibinfo {author} {\bibfnamefont {A.~A.}\ \bibnamefont
  {Chael}},\ and\ \bibinfo {author} {\bibfnamefont {M.~D.}\ \bibnamefont
  {Johnson}},\ }\bibfield  {title} {\bibinfo {title} {{Demonstrating Photon
  Ring Existence with Single-baseline Polarimetry}},\ }\href
  {https://doi.org/10.3847/2041-8213/ace630} {\bibfield  {journal} {\bibinfo
  {journal} {Astrophys. J. Lett.}\ }\textbf {\bibinfo {volume} {952}},\
  \bibinfo {pages} {L31} (\bibinfo {year} {2023})},\ \Eprint
  {https://arxiv.org/abs/2307.05293} {arXiv:2307.05293 [astro-ph.HE]}
  \BibitemShut {NoStop}%
\bibitem [{\citenamefont {Akiyama}\ \emph {et~al.}(2023)\citenamefont {Akiyama}
  \emph {et~al.}}]{EventHorizonTelescope:2023fox}%
  \BibitemOpen
  \bibfield  {author} {\bibinfo {author} {\bibfnamefont {K.}~\bibnamefont
  {Akiyama}} \emph {et~al.} (\bibinfo {collaboration} {Event Horizon
  Telescope}),\ }\bibfield  {title} {\bibinfo {title} {{First M87 Event Horizon
  Telescope Results. IX. Detection of Near-horizon Circular Polarization}},\
  }\href {https://doi.org/10.3847/2041-8213/acff70} {\bibfield  {journal}
  {\bibinfo  {journal} {Astrophys. J. Lett.}\ }\textbf {\bibinfo {volume}
  {957}},\ \bibinfo {pages} {L20} (\bibinfo {year} {2023})}\BibitemShut
  {NoStop}%
\bibitem [{\citenamefont {Blackburn}\ \emph {et~al.}(2019)\citenamefont
  {Blackburn} \emph {et~al.}}]{Blackburn:2019bly}%
  \BibitemOpen
  \bibfield  {author} {\bibinfo {author} {\bibfnamefont {L.}~\bibnamefont
  {Blackburn}} \emph {et~al.},\ }\bibfield  {title} {\bibinfo {title}
  {{Studying Black Holes on Horizon Scales with VLBI Ground Arrays}},\
  }\href@noop {} {\  (\bibinfo {year} {2019})},\ \Eprint
  {https://arxiv.org/abs/1909.01411} {arXiv:1909.01411 [astro-ph.IM]}
  \BibitemShut {NoStop}%
\bibitem [{\citenamefont {Johnson}\ \emph {et~al.}(2023)\citenamefont {Johnson}
  \emph {et~al.}}]{Johnson:2023ynn}%
  \BibitemOpen
  \bibfield  {author} {\bibinfo {author} {\bibfnamefont {M.~D.}\ \bibnamefont
  {Johnson}} \emph {et~al.},\ }\bibfield  {title} {\bibinfo {title} {{Key
  Science Goals for the Next-Generation Event Horizon Telescope}},\ }\href
  {https://doi.org/10.3390/galaxies11030061} {\bibfield  {journal} {\bibinfo
  {journal} {Galaxies}\ }\textbf {\bibinfo {volume} {11}},\ \bibinfo {pages}
  {61} (\bibinfo {year} {2023})},\ \Eprint {https://arxiv.org/abs/2304.11188}
  {arXiv:2304.11188 [astro-ph.HE]} \BibitemShut {NoStop}%
\bibitem [{\citenamefont {Akiyama}\ \emph
  {et~al.}(2019{\natexlab{b}})\citenamefont {Akiyama} \emph
  {et~al.}}]{EventHorizonTelescope:2019ggy}%
  \BibitemOpen
  \bibfield  {author} {\bibinfo {author} {\bibfnamefont {K.}~\bibnamefont
  {Akiyama}} \emph {et~al.} (\bibinfo {collaboration} {Event Horizon
  Telescope}),\ }\bibfield  {title} {\bibinfo {title} {{First M87 Event Horizon
  Telescope Results. VI. The Shadow and Mass of the Central Black Hole}},\
  }\href {https://doi.org/10.3847/2041-8213/ab1141} {\bibfield  {journal}
  {\bibinfo  {journal} {Astrophys. J. Lett.}\ }\textbf {\bibinfo {volume}
  {875}},\ \bibinfo {pages} {L6} (\bibinfo {year} {2019}{\natexlab{b}})},\
  \Eprint {https://arxiv.org/abs/1906.11243} {arXiv:1906.11243 [astro-ph.GA]}
  \BibitemShut {NoStop}%
\bibitem [{\citenamefont {{Blakeslee}}\ \emph {et~al.}(2009)\citenamefont
  {{Blakeslee}}, \citenamefont {{Jord{\'a}n}}, \citenamefont {{Mei}},
  \citenamefont {{C{\^o}t{\'e}}}, \citenamefont {{Ferrarese}}, \citenamefont
  {{Infante}}, \citenamefont {{Peng}}, \citenamefont {{Tonry}},\ and\
  \citenamefont {{West}}}]{2009ApJ...694..556B}%
  \BibitemOpen
  \bibfield  {author} {\bibinfo {author} {\bibfnamefont {J.~P.}\ \bibnamefont
  {{Blakeslee}}}, \bibinfo {author} {\bibfnamefont {A.}~\bibnamefont
  {{Jord{\'a}n}}}, \bibinfo {author} {\bibfnamefont {S.}~\bibnamefont {{Mei}}},
  \bibinfo {author} {\bibfnamefont {P.}~\bibnamefont {{C{\^o}t{\'e}}}},
  \bibinfo {author} {\bibfnamefont {L.}~\bibnamefont {{Ferrarese}}}, \bibinfo
  {author} {\bibfnamefont {L.}~\bibnamefont {{Infante}}}, \bibinfo {author}
  {\bibfnamefont {E.~W.}\ \bibnamefont {{Peng}}}, \bibinfo {author}
  {\bibfnamefont {J.~L.}\ \bibnamefont {{Tonry}}},\ and\ \bibinfo {author}
  {\bibfnamefont {M.~J.}\ \bibnamefont {{West}}},\ }\bibfield  {title}
  {\bibinfo {title} {{The ACS Fornax Cluster Survey. V. Measurement and
  Recalibration of Surface Brightness Fluctuations and a Precise Value of the
  Fornax-Virgo Relative Distance}},\ }\href
  {https://doi.org/10.1088/0004-637X/694/1/556} {\bibfield  {journal} {\bibinfo
   {journal} {\apj}\ }\textbf {\bibinfo {volume} {694}},\ \bibinfo {pages}
  {556} (\bibinfo {year} {2009})},\ \Eprint {https://arxiv.org/abs/0901.1138}
  {arXiv:0901.1138 [astro-ph.CO]} \BibitemShut {NoStop}%
\bibitem [{\citenamefont {{Bird}}\ \emph {et~al.}(2010)\citenamefont {{Bird}},
  \citenamefont {{Harris}}, \citenamefont {{Blakeslee}},\ and\ \citenamefont
  {{Flynn}}}]{2010A&A...524A..71B}%
  \BibitemOpen
  \bibfield  {author} {\bibinfo {author} {\bibfnamefont {S.}~\bibnamefont
  {{Bird}}}, \bibinfo {author} {\bibfnamefont {W.~E.}\ \bibnamefont
  {{Harris}}}, \bibinfo {author} {\bibfnamefont {J.~P.}\ \bibnamefont
  {{Blakeslee}}},\ and\ \bibinfo {author} {\bibfnamefont {C.}~\bibnamefont
  {{Flynn}}},\ }\bibfield  {title} {\bibinfo {title} {{The inner halo of M 87:
  a first direct view of the red-giant population}},\ }\href
  {https://doi.org/10.1051/0004-6361/201014876} {\bibfield  {journal} {\bibinfo
   {journal} {Astronomy and Astrophysics}\ }\textbf {\bibinfo {volume} {524}},\
  \bibinfo {eid} {A71} (\bibinfo {year} {2010})},\ \Eprint
  {https://arxiv.org/abs/1009.3202} {arXiv:1009.3202 [astro-ph.GA]}
  \BibitemShut {NoStop}%
\bibitem [{\citenamefont {{Cantiello}}\ \emph {et~al.}(2018)\citenamefont
  {{Cantiello}}, \citenamefont {{Blakeslee}}, \citenamefont {{Ferrarese}},
  \citenamefont {{C{\^o}t{\'e}}}, \citenamefont {{Roediger}}, \citenamefont
  {{Raimondo}}, \citenamefont {{Peng}}, \citenamefont {{Gwyn}}, \citenamefont
  {{Durrell}},\ and\ \citenamefont {{Cuillandre}}}]{2018ApJ...856..126C}%
  \BibitemOpen
  \bibfield  {author} {\bibinfo {author} {\bibfnamefont {M.}~\bibnamefont
  {{Cantiello}}}, \bibinfo {author} {\bibfnamefont {J.~P.}\ \bibnamefont
  {{Blakeslee}}}, \bibinfo {author} {\bibfnamefont {L.}~\bibnamefont
  {{Ferrarese}}}, \bibinfo {author} {\bibfnamefont {P.}~\bibnamefont
  {{C{\^o}t{\'e}}}}, \bibinfo {author} {\bibfnamefont {J.~C.}\ \bibnamefont
  {{Roediger}}}, \bibinfo {author} {\bibfnamefont {G.}~\bibnamefont
  {{Raimondo}}}, \bibinfo {author} {\bibfnamefont {E.~W.}\ \bibnamefont
  {{Peng}}}, \bibinfo {author} {\bibfnamefont {S.}~\bibnamefont {{Gwyn}}},
  \bibinfo {author} {\bibfnamefont {P.~R.}\ \bibnamefont {{Durrell}}},\ and\
  \bibinfo {author} {\bibfnamefont {J.-C.}\ \bibnamefont {{Cuillandre}}},\
  }\bibfield  {title} {\bibinfo {title} {{The Next Generation Virgo Cluster
  Survey (NGVS). XVIII. Measurement and Calibration of Surface Brightness
  Fluctuation Distances for Bright Galaxies in Virgo (and Beyond)}},\ }\href
  {https://doi.org/10.3847/1538-4357/aab043} {\bibfield  {journal} {\bibinfo
  {journal} {\apj}\ }\textbf {\bibinfo {volume} {856}},\ \bibinfo {eid} {126}
  (\bibinfo {year} {2018})},\ \Eprint {https://arxiv.org/abs/1802.05526}
  {arXiv:1802.05526 [astro-ph.GA]} \BibitemShut {NoStop}%
\bibitem [{\citenamefont {Papoutsis}\ \emph {et~al.}(2023)\citenamefont
  {Papoutsis}, \citenamefont {Baub\"ock}, \citenamefont {Chang},\ and\
  \citenamefont {Gammie}}]{Papoutsis:2022kzp}%
  \BibitemOpen
  \bibfield  {author} {\bibinfo {author} {\bibfnamefont {E.}~\bibnamefont
  {Papoutsis}}, \bibinfo {author} {\bibfnamefont {M.}~\bibnamefont
  {Baub\"ock}}, \bibinfo {author} {\bibfnamefont {D.}~\bibnamefont {Chang}},\
  and\ \bibinfo {author} {\bibfnamefont {C.~F.}\ \bibnamefont {Gammie}},\
  }\bibfield  {title} {\bibinfo {title} {{Jets and Rings in Images of Spinning
  Black Holes}},\ }\href {https://doi.org/10.3847/1538-4357/acafe3} {\bibfield
  {journal} {\bibinfo  {journal} {Astrophys. J.}\ }\textbf {\bibinfo {volume}
  {944}},\ \bibinfo {pages} {55} (\bibinfo {year} {2023})},\ \Eprint
  {https://arxiv.org/abs/2212.06281} {arXiv:2212.06281 [astro-ph.HE]}
  \BibitemShut {NoStop}%
\bibitem [{\citenamefont {Younsi}\ \emph {et~al.}(2016)\citenamefont {Younsi},
  \citenamefont {Zhidenko}, \citenamefont {Rezzolla}, \citenamefont
  {Konoplya},\ and\ \citenamefont {Mizuno}}]{Younsi:2016azx}%
  \BibitemOpen
  \bibfield  {author} {\bibinfo {author} {\bibfnamefont {Z.}~\bibnamefont
  {Younsi}}, \bibinfo {author} {\bibfnamefont {A.}~\bibnamefont {Zhidenko}},
  \bibinfo {author} {\bibfnamefont {L.}~\bibnamefont {Rezzolla}}, \bibinfo
  {author} {\bibfnamefont {R.}~\bibnamefont {Konoplya}},\ and\ \bibinfo
  {author} {\bibfnamefont {Y.}~\bibnamefont {Mizuno}},\ }\bibfield  {title}
  {\bibinfo {title} {{New method for shadow calculations: Application to
  parametrized axisymmetric black holes}},\ }\href
  {https://doi.org/10.1103/PhysRevD.94.084025} {\bibfield  {journal} {\bibinfo
  {journal} {Phys. Rev. D}\ }\textbf {\bibinfo {volume} {94}},\ \bibinfo
  {pages} {084025} (\bibinfo {year} {2016})},\ \Eprint
  {https://arxiv.org/abs/1607.05767} {arXiv:1607.05767 [gr-qc]} \BibitemShut
  {NoStop}%
\bibitem [{\citenamefont {Vincent}\ \emph {et~al.}(2011)\citenamefont
  {Vincent}, \citenamefont {Paumard}, \citenamefont {Gourgoulhon},\ and\
  \citenamefont {Perrin}}]{Vincent:2011wz}%
  \BibitemOpen
  \bibfield  {author} {\bibinfo {author} {\bibfnamefont {F.~H.}\ \bibnamefont
  {Vincent}}, \bibinfo {author} {\bibfnamefont {T.}~\bibnamefont {Paumard}},
  \bibinfo {author} {\bibfnamefont {E.}~\bibnamefont {Gourgoulhon}},\ and\
  \bibinfo {author} {\bibfnamefont {G.}~\bibnamefont {Perrin}},\ }\bibfield
  {title} {\bibinfo {title} {{GYOTO: a new general relativistic ray-tracing
  code}},\ }\href {https://doi.org/10.1088/0264-9381/28/22/225011} {\bibfield
  {journal} {\bibinfo  {journal} {Class. Quant. Grav.}\ }\textbf {\bibinfo
  {volume} {28}},\ \bibinfo {pages} {225011} (\bibinfo {year} {2011})},\
  \Eprint {https://arxiv.org/abs/1109.4769} {arXiv:1109.4769 [gr-qc]}
  \BibitemShut {NoStop}%
\bibitem [{\citenamefont {Sharma}\ \emph {et~al.}(2023)\citenamefont {Sharma},
  \citenamefont {Medeiros}, \citenamefont {Chan}, \citenamefont {Halevi},
  \citenamefont {Mullen}, \citenamefont {Stone},\ and\ \citenamefont
  {Wong}}]{Sharma:2023nbk}%
  \BibitemOpen
  \bibfield  {author} {\bibinfo {author} {\bibfnamefont {A.}~\bibnamefont
  {Sharma}}, \bibinfo {author} {\bibfnamefont {L.}~\bibnamefont {Medeiros}},
  \bibinfo {author} {\bibfnamefont {C.-k.}\ \bibnamefont {Chan}}, \bibinfo
  {author} {\bibfnamefont {G.}~\bibnamefont {Halevi}}, \bibinfo {author}
  {\bibfnamefont {P.~D.}\ \bibnamefont {Mullen}}, \bibinfo {author}
  {\bibfnamefont {J.~M.}\ \bibnamefont {Stone}},\ and\ \bibinfo {author}
  {\bibfnamefont {G.~N.}\ \bibnamefont {Wong}},\ }\bibfield  {title} {\bibinfo
  {title} {{Mahakala: a Python-based Modular Ray-tracing and Radiative Transfer
  Algorithm for Curved Space-times}},\ }\href@noop {} {\  (\bibinfo {year}
  {2023})},\ \Eprint {https://arxiv.org/abs/2304.03804} {arXiv:2304.03804
  [astro-ph.HE]} \BibitemShut {NoStop}%
\bibitem [{\citenamefont {Bozzola}\ \emph {et~al.}(2023)\citenamefont
  {Bozzola}, \citenamefont {Chan},\ and\ \citenamefont
  {Paschalidis}}]{Bozzola:2023daz}%
  \BibitemOpen
  \bibfield  {author} {\bibinfo {author} {\bibfnamefont {G.}~\bibnamefont
  {Bozzola}}, \bibinfo {author} {\bibfnamefont {C.-k.}\ \bibnamefont {Chan}},\
  and\ \bibinfo {author} {\bibfnamefont {V.}~\bibnamefont {Paschalidis}},\
  }\bibfield  {title} {\bibinfo {title} {{Not all spacetime coordinates for
  general-relativistic ray tracing are created equal}},\ }\href
  {https://doi.org/10.1103/PhysRevD.108.084004} {\bibfield  {journal} {\bibinfo
   {journal} {Phys. Rev. D}\ }\textbf {\bibinfo {volume} {108}},\ \bibinfo
  {pages} {084004} (\bibinfo {year} {2023})},\ \Eprint
  {https://arxiv.org/abs/2310.02321} {arXiv:2310.02321 [gr-qc]} \BibitemShut
  {NoStop}%
\bibitem [{\citenamefont {Tahura}\ \emph {et~al.}(2023)\citenamefont {Tahura},
  \citenamefont {Khalvati},\ and\ \citenamefont {Yang}}]{Tahura:2023qqt}%
  \BibitemOpen
  \bibfield  {author} {\bibinfo {author} {\bibfnamefont {S.}~\bibnamefont
  {Tahura}}, \bibinfo {author} {\bibfnamefont {H.}~\bibnamefont {Khalvati}},\
  and\ \bibinfo {author} {\bibfnamefont {H.}~\bibnamefont {Yang}},\ }\bibfield
  {title} {\bibinfo {title} {{Vacuum Spacetime With Multipole Moments: The
  Minimal Size Conjecture, Black Hole Shadow, and Gravitational Wave
  Observables}},\ }\href@noop {} {\  (\bibinfo {year} {2023})},\ \Eprint
  {https://arxiv.org/abs/2309.11491} {arXiv:2309.11491 [gr-qc]} \BibitemShut
  {NoStop}%
\bibitem [{\citenamefont {Collins}\ and\ \citenamefont
  {Hughes}(2004)}]{Collins:2004ex}%
  \BibitemOpen
  \bibfield  {author} {\bibinfo {author} {\bibfnamefont {N.~A.}\ \bibnamefont
  {Collins}}\ and\ \bibinfo {author} {\bibfnamefont {S.~A.}\ \bibnamefont
  {Hughes}},\ }\bibfield  {title} {\bibinfo {title} {{Towards a formalism for
  mapping the space-times of massive compact objects: Bumpy black holes and
  their orbits}},\ }\href {https://doi.org/10.1103/PhysRevD.69.124022}
  {\bibfield  {journal} {\bibinfo  {journal} {Phys. Rev. D}\ }\textbf {\bibinfo
  {volume} {69}},\ \bibinfo {pages} {124022} (\bibinfo {year} {2004})},\
  \Eprint {https://arxiv.org/abs/gr-qc/0402063} {arXiv:gr-qc/0402063}
  \BibitemShut {NoStop}%
\bibitem [{\citenamefont {{Foreman-Mackey}}\ \emph {et~al.}(2013)\citenamefont
  {{Foreman-Mackey}}, \citenamefont {{Hogg}}, \citenamefont {{Lang}},\ and\
  \citenamefont {{Goodman}}}]{emcee}%
  \BibitemOpen
  \bibfield  {author} {\bibinfo {author} {\bibfnamefont {D.}~\bibnamefont
  {{Foreman-Mackey}}}, \bibinfo {author} {\bibfnamefont {D.~W.}\ \bibnamefont
  {{Hogg}}}, \bibinfo {author} {\bibfnamefont {D.}~\bibnamefont {{Lang}}},\
  and\ \bibinfo {author} {\bibfnamefont {J.}~\bibnamefont {{Goodman}}},\
  }\bibfield  {title} {\bibinfo {title} {{emcee: The MCMC Hammer}},\ }\href
  {https://doi.org/10.1086/670067} {\bibfield  {journal} {\bibinfo  {journal}
  {Publications of the ASP}\ }\textbf {\bibinfo {volume} {125}},\ \bibinfo
  {pages} {306} (\bibinfo {year} {2013})},\ \Eprint
  {https://arxiv.org/abs/1202.3665} {arXiv:1202.3665 [astro-ph.IM]}
  \BibitemShut {NoStop}%
\bibitem [{\citenamefont {Broderick}\ \emph
  {et~al.}(2022{\natexlab{b}})\citenamefont {Broderick}, \citenamefont {Tiede},
  \citenamefont {Pesce},\ and\ \citenamefont {Gold}}]{Broderick:2021ohx}%
  \BibitemOpen
  \bibfield  {author} {\bibinfo {author} {\bibfnamefont {A.~E.}\ \bibnamefont
  {Broderick}}, \bibinfo {author} {\bibfnamefont {P.}~\bibnamefont {Tiede}},
  \bibinfo {author} {\bibfnamefont {D.~W.}\ \bibnamefont {Pesce}},\ and\
  \bibinfo {author} {\bibfnamefont {R.}~\bibnamefont {Gold}},\ }\bibfield
  {title} {\bibinfo {title} {{Measuring Spin from Relative Photon-ring
  Sizes}},\ }\href {https://doi.org/10.3847/1538-4357/ac4970} {\bibfield
  {journal} {\bibinfo  {journal} {Astrophys. J.}\ }\textbf {\bibinfo {volume}
  {927}},\ \bibinfo {pages} {6} (\bibinfo {year} {2022}{\natexlab{b}})},\
  \Eprint {https://arxiv.org/abs/2105.09962} {arXiv:2105.09962 [astro-ph.HE]}
  \BibitemShut {NoStop}%
\bibitem [{\citenamefont {Berti}\ \emph {et~al.}(2015)\citenamefont {Berti}
  \emph {et~al.}}]{Berti:2015itd}%
  \BibitemOpen
  \bibfield  {author} {\bibinfo {author} {\bibfnamefont {E.}~\bibnamefont
  {Berti}} \emph {et~al.},\ }\bibfield  {title} {\bibinfo {title} {{Testing
  General Relativity with Present and Future Astrophysical Observations}},\
  }\href {https://doi.org/10.1088/0264-9381/32/24/243001} {\bibfield  {journal}
  {\bibinfo  {journal} {Class. Quant. Grav.}\ }\textbf {\bibinfo {volume}
  {32}},\ \bibinfo {pages} {243001} (\bibinfo {year} {2015})},\ \Eprint
  {https://arxiv.org/abs/1501.07274} {arXiv:1501.07274 [gr-qc]} \BibitemShut
  {NoStop}%
\bibitem [{\citenamefont {Yunes}\ and\ \citenamefont
  {Pretorius}(2009)}]{Yunes:2009ke}%
  \BibitemOpen
  \bibfield  {author} {\bibinfo {author} {\bibfnamefont {N.}~\bibnamefont
  {Yunes}}\ and\ \bibinfo {author} {\bibfnamefont {F.}~\bibnamefont
  {Pretorius}},\ }\bibfield  {title} {\bibinfo {title} {{Fundamental
  Theoretical Bias in Gravitational Wave Astrophysics and the Parameterized
  Post-Einsteinian Framework}},\ }\href
  {https://doi.org/10.1103/PhysRevD.80.122003} {\bibfield  {journal} {\bibinfo
  {journal} {Phys. Rev. D}\ }\textbf {\bibinfo {volume} {80}},\ \bibinfo
  {pages} {122003} (\bibinfo {year} {2009})},\ \Eprint
  {https://arxiv.org/abs/0909.3328} {arXiv:0909.3328 [gr-qc]} \BibitemShut
  {NoStop}%
\bibitem [{\citenamefont {Baibhav}\ \emph {et~al.}(2023)\citenamefont
  {Baibhav}, \citenamefont {Cheung}, \citenamefont {Berti}, \citenamefont
  {Cardoso}, \citenamefont {Carullo}, \citenamefont {Cotesta}, \citenamefont
  {Del~Pozzo},\ and\ \citenamefont {Duque}}]{Baibhav:2023clw}%
  \BibitemOpen
  \bibfield  {author} {\bibinfo {author} {\bibfnamefont {V.}~\bibnamefont
  {Baibhav}}, \bibinfo {author} {\bibfnamefont {M.~H.-Y.}\ \bibnamefont
  {Cheung}}, \bibinfo {author} {\bibfnamefont {E.}~\bibnamefont {Berti}},
  \bibinfo {author} {\bibfnamefont {V.}~\bibnamefont {Cardoso}}, \bibinfo
  {author} {\bibfnamefont {G.}~\bibnamefont {Carullo}}, \bibinfo {author}
  {\bibfnamefont {R.}~\bibnamefont {Cotesta}}, \bibinfo {author} {\bibfnamefont
  {W.}~\bibnamefont {Del~Pozzo}},\ and\ \bibinfo {author} {\bibfnamefont
  {F.}~\bibnamefont {Duque}},\ }\bibfield  {title} {\bibinfo {title} {{Agnostic
  black hole spectroscopy: quasinormal mode content of numerical relativity
  waveforms and limits of validity of linear perturbation theory}},\
  }\href@noop {} {\  (\bibinfo {year} {2023})},\ \Eprint
  {https://arxiv.org/abs/2302.03050} {arXiv:2302.03050 [gr-qc]} \BibitemShut
  {NoStop}%
\bibitem [{\citenamefont {Carter}(1970)}]{Carter:1970ea}%
  \BibitemOpen
  \bibfield  {author} {\bibinfo {author} {\bibfnamefont {B.}~\bibnamefont
  {Carter}},\ }\bibfield  {title} {\bibinfo {title} {{The commutation property
  of a stationary, axisymmetric system}},\ }\href
  {https://doi.org/10.1007/BF01647092} {\bibfield  {journal} {\bibinfo
  {journal} {Commun. Math. Phys.}\ }\textbf {\bibinfo {volume} {17}},\ \bibinfo
  {pages} {233} (\bibinfo {year} {1970})}\BibitemShut {NoStop}%
\bibitem [{\citenamefont {{Gourgoulhon}}\ and\ \citenamefont
  {{Bonazzola}}(1993)}]{1993PhRvD..48.2635G}%
  \BibitemOpen
  \bibfield  {author} {\bibinfo {author} {\bibfnamefont {E.}~\bibnamefont
  {{Gourgoulhon}}}\ and\ \bibinfo {author} {\bibfnamefont {S.}~\bibnamefont
  {{Bonazzola}}},\ }\bibfield  {title} {\bibinfo {title} {{Noncircular
  axisymmetric stationary spacetimes}},\ }\href
  {https://doi.org/10.1103/PhysRevD.48.2635} {\bibfield  {journal} {\bibinfo
  {journal} {\prd}\ }\textbf {\bibinfo {volume} {48}},\ \bibinfo {pages} {2635}
  (\bibinfo {year} {1993})}\BibitemShut {NoStop}%
\bibitem [{\citenamefont {Bondi}\ \emph {et~al.}(1962)\citenamefont {Bondi},
  \citenamefont {van~der Burg},\ and\ \citenamefont {Metzner}}]{Bondi:1962px}%
  \BibitemOpen
  \bibfield  {author} {\bibinfo {author} {\bibfnamefont {H.}~\bibnamefont
  {Bondi}}, \bibinfo {author} {\bibfnamefont {M.~G.~J.}\ \bibnamefont {van~der
  Burg}},\ and\ \bibinfo {author} {\bibfnamefont {A.~W.~K.}\ \bibnamefont
  {Metzner}},\ }\bibfield  {title} {\bibinfo {title} {{Gravitational waves in
  general relativity. 7. Waves from axisymmetric isolated systems}},\ }\href
  {https://doi.org/10.1098/rspa.1962.0161} {\bibfield  {journal} {\bibinfo
  {journal} {Proc. Roy. Soc. Lond. A}\ }\textbf {\bibinfo {volume} {269}},\
  \bibinfo {pages} {21} (\bibinfo {year} {1962})}\BibitemShut {NoStop}%
\bibitem [{\citenamefont {Sachs}(1962)}]{Sachs:1962wk}%
  \BibitemOpen
  \bibfield  {author} {\bibinfo {author} {\bibfnamefont {R.~K.}\ \bibnamefont
  {Sachs}},\ }\bibfield  {title} {\bibinfo {title} {{Gravitational waves in
  general relativity. 8. Waves in asymptotically flat space-times}},\ }\href
  {https://doi.org/10.1098/rspa.1962.0206} {\bibfield  {journal} {\bibinfo
  {journal} {Proc. Roy. Soc. Lond. A}\ }\textbf {\bibinfo {volume} {270}},\
  \bibinfo {pages} {103} (\bibinfo {year} {1962})}\BibitemShut {NoStop}%
\bibitem [{\citenamefont {Ay\'on-Beato}\ and\ \citenamefont
  {Vel\'azquez-Rodr\'\i{}guez}(2016)}]{Ayon-Beato:2015xsz}%
  \BibitemOpen
  \bibfield  {author} {\bibinfo {author} {\bibfnamefont {E.}~\bibnamefont
  {Ay\'on-Beato}}\ and\ \bibinfo {author} {\bibfnamefont {G.}~\bibnamefont
  {Vel\'azquez-Rodr\'\i{}guez}},\ }\bibfield  {title} {\bibinfo {title}
  {{Residual symmetries of the gravitational field}},\ }\href
  {https://doi.org/10.1103/PhysRevD.93.044040} {\bibfield  {journal} {\bibinfo
  {journal} {Phys. Rev. D}\ }\textbf {\bibinfo {volume} {93}},\ \bibinfo
  {pages} {044040} (\bibinfo {year} {2016})},\ \bibinfo {note} {[Addendum:
  Phys.Rev.D 96, 049904 (2017)]},\ \Eprint {https://arxiv.org/abs/1511.07461}
  {arXiv:1511.07461 [gr-qc]} \BibitemShut {NoStop}%
\bibitem [{\citenamefont {Delaporte}\ \emph {et~al.}(2022)\citenamefont
  {Delaporte}, \citenamefont {Eichhorn},\ and\ \citenamefont
  {Held}}]{Delaporte:2022acp}%
  \BibitemOpen
  \bibfield  {author} {\bibinfo {author} {\bibfnamefont {H.}~\bibnamefont
  {Delaporte}}, \bibinfo {author} {\bibfnamefont {A.}~\bibnamefont
  {Eichhorn}},\ and\ \bibinfo {author} {\bibfnamefont {A.}~\bibnamefont
  {Held}},\ }\bibfield  {title} {\bibinfo {title} {{Parameterizations of
  black-hole spacetimes beyond circularity}},\ }\href
  {https://doi.org/10.1088/1361-6382/ac7027} {\bibfield  {journal} {\bibinfo
  {journal} {Class. Quant. Grav.}\ }\textbf {\bibinfo {volume} {39}},\ \bibinfo
  {pages} {134002} (\bibinfo {year} {2022})},\ \Eprint
  {https://arxiv.org/abs/2203.00105} {arXiv:2203.00105 [gr-qc]} \BibitemShut
  {NoStop}%
\bibitem [{\citenamefont {Papapetrou}(1966)}]{Papapetrou:1966zz}%
  \BibitemOpen
  \bibfield  {author} {\bibinfo {author} {\bibfnamefont {A.}~\bibnamefont
  {Papapetrou}},\ }\bibfield  {title} {\bibinfo {title} {{Champs
  gravitationnels stationnaires a symetrie axiale}},\ }\href@noop {} {\bibfield
   {journal} {\bibinfo  {journal} {Ann. Inst. H. Poincare Phys. Theor.}\
  }\textbf {\bibinfo {volume} {4}},\ \bibinfo {pages} {83} (\bibinfo {year}
  {1966})}\BibitemShut {NoStop}%
\bibitem [{\citenamefont {Kundt}\ and\ \citenamefont
  {Trumper}(1966)}]{Kundt:1966zz}%
  \BibitemOpen
  \bibfield  {author} {\bibinfo {author} {\bibfnamefont {W.}~\bibnamefont
  {Kundt}}\ and\ \bibinfo {author} {\bibfnamefont {M.}~\bibnamefont
  {Trumper}},\ }\bibfield  {title} {\bibinfo {title} {{Orthogonal decomposition
  of axi-symmetric stationary spacetimes}},\ }\href
  {https://doi.org/10.1007/BF01325677} {\bibfield  {journal} {\bibinfo
  {journal} {Z. Phys.}\ }\textbf {\bibinfo {volume} {192}},\ \bibinfo {pages}
  {419} (\bibinfo {year} {1966})}\BibitemShut {NoStop}%
\bibitem [{\citenamefont {Wald}(1984)}]{Wald:1984rg}%
  \BibitemOpen
  \bibfield  {author} {\bibinfo {author} {\bibfnamefont {R.~M.}\ \bibnamefont
  {Wald}},\ }\href {https://doi.org/10.7208/chicago/9780226870373.001.0001}
  {\emph {\bibinfo {title} {{General Relativity}}}}\ (\bibinfo  {publisher}
  {Chicago Univ. Pr.},\ \bibinfo {address} {Chicago, USA},\ \bibinfo {year}
  {1984})\BibitemShut {NoStop}%
\bibitem [{\citenamefont {Weyl}(1917)}]{Weyl:1917gp}%
  \BibitemOpen
  \bibfield  {author} {\bibinfo {author} {\bibfnamefont {H.}~\bibnamefont
  {Weyl}},\ }\bibfield  {title} {\bibinfo {title} {{The theory of
  gravitation}},\ }\href {https://doi.org/10.1007/s10714-011-1310-7} {\bibfield
   {journal} {\bibinfo  {journal} {Annalen Phys.}\ }\textbf {\bibinfo {volume}
  {54}},\ \bibinfo {pages} {117} (\bibinfo {year} {1917})}\BibitemShut
  {NoStop}%
\bibitem [{\citenamefont {Carter}(1968)}]{Carter:1968rr}%
  \BibitemOpen
  \bibfield  {author} {\bibinfo {author} {\bibfnamefont {B.}~\bibnamefont
  {Carter}},\ }\bibfield  {title} {\bibinfo {title} {{Global structure of the
  Kerr family of gravitational fields}},\ }\href
  {https://doi.org/10.1103/PhysRev.174.1559} {\bibfield  {journal} {\bibinfo
  {journal} {Phys. Rev.}\ }\textbf {\bibinfo {volume} {174}},\ \bibinfo {pages}
  {1559} (\bibinfo {year} {1968})}\BibitemShut {NoStop}%
\bibitem [{\citenamefont {Benenti}\ and\ \citenamefont
  {Francaviglia}(1979)}]{Benenti:1979erw}%
  \BibitemOpen
  \bibfield  {author} {\bibinfo {author} {\bibfnamefont {S.}~\bibnamefont
  {Benenti}}\ and\ \bibinfo {author} {\bibfnamefont {M.}~\bibnamefont
  {Francaviglia}},\ }\bibfield  {title} {\bibinfo {title} {{Remarks on certain
  separability structures and their applications to general relativity}},\
  }\href {https://doi.org/10.1007/bf00757025} {\bibfield  {journal} {\bibinfo
  {journal} {Gen. Rel. Grav.}\ }\textbf {\bibinfo {volume} {10}},\ \bibinfo
  {pages} {79} (\bibinfo {year} {1979})}\BibitemShut {NoStop}%
\bibitem [{\citenamefont {Chen}(2020)}]{Chen:2020aix}%
  \BibitemOpen
  \bibfield  {author} {\bibinfo {author} {\bibfnamefont {C.-Y.}\ \bibnamefont
  {Chen}},\ }\bibfield  {title} {\bibinfo {title} {{Rotating black holes
  without $\mathbb{Z}_2$ symmetry and their shadow images}},\ }\href
  {https://doi.org/10.1088/1475-7516/2020/05/040} {\bibfield  {journal}
  {\bibinfo  {journal} {JCAP}\ }\textbf {\bibinfo {volume} {05}},\ \bibinfo
  {pages} {040}},\ \Eprint {https://arxiv.org/abs/2004.01440} {arXiv:2004.01440
  [gr-qc]} \BibitemShut {NoStop}%
\bibitem [{\citenamefont {Vigeland}(2010)}]{Vigeland:2010xe}%
  \BibitemOpen
  \bibfield  {author} {\bibinfo {author} {\bibfnamefont {S.~J.}\ \bibnamefont
  {Vigeland}},\ }\bibfield  {title} {\bibinfo {title} {{Multipole moments of
  bumpy black holes}},\ }\href {https://doi.org/10.1103/PhysRevD.82.104041}
  {\bibfield  {journal} {\bibinfo  {journal} {Phys. Rev. D}\ }\textbf {\bibinfo
  {volume} {82}},\ \bibinfo {pages} {104041} (\bibinfo {year} {2010})},\
  \Eprint {https://arxiv.org/abs/1008.1278} {arXiv:1008.1278 [gr-qc]}
  \BibitemShut {NoStop}%
\bibitem [{\citenamefont {Newman}\ and\ \citenamefont
  {Janis}(1965)}]{Newman:1965tw}%
  \BibitemOpen
  \bibfield  {author} {\bibinfo {author} {\bibfnamefont {E.~T.}\ \bibnamefont
  {Newman}}\ and\ \bibinfo {author} {\bibfnamefont {A.~I.}\ \bibnamefont
  {Janis}},\ }\bibfield  {title} {\bibinfo {title} {{Note on the Kerr spinning
  particle metric}},\ }\href {https://doi.org/10.1063/1.1704350} {\bibfield
  {journal} {\bibinfo  {journal} {J. Math. Phys.}\ }\textbf {\bibinfo {volume}
  {6}},\ \bibinfo {pages} {915} (\bibinfo {year} {1965})}\BibitemShut {NoStop}%
\bibitem [{\citenamefont {Vigeland}\ and\ \citenamefont
  {Hughes}(2010)}]{Vigeland:2009pr}%
  \BibitemOpen
  \bibfield  {author} {\bibinfo {author} {\bibfnamefont {S.~J.}\ \bibnamefont
  {Vigeland}}\ and\ \bibinfo {author} {\bibfnamefont {S.~A.}\ \bibnamefont
  {Hughes}},\ }\bibfield  {title} {\bibinfo {title} {{Spacetime and orbits of
  bumpy black holes}},\ }\href {https://doi.org/10.1103/PhysRevD.81.024030}
  {\bibfield  {journal} {\bibinfo  {journal} {Phys. Rev. D}\ }\textbf {\bibinfo
  {volume} {81}},\ \bibinfo {pages} {024030} (\bibinfo {year} {2010})},\
  \Eprint {https://arxiv.org/abs/0911.1756} {arXiv:0911.1756 [gr-qc]}
  \BibitemShut {NoStop}%
\bibitem [{\citenamefont {Vigeland}\ \emph {et~al.}(2011)\citenamefont
  {Vigeland}, \citenamefont {Yunes},\ and\ \citenamefont
  {Stein}}]{Vigeland:2011ji}%
  \BibitemOpen
  \bibfield  {author} {\bibinfo {author} {\bibfnamefont {S.}~\bibnamefont
  {Vigeland}}, \bibinfo {author} {\bibfnamefont {N.}~\bibnamefont {Yunes}},\
  and\ \bibinfo {author} {\bibfnamefont {L.}~\bibnamefont {Stein}},\ }\bibfield
   {title} {\bibinfo {title} {{Bumpy Black Holes in Alternate Theories of
  Gravity}},\ }\href {https://doi.org/10.1103/PhysRevD.83.104027} {\bibfield
  {journal} {\bibinfo  {journal} {Phys. Rev. D}\ }\textbf {\bibinfo {volume}
  {83}},\ \bibinfo {pages} {104027} (\bibinfo {year} {2011})},\ \Eprint
  {https://arxiv.org/abs/1102.3706} {arXiv:1102.3706 [gr-qc]} \BibitemShut
  {NoStop}%
\bibitem [{\citenamefont {Johannsen}\ and\ \citenamefont
  {Psaltis}(2011)}]{Johannsen:2011dh}%
  \BibitemOpen
  \bibfield  {author} {\bibinfo {author} {\bibfnamefont {T.}~\bibnamefont
  {Johannsen}}\ and\ \bibinfo {author} {\bibfnamefont {D.}~\bibnamefont
  {Psaltis}},\ }\bibfield  {title} {\bibinfo {title} {{A Metric for Rapidly
  Spinning Black Holes Suitable for Strong-Field Tests of the No-Hair
  Theorem}},\ }\href {https://doi.org/10.1103/PhysRevD.83.124015} {\bibfield
  {journal} {\bibinfo  {journal} {Phys. Rev. D}\ }\textbf {\bibinfo {volume}
  {83}},\ \bibinfo {pages} {124015} (\bibinfo {year} {2011})},\ \Eprint
  {https://arxiv.org/abs/1105.3191} {arXiv:1105.3191 [gr-qc]} \BibitemShut
  {NoStop}%
\bibitem [{\citenamefont {Johannsen}(2013)}]{Johannsen:2013szh}%
  \BibitemOpen
  \bibfield  {author} {\bibinfo {author} {\bibfnamefont {T.}~\bibnamefont
  {Johannsen}},\ }\bibfield  {title} {\bibinfo {title} {{Regular Black Hole
  Metric with Three Constants of Motion}},\ }\href
  {https://doi.org/10.1103/PhysRevD.88.044002} {\bibfield  {journal} {\bibinfo
  {journal} {Phys. Rev. D}\ }\textbf {\bibinfo {volume} {88}},\ \bibinfo
  {pages} {044002} (\bibinfo {year} {2013})},\ \Eprint
  {https://arxiv.org/abs/1501.02809} {arXiv:1501.02809 [gr-qc]} \BibitemShut
  {NoStop}%
\bibitem [{\citenamefont {Carson}\ and\ \citenamefont
  {Yagi}(2020)}]{Carson:2020dez}%
  \BibitemOpen
  \bibfield  {author} {\bibinfo {author} {\bibfnamefont {Z.}~\bibnamefont
  {Carson}}\ and\ \bibinfo {author} {\bibfnamefont {K.}~\bibnamefont {Yagi}},\
  }\bibfield  {title} {\bibinfo {title} {{Asymptotically flat, parameterized
  black hole metric preserving Kerr symmetries}},\ }\href
  {https://doi.org/10.1103/PhysRevD.101.084030} {\bibfield  {journal} {\bibinfo
   {journal} {Phys. Rev. D}\ }\textbf {\bibinfo {volume} {101}},\ \bibinfo
  {pages} {084030} (\bibinfo {year} {2020})},\ \Eprint
  {https://arxiv.org/abs/2002.01028} {arXiv:2002.01028 [gr-qc]} \BibitemShut
  {NoStop}%
\bibitem [{\citenamefont {Yagi}\ \emph {et~al.}(2023)\citenamefont {Yagi},
  \citenamefont {Lomuscio}, \citenamefont {Lowrey},\ and\ \citenamefont
  {Carson}}]{Yagi:2023eap}%
  \BibitemOpen
  \bibfield  {author} {\bibinfo {author} {\bibfnamefont {K.}~\bibnamefont
  {Yagi}}, \bibinfo {author} {\bibfnamefont {S.}~\bibnamefont {Lomuscio}},
  \bibinfo {author} {\bibfnamefont {T.}~\bibnamefont {Lowrey}},\ and\ \bibinfo
  {author} {\bibfnamefont {Z.}~\bibnamefont {Carson}},\ }\bibfield  {title}
  {\bibinfo {title} {{Regularizing Parameterized Black Hole Spacetimes with
  Kerr Symmetries}},\ }\href@noop {} {\  (\bibinfo {year} {2023})},\ \Eprint
  {https://arxiv.org/abs/2311.08659} {arXiv:2311.08659 [gr-qc]} \BibitemShut
  {NoStop}%
\bibitem [{\citenamefont {Rezzolla}\ and\ \citenamefont
  {Zhidenko}(2014)}]{Rezzolla:2014mua}%
  \BibitemOpen
  \bibfield  {author} {\bibinfo {author} {\bibfnamefont {L.}~\bibnamefont
  {Rezzolla}}\ and\ \bibinfo {author} {\bibfnamefont {A.}~\bibnamefont
  {Zhidenko}},\ }\bibfield  {title} {\bibinfo {title} {{New parametrization for
  spherically symmetric black holes in metric theories of gravity}},\ }\href
  {https://doi.org/10.1103/PhysRevD.90.084009} {\bibfield  {journal} {\bibinfo
  {journal} {Phys. Rev. D}\ }\textbf {\bibinfo {volume} {90}},\ \bibinfo
  {pages} {084009} (\bibinfo {year} {2014})},\ \Eprint
  {https://arxiv.org/abs/1407.3086} {arXiv:1407.3086 [gr-qc]} \BibitemShut
  {NoStop}%
\bibitem [{\citenamefont {Konoplya}\ \emph {et~al.}(2016)\citenamefont
  {Konoplya}, \citenamefont {Rezzolla},\ and\ \citenamefont
  {Zhidenko}}]{Konoplya:2016jvv}%
  \BibitemOpen
  \bibfield  {author} {\bibinfo {author} {\bibfnamefont {R.}~\bibnamefont
  {Konoplya}}, \bibinfo {author} {\bibfnamefont {L.}~\bibnamefont {Rezzolla}},\
  and\ \bibinfo {author} {\bibfnamefont {A.}~\bibnamefont {Zhidenko}},\
  }\bibfield  {title} {\bibinfo {title} {{General parametrization of
  axisymmetric black holes in metric theories of gravity}},\ }\href
  {https://doi.org/10.1103/PhysRevD.93.064015} {\bibfield  {journal} {\bibinfo
  {journal} {Phys. Rev. D}\ }\textbf {\bibinfo {volume} {93}},\ \bibinfo
  {pages} {064015} (\bibinfo {year} {2016})},\ \Eprint
  {https://arxiv.org/abs/1602.02378} {arXiv:1602.02378 [gr-qc]} \BibitemShut
  {NoStop}%
\bibitem [{\citenamefont {Mertens}\ \emph {et~al.}(2016)\citenamefont
  {Mertens}, \citenamefont {Lobanov}, \citenamefont {Walker},\ and\
  \citenamefont {Hardee}}]{mertens2016kinematics}%
  \BibitemOpen
  \bibfield  {author} {\bibinfo {author} {\bibfnamefont {F.}~\bibnamefont
  {Mertens}}, \bibinfo {author} {\bibfnamefont {A.}~\bibnamefont {Lobanov}},
  \bibinfo {author} {\bibfnamefont {R.}~\bibnamefont {Walker}},\ and\ \bibinfo
  {author} {\bibfnamefont {P.}~\bibnamefont {Hardee}},\ }\bibfield  {title}
  {\bibinfo {title} {Kinematics of the jet in \rm{M}87 on scales of 100--1000
  schwarzschild radii},\ }\href@noop {} {\bibfield  {journal} {\bibinfo
  {journal} {Astronomy \& Astrophysics}\ }\textbf {\bibinfo {volume} {595}},\
  \bibinfo {pages} {A54} (\bibinfo {year} {2016})}\BibitemShut {NoStop}%
\bibitem [{\citenamefont {{Walker}}\ \emph {et~al.}(2018)\citenamefont
  {{Walker}}, \citenamefont {{Hardee}}, \citenamefont {{Davies}}, \citenamefont
  {{Ly}},\ and\ \citenamefont {{Junor}}}]{2018ApJ...855..128W}%
  \BibitemOpen
  \bibfield  {author} {\bibinfo {author} {\bibfnamefont {R.~C.}\ \bibnamefont
  {{Walker}}}, \bibinfo {author} {\bibfnamefont {P.~E.}\ \bibnamefont
  {{Hardee}}}, \bibinfo {author} {\bibfnamefont {F.~B.}\ \bibnamefont
  {{Davies}}}, \bibinfo {author} {\bibfnamefont {C.}~\bibnamefont {{Ly}}},\
  and\ \bibinfo {author} {\bibfnamefont {W.}~\bibnamefont {{Junor}}},\
  }\bibfield  {title} {\bibinfo {title} {{The Structure and Dynamics of the
  Subparsec Jet in M87 Based on 50 VLBA Observations over 17 Years at 43
  GHz}},\ }\href {https://doi.org/10.3847/1538-4357/aaafcc} {\bibfield
  {journal} {\bibinfo  {journal} {\apj}\ }\textbf {\bibinfo {volume} {855}},\
  \bibinfo {eid} {128} (\bibinfo {year} {2018})},\ \Eprint
  {https://arxiv.org/abs/1802.06166} {arXiv:1802.06166 [astro-ph.HE]}
  \BibitemShut {NoStop}%
\bibitem [{\citenamefont {Broderick}\ \emph {et~al.}(2023)\citenamefont
  {Broderick}, \citenamefont {Salehi},\ and\ \citenamefont
  {Georgiev}}]{Broderick:2023jfl}%
  \BibitemOpen
  \bibfield  {author} {\bibinfo {author} {\bibfnamefont {A.~E.}\ \bibnamefont
  {Broderick}}, \bibinfo {author} {\bibfnamefont {K.}~\bibnamefont {Salehi}},\
  and\ \bibinfo {author} {\bibfnamefont {B.}~\bibnamefont {Georgiev}},\
  }\bibfield  {title} {\bibinfo {title} {{Shadow Implications: What Does
  Measuring the Photon Ring Imply for Gravity?}},\ }\href
  {https://doi.org/10.3847/1538-4357/acf9f6} {\bibfield  {journal} {\bibinfo
  {journal} {Astrophys. J.}\ }\textbf {\bibinfo {volume} {958}},\ \bibinfo
  {pages} {114} (\bibinfo {year} {2023})},\ \Eprint
  {https://arxiv.org/abs/2307.15120} {arXiv:2307.15120 [astro-ph.HE]}
  \BibitemShut {NoStop}%
\bibitem [{\citenamefont {Cardenas-Avendano}\ \emph {et~al.}(2020)\citenamefont
  {Cardenas-Avendano}, \citenamefont {Nampalliwar},\ and\ \citenamefont
  {Yunes}}]{Cardenas-Avendano:2019zxd}%
  \BibitemOpen
  \bibfield  {author} {\bibinfo {author} {\bibfnamefont {A.}~\bibnamefont
  {Cardenas-Avendano}}, \bibinfo {author} {\bibfnamefont {S.}~\bibnamefont
  {Nampalliwar}},\ and\ \bibinfo {author} {\bibfnamefont {N.}~\bibnamefont
  {Yunes}},\ }\bibfield  {title} {\bibinfo {title} {{Gravitational-wave versus
  X-ray tests of strong-field gravity}},\ }\href
  {https://doi.org/10.1088/1361-6382/ab8f64} {\bibfield  {journal} {\bibinfo
  {journal} {Class. Quant. Grav.}\ }\textbf {\bibinfo {volume} {37}},\ \bibinfo
  {pages} {135008} (\bibinfo {year} {2020})},\ \Eprint
  {https://arxiv.org/abs/1912.08062} {arXiv:1912.08062 [gr-qc]} \BibitemShut
  {NoStop}%
\bibitem [{\citenamefont {Broderick}\ and\ \citenamefont
  {Loeb}(2006)}]{Broderick:2005jj}%
  \BibitemOpen
  \bibfield  {author} {\bibinfo {author} {\bibfnamefont {A.~E.}\ \bibnamefont
  {Broderick}}\ and\ \bibinfo {author} {\bibfnamefont {A.}~\bibnamefont
  {Loeb}},\ }\bibfield  {title} {\bibinfo {title} {{Imaging optically-thin hot
  spots near the black hole horizon of Sgr A* at radio and near-infrared
  wavelengths}},\ }\href {https://doi.org/10.1111/j.1365-2966.2006.10152.x}
  {\bibfield  {journal} {\bibinfo  {journal} {Mon. Not. Roy. Astron. Soc.}\
  }\textbf {\bibinfo {volume} {367}},\ \bibinfo {pages} {905} (\bibinfo {year}
  {2006})},\ \Eprint {https://arxiv.org/abs/astro-ph/0509237}
  {arXiv:astro-ph/0509237} \BibitemShut {NoStop}%
\bibitem [{\citenamefont {{Tiede}}\ \emph {et~al.}(2020)\citenamefont
  {{Tiede}}, \citenamefont {{Pu}}, \citenamefont {{Broderick}}, \citenamefont
  {{Gold}}, \citenamefont {{Karami}},\ and\ \citenamefont
  {{Preciado-L{\'o}pez}}}]{2020ApJ...892..132T}%
  \BibitemOpen
  \bibfield  {author} {\bibinfo {author} {\bibfnamefont {P.}~\bibnamefont
  {{Tiede}}}, \bibinfo {author} {\bibfnamefont {H.-Y.}\ \bibnamefont {{Pu}}},
  \bibinfo {author} {\bibfnamefont {A.~E.}\ \bibnamefont {{Broderick}}},
  \bibinfo {author} {\bibfnamefont {R.}~\bibnamefont {{Gold}}}, \bibinfo
  {author} {\bibfnamefont {M.}~\bibnamefont {{Karami}}},\ and\ \bibinfo
  {author} {\bibfnamefont {J.~A.}\ \bibnamefont {{Preciado-L{\'o}pez}}},\
  }\bibfield  {title} {\bibinfo {title} {{Spacetime Tomography Using the Event
  Horizon Telescope}},\ }\href {https://doi.org/10.3847/1538-4357/ab744c}
  {\bibfield  {journal} {\bibinfo  {journal} {\apj}\ }\textbf {\bibinfo
  {volume} {892}},\ \bibinfo {eid} {132} (\bibinfo {year} {2020})},\ \Eprint
  {https://arxiv.org/abs/2002.05735} {arXiv:2002.05735 [astro-ph.HE]}
  \BibitemShut {NoStop}%
\bibitem [{\citenamefont {Eichhorn}\ and\ \citenamefont
  {Held}(2021{\natexlab{b}})}]{Eichhorn:2021etc}%
  \BibitemOpen
  \bibfield  {author} {\bibinfo {author} {\bibfnamefont {A.}~\bibnamefont
  {Eichhorn}}\ and\ \bibinfo {author} {\bibfnamefont {A.}~\bibnamefont
  {Held}},\ }\bibfield  {title} {\bibinfo {title} {{Image features of spinning
  regular black holes based on a locality principle}},\ }\href
  {https://doi.org/10.1140/epjc/s10052-021-09716-2} {\bibfield  {journal}
  {\bibinfo  {journal} {Eur. Phys. J. C}\ }\textbf {\bibinfo {volume} {81}},\
  \bibinfo {pages} {933} (\bibinfo {year} {2021}{\natexlab{b}})},\ \Eprint
  {https://arxiv.org/abs/2103.07473} {arXiv:2103.07473 [gr-qc]} \BibitemShut
  {NoStop}%
\bibitem [{\citenamefont {Eichhorn}\ \emph
  {et~al.}(2023{\natexlab{b}})\citenamefont {Eichhorn}, \citenamefont {Held},\
  and\ \citenamefont {Johannsen}}]{Eichhorn:2022oma}%
  \BibitemOpen
  \bibfield  {author} {\bibinfo {author} {\bibfnamefont {A.}~\bibnamefont
  {Eichhorn}}, \bibinfo {author} {\bibfnamefont {A.}~\bibnamefont {Held}},\
  and\ \bibinfo {author} {\bibfnamefont {P.-V.}\ \bibnamefont {Johannsen}},\
  }\bibfield  {title} {\bibinfo {title} {{Universal signatures of
  singularity-resolving physics in photon rings of black holes and horizonless
  objects}},\ }\href {https://doi.org/10.1088/1475-7516/2023/01/043} {\bibfield
   {journal} {\bibinfo  {journal} {JCAP}\ }\textbf {\bibinfo {volume} {01}},\
  \bibinfo {pages} {043}},\ \Eprint {https://arxiv.org/abs/2204.02429}
  {arXiv:2204.02429 [gr-qc]} \BibitemShut {NoStop}%
\bibitem [{\citenamefont {Saurabh}\ and\ \citenamefont
  {Nampalliwar}(2023)}]{Saurabh:2022oho}%
  \BibitemOpen
  \bibfield  {author} {\bibinfo {author} {\bibnamefont {Saurabh}}\ and\
  \bibinfo {author} {\bibfnamefont {S.}~\bibnamefont {Nampalliwar}},\
  }\bibfield  {title} {\bibinfo {title} {{GALLIFRAY\textemdash{}A Geometric
  Modeling and Parameter Estimation Framework for Black Hole Images Using
  Bayesian Techniques}},\ }\href {https://doi.org/10.3847/1538-4357/acc6d3}
  {\bibfield  {journal} {\bibinfo  {journal} {Astrophys. J.}\ }\textbf
  {\bibinfo {volume} {947}},\ \bibinfo {pages} {43} (\bibinfo {year} {2023})},\
  \Eprint {https://arxiv.org/abs/2212.06827} {arXiv:2212.06827 [astro-ph.IM]}
  \BibitemShut {NoStop}%
\bibitem [{\citenamefont {Harris}\ \emph {et~al.}(2020)\citenamefont {Harris}
  \emph {et~al.}}]{Harris:2020xlr}%
  \BibitemOpen
  \bibfield  {author} {\bibinfo {author} {\bibfnamefont {C.~R.}\ \bibnamefont
  {Harris}} \emph {et~al.},\ }\bibfield  {title} {\bibinfo {title} {{Array
  programming with NumPy}},\ }\href {https://doi.org/10.1038/s41586-020-2649-2}
  {\bibfield  {journal} {\bibinfo  {journal} {Nature}\ }\textbf {\bibinfo
  {volume} {585}},\ \bibinfo {pages} {357} (\bibinfo {year} {2020})},\ \Eprint
  {https://arxiv.org/abs/2006.10256} {arXiv:2006.10256 [cs.MS]} \BibitemShut
  {NoStop}%
\bibitem [{\citenamefont {Virtanen}\ \emph {et~al.}(2020)\citenamefont
  {Virtanen} \emph {et~al.}}]{Virtanen:2019joe}%
  \BibitemOpen
  \bibfield  {author} {\bibinfo {author} {\bibfnamefont {P.}~\bibnamefont
  {Virtanen}} \emph {et~al.},\ }\bibfield  {title} {\bibinfo {title} {{SciPy
  1.0--Fundamental Algorithms for Scientific Computing in Python}},\ }\href
  {https://doi.org/10.1038/s41592-019-0686-2} {\bibfield  {journal} {\bibinfo
  {journal} {Nature Meth.}\ }\textbf {\bibinfo {volume} {17}},\ \bibinfo
  {pages} {261} (\bibinfo {year} {2020})},\ \Eprint
  {https://arxiv.org/abs/1907.10121} {arXiv:1907.10121 [cs.MS]} \BibitemShut
  {NoStop}%
\bibitem [{\citenamefont {Hunter}(2007)}]{Hunter:2007}%
  \BibitemOpen
  \bibfield  {author} {\bibinfo {author} {\bibfnamefont {J.~D.}\ \bibnamefont
  {Hunter}},\ }\bibfield  {title} {\bibinfo {title} {Matplotlib: A 2d graphics
  environment},\ }\href {https://doi.org/10.1109/MCSE.2007.55} {\bibfield
  {journal} {\bibinfo  {journal} {Computing in Science \& Engineering}\
  }\textbf {\bibinfo {volume} {9}},\ \bibinfo {pages} {90} (\bibinfo {year}
  {2007})}\BibitemShut {NoStop}%
\bibitem [{\citenamefont {Nampalliwar}\ \emph {et~al.}(2020)\citenamefont
  {Nampalliwar}, \citenamefont {Xin}, \citenamefont {Srivastava}, \citenamefont
  {Abdikamalov}, \citenamefont {Ayzenberg}, \citenamefont {Bambi},
  \citenamefont {Dauser}, \citenamefont {Garcia},\ and\ \citenamefont
  {Tripathi}}]{Nampalliwar:2019iti}%
  \BibitemOpen
  \bibfield  {author} {\bibinfo {author} {\bibfnamefont {S.}~\bibnamefont
  {Nampalliwar}}, \bibinfo {author} {\bibfnamefont {S.}~\bibnamefont {Xin}},
  \bibinfo {author} {\bibfnamefont {S.}~\bibnamefont {Srivastava}}, \bibinfo
  {author} {\bibfnamefont {A.~B.}\ \bibnamefont {Abdikamalov}}, \bibinfo
  {author} {\bibfnamefont {D.}~\bibnamefont {Ayzenberg}}, \bibinfo {author}
  {\bibfnamefont {C.}~\bibnamefont {Bambi}}, \bibinfo {author} {\bibfnamefont
  {T.}~\bibnamefont {Dauser}}, \bibinfo {author} {\bibfnamefont {J.~A.}\
  \bibnamefont {Garcia}},\ and\ \bibinfo {author} {\bibfnamefont
  {A.}~\bibnamefont {Tripathi}},\ }\bibfield  {title} {\bibinfo {title}
  {{Testing General Relativity with X-ray reflection spectroscopy: The
  Konoplya-Rezzolla-Zhidenko parametrization}},\ }\href
  {https://doi.org/10.1103/PhysRevD.102.124071} {\bibfield  {journal} {\bibinfo
   {journal} {Phys. Rev. D}\ }\textbf {\bibinfo {volume} {102}},\ \bibinfo
  {pages} {124071} (\bibinfo {year} {2020})},\ \Eprint
  {https://arxiv.org/abs/1903.12119} {arXiv:1903.12119 [gr-qc]} \BibitemShut
  {NoStop}%
\bibitem [{\citenamefont {Ni}\ \emph {et~al.}(2016)\citenamefont {Ni},
  \citenamefont {Jiang},\ and\ \citenamefont {Bambi}}]{Ni:2016uik}%
  \BibitemOpen
  \bibfield  {author} {\bibinfo {author} {\bibfnamefont {Y.}~\bibnamefont
  {Ni}}, \bibinfo {author} {\bibfnamefont {J.}~\bibnamefont {Jiang}},\ and\
  \bibinfo {author} {\bibfnamefont {C.}~\bibnamefont {Bambi}},\ }\bibfield
  {title} {\bibinfo {title} {{Testing the Kerr metric with the iron line and
  the KRZ parametrization}},\ }\href
  {https://doi.org/10.1088/1475-7516/2016/09/014} {\bibfield  {journal}
  {\bibinfo  {journal} {JCAP}\ }\textbf {\bibinfo {volume} {09}},\ \bibinfo
  {pages} {014}},\ \Eprint {https://arxiv.org/abs/1607.04893} {arXiv:1607.04893
  [gr-qc]} \BibitemShut {NoStop}%
\bibitem [{\citenamefont {Brandt}\ and\ \citenamefont
  {Seidel}(1996)}]{Brandt:1996si}%
  \BibitemOpen
  \bibfield  {author} {\bibinfo {author} {\bibfnamefont {S.~R.}\ \bibnamefont
  {Brandt}}\ and\ \bibinfo {author} {\bibfnamefont {E.}~\bibnamefont
  {Seidel}},\ }\bibfield  {title} {\bibinfo {title} {{The Evolution of
  distorted rotating black holes. 3: Initial data}},\ }\href
  {https://doi.org/10.1103/PhysRevD.54.1403} {\bibfield  {journal} {\bibinfo
  {journal} {Phys. Rev. D}\ }\textbf {\bibinfo {volume} {54}},\ \bibinfo
  {pages} {1403} (\bibinfo {year} {1996})},\ \Eprint
  {https://arxiv.org/abs/gr-qc/9601010} {arXiv:gr-qc/9601010} \BibitemShut
  {NoStop}%
\bibitem [{\citenamefont {Will}\ and\ \citenamefont
  {Nordtvedt}(1972)}]{Will:1972zz}%
  \BibitemOpen
  \bibfield  {author} {\bibinfo {author} {\bibfnamefont {C.~M.}\ \bibnamefont
  {Will}}\ and\ \bibinfo {author} {\bibfnamefont {K.}~\bibnamefont {Nordtvedt},
  \bibfnamefont {Jr.}},\ }\bibfield  {title} {\bibinfo {title} {{Conservation
  Laws and Preferred Frames in Relativistic Gravity. I. Preferred-Frame
  Theories and an Extended PPN Formalism}},\ }\href
  {https://doi.org/10.1086/151754} {\bibfield  {journal} {\bibinfo  {journal}
  {Astrophys. J.}\ }\textbf {\bibinfo {volume} {177}},\ \bibinfo {pages} {757}
  (\bibinfo {year} {1972})}\BibitemShut {NoStop}%
\bibitem [{\citenamefont {Visser}(2007)}]{Visser:2007fj}%
  \BibitemOpen
  \bibfield  {author} {\bibinfo {author} {\bibfnamefont {M.}~\bibnamefont
  {Visser}},\ }\bibfield  {title} {\bibinfo {title} {{The Kerr spacetime: A
  Brief introduction}},\ }in\ \href@noop {} {\emph {\bibinfo {booktitle} {{Kerr
  Fest: Black Holes in Astrophysics, General Relativity and Quantum
  Gravity}}}}\ (\bibinfo {year} {2007})\ \Eprint
  {https://arxiv.org/abs/0706.0622} {arXiv:0706.0622 [gr-qc]} \BibitemShut
  {NoStop}%
\end{thebibliography}%

\end{document}